\documentclass[%
 amsmath,amssymb,
 aps,
]{revtex4-1}
\usepackage{graphicx}
\usepackage{dcolumn}
\usepackage{bm}
\usepackage{epstopdf}

\begin{document}

\preprint{APS/123-QED}

\title{Investigation of Nuclear Phase Transition by Solvababe supersymmetric algebraic model and its application in Ru-Rh and Zn-Cu Isotopes
}

\author{ M. A. Jafarizadeh}
\email{jafarizadeh@tabrizu.ac.ir}
\affiliation{Department of Theoretical Physics and Astrophysics,
University of Tabriz, Tabriz 51664, Iran.}
\affiliation{Research Institute for Fundamental Sciences, Tabriz 51664, Iran}
\author{ M.Ghapanvari}
\email{M.ghapanvari@tabrizu.ac.ir}
\affiliation{Department of Nuclear Physics, University of Tabriz, Tabriz 51664, Iran.}
\affiliation{The Plasma Physics and Fusion Research School, Tehran, Iran.}

\author{ N.Fouladi}
\email{fooladi@tabrizu.ac.ir}
\affiliation{Department of Nuclear Physics, University of Tabriz, Tabriz 51664, Iran.}

\author{Z.Ranjbar}
\email{Z.ranjbar@tabrizu.ac.ir}
\affiliation{Department of Nuclear Physics, University of Tabriz, Tabriz 51664, Iran.}

\author{ A.Sadighzadeh}
\affiliation{The Plasma Physics and Fusion Research School, Tehran, Iran.}
 \email{A.Sadighzadeh@aeoi.ir}

\date{\today}

\begin{abstract}
Solvable supersymmetric algebraic model for descriptions of the spherical to $ \gamma-unstable$ shape-phase transition in even and odd mass nuclei is proposed. This model is based on dual algebraic structure and Richardson - Gaudin method, where the duality relations between the unitary and quasispin algebraic structures for the boson and fermion systems are extended to mixed boson-fermion system.
The structure of two type of nuclear supersymmetry schemes, based on the U(6/2) and U(6/4) supergroups, is discussed. We investigate the change in level structure induced by the phase transition by doing a quantal analysis. By using the generalized quasispin algebra, it is shown that the nuclear supersymmetry concept can be also used for transitional regions in addition to dynamical symmetry limits. Experimental evidence for the U(5)-O(6) transition in Ru-Rh and Zn-Cu supermultiplets is presented. The low-states energy spectra and B(E2)values for these nuclei have been calculated and compared with the experimental data.
\end{abstract}

\maketitle
\section{Introduction}
Symmetry is one of the fundamental concepts in physics. In the nuclei, symmetry is in the form of dynamical symmetry. The dynamical symmetries were introduced in nuclear physics with the development of the interacting boson model and its extensions\cite{1,2}.
The IBM describes the collective  excitations  of
even-even  nuclei in terms of correlated pairs of nucleons with L = 0, 2 treated as bosons (s, d bosons)\cite{1}. The IBM represents a simple description of nuclear collective excitations based of an algebraic theory where symmetry transformations play a significant role.  The  N-boson system space is spanned by the irrep $ \mathrm{[N]} $ of $ \mathrm{U^{B}(6)} $ \cite{1}. The interacting boson - fermion model also explains odd-A nuclei in terms of correlated pairs,  s and d bosons , and unpaired particles of angular momentum j (j fermions)\cite{2}. The states of the boson - fermion system can be classified according to the irreducible representation  $ \mathrm{[N] \times [1]} $ of $ \mathrm{U^{B}(6) \times U^{F}(M)} $ where M is the dimension of the single particle space. The IBM and IBFM can be unified into a supersymmetry (SUSY) approach that was discovered into nuclear structure physics in the early 1980 \cite{3,4}.The experiments performed at various laboratories have confirmed
the predictions made using a SUSY scheme \cite{5}
. Originally, nuclear supersymmetry was considered as symmetry among pairs of nuclei consisting of an even-even and an odd-even nucleus\cite{3,4}. The supersymmetric representations $[\mathcal{N}\}$, $\mathcal{N}=\mathrm{N_{B}+N_{F}} $, of $\mathrm{U(6/ M)}$
spanned a space that explains the lowest states of an even-even nucleus with $\mathcal{N}$ bosons and an odd-A nucleus with $\mathcal{N}-1$ bosons and a unpaired fermion\cite{2}. In supersymmetry approach, the core-particle interaction entirely are determined by the even-even Hamiltonian and most of the parameters in the core-particle interaction can be taken from the neighboring even-even nucleus thus the nuclear properties of neighboring even-even and odd-A nuclei can be described with a single Hamiltonian and a single set of transition and transfer operators\cite{2,6}. The nuclear supersymmetry have been used successfully in description dynamical symmetry limits of the even-even and odd-A nuclei\cite{2,3,4,7}. The IBFM could be an ideal ground for testing supersymmetry\cite{7}. Virtually simultaneously,  with the introduction of the nuclear supersymmetry, the idea of spherical-deformed phase transitions at low energy in finite nuclei germinated\cite{8,9}. Studies of QPTs in odd-even nuclei with supersymmetric scheme had implicitly been initiated years before by A. Frank et al.\cite{6}. They have studied successfully a combination of $\mathrm{U^{BF}(5)}$ and $\mathrm{SO^{BF}(6)}$ symmetry by using $\mathrm{U(6/12)}$ supersymmetry for the Ru and Rh isotopes. Iachello \cite{10}  extended the concept of critical symmetry to critical supersymmetry and provided a benchmark for the study of shape phase transition in odd-even nuclei and also J.Jolie et al.\cite{11}  studied QPTs in odd-even nuclei using a supersymmetric approach in interacting Boson-Fermion model.

In this paper, concept of supersymmetry and phase transitions  are brought together by using the generalized quasi-spin algebra and Richardson - Gaudin method.  Exactly solvable solution for the spherical to gamma - unstable transition in transitional nuclei
based on dual algebraic structure and nuclear supersymmetry concept is proposed. Two separate cases  are considered with the
fermions lying in a j=1/2 shell or in a j=3/2 shell. The existence of duality symmetries has proven to be a
powerful tool in relating  the Hamiltonians with the number-conserving unitary and number-nonconserving quasispin algebras for system with pairing interactions that first introduced in nuclear physics in Ref.\cite{12}
and further developed in Refs.\cite{13,14}. These
relations were obtained for both bosonic and fermionic systems\cite{13,14}. We have established the duality relations for
mixed boson-fermion system. We evaluate exact solutions for eigenstate and energy eigenvalues for transitional region and dynamical-symmetry limits for even-even and odd-A nuclei in Supersymmetry scheme by using Richardson - Gaudin method and changing the control parameters that based on affine $ \mathrm{\widehat{GQA }} $ generalized quasi-spin algebra. In order to investigation of phase transition,we calculate observables such as level crossing, expectation values of the d-boson number operator and expectation values of the fermion number operator. The low-lying states of even-mass ruthenium and the odd-mass rhodium isotopes and Zn-Cu supermultiplets have been studied within suggested model. The results of calculations for these nuclei will present for energy levels and transitions probabilities, two neutron separation energies and will compare with the corresponding the experimental data.

This paper is organized as follows: Section 2 briefly summarizes theoretical aspects of transitional Hamiltonian and generalized quasi-spin algebraic technique. Sections 3 and 4 include the quantal analysis and experimental evidence and sect. 5 is devoted to the summary and some conclusions.

\section{Theoretical framework}
\subsection{The Model}
The quasispin algebras have been explained in detail in Refs \cite{12,13,14}.
The  SU(1,1) algebra is produced by $ \mathrm{S^{\nu}}$,
 $ \mathrm{{\nu=0}} $ and $\mathrm{{\pm}} $ , which satisfies the following
commutation relations\cite{12,13,14}
\begin{equation}
\label{1}
 [S_{B}^{0},S_{B}^{ \pm }]={\pm}S_{B}^{\pm}      \quad     ,    \,\,\,    [S_{B}^{+},S_{B}^{-} ]=-2S_{B}^{0}
\end{equation}
In IBM , the generators of $ \mathrm{SU^{d} (1,1)}$ generated by the
d-boson pairing algebra\cite{12}
\begin{equation}
\label{2}
S^{+}(d)=\frac{1}{2} (d^{+}.d^{+} ) \quad\quad,S^{-}(d)=\frac{1}{2}
(\widetilde{d}.\widetilde{d} ), \quad\quad S^{0}(d)=\frac{1}{4}{\sum_{\nu}
({d_{\nu}^{+}d_{\nu}+d_{\nu}d_{\nu}^{+}}})=\frac{1}{4}(2\hat{n } _{d}+5)
\end{equation}
Similarly, s- boson pairing algebra forms another $ \mathrm{SU^{s} (1,1)}$
algebra generated by \cite{12}
\begin{equation}
\label{3}
 S^{+}(s)=\frac{1}{2} s^{+^2}  \quad , \quad S^{-}(s)=\frac{1}{2} s^{2}
    ,\quad S^{0}(s)=\frac{1}{4} (s^{+} s+ss^{+} )=\frac{1}{4}(2\hat{n } _{s}+1)
\end{equation}
$ \mathrm{SU^{sd} (1,1)}$is the s and d boson pairing algebras generated by\cite{12}
\begin{equation}
\label{4}
 S^{+}(sd)=\frac{1}{2} (d^{+}.d^{+}\pm s^{+^2} ) \quad ,
 \quad S^{-}(sd)=\frac{1}{2} (\widetilde{d}.\widetilde{d} \pm s^{2})
 \quad , S^{0}(sd)=\frac{1}{4}{\sum_{\nu}({d_{\nu}^{+}d_{\nu}+d_{\nu}d_{\nu}^{+}})}
  +\frac{1}{4} (s^{+} s+ss^{+} )
  \end{equation}
The fermionic quasispin operators are\cite{13,14}
\begin{equation}
\label{5}
S_{F}^{+}=\frac{1}{2} \sum\limits_{m'}(-1)^{j\mp m'} a_{jm'}^{+} a_{j-m'}^{+}=\frac{1}{2}(a_{j}^{+} \cdot a_{j}^{+})
\end{equation}
\begin{equation}
\label{6}
S_{F}^{-}=\frac{1}{2} \sum\limits_{m'}(-1)^{j\mp m'} \widetilde{a}_{jm'} \widetilde{a}_{j-m'}=\frac{1}{2}(\widetilde{a}_{j} \cdot \widetilde{a}_{j})
\end{equation}
\begin{equation}
\label{7}
S_{F}^{0}=\frac {n_{f}}{2}-\frac{2j+1}{4}
\end{equation}
satisfy the commutation relations of the fermion quasispin SU(2) algebra\cite{13,14}.
\begin{equation}
\label{8}
 [S_{F}^{0},S_{F}^{ \pm }]={\pm}S_{F}^{\pm}      \quad     ,    \,\,\,    [S_{F}^{+},S_{F}^{-} ]=2S_{F}^{0}
\end{equation}
The generalized quasispin algebra contain both bosonic (B) and fermionic (F) operators defined as \cite{15}
\begin{equation}
\label{9}
S_{BF}^{0}=S_{B}^{0}+S_{F}^{0}  , \quad S_{BF}^{+}=S_{B}^{+}+S_{F}^{+}, \quad S_{BF}^{-}=S_{B}^{-}+S_{F}^{-}
\end{equation}
Thus
\begin{equation}
\label{10}
S_{BF}^{0}=\frac{1}{4}(N-M)+\frac{1}{2}(n_{b}+n_{f})
\end{equation}
\begin{eqnarray}
\label{11}
S_{BF}^{+}=\frac{1}{2} \sum\limits_{m}(-1)^{2\mp m}b_{m}^{+}b_{-m}^{+}\mp \frac{1}{2} \sum\limits_{m'}(-1)^{j\mp
 m'}a_{jm'}^{+}a_{j-m'}^{+}=\frac{1}{2}(b^{+}\cdot b^{+} )\mp (a_{j}^{+} \cdot a_{j}^{+} )
\end{eqnarray}
\begin{equation}
\label{12}
S_{BF}^{-}=\frac{1}{2} ( \widetilde{b}.\widetilde{b}  ) \pm \frac{1}{2} ( \widetilde{a}_{j}.\widetilde{a}_{j}  )
\end{equation}
Where $\mathrm{n_{b}}$ and $\mathrm{n_{f}}$ are the boson and fermion number operator, respectively. Thus the operators $\mathrm{(S_{BF}^{\pm}}$,$\mathrm{S_{BF}^{0})}$ form a generalization of the usual fermionic and bosonic quasispin algebras with commutation relations given as\cite{15}
\begin{equation}
\label{13}
[S_{BF}^{0},S_{BF}^{\pm} ]=\pm S_{BF}^{\pm}   \quad                   ,  \quad  [S_{BF}^{+},S_{BF}^{-} ]=-2S_{BF}^{0}
\end{equation}
The generalized quasispin algebra(GQA) may be labeled in terms of the eigenvalues of the second-order Casimir invariant and those of the quasispin operator $\mathrm{S_{BF}^{0}}$ \cite{15}. The second-order Casimir invariant for the GQA is defined as
\begin{equation}
\label{14}
C_{2} (GQA)=S_{BF}^{0} (S_{BF}^{0}-1)-S_{BF}^{+} S_{BF}^{-}
\end{equation}
The basis states of an irreducible representation (irrep)GQA, $\mathrm{|k,\mu \rangle}$ ,are determined by a single number $ k$ , where can be any positive number and $ \mathrm{\mu=k,k+1,....}$  Therefore,
\begin{equation}
\label{15}
S_{BF}^{0}|k,\mu \rangle =(\frac{N-M}{4}+\frac{(N_{B}+N_{F} )}{2}) |k,\mu \rangle
\end{equation}
\begin{equation}
\label{16}
C_{2} (su(1,1))|k,\mu \rangle =k(k-1)|k,\mu \rangle =(\frac{N-M}{4}+\frac{(\nu_{B}+\nu_{F} )}{2})(\frac{N-M}{4}+\frac{(\nu_{B}+\nu_{F} )}{2}-1) |k,\mu \rangle
\end{equation}
\begin{equation}
\label{17}
|k,\mu \rangle=|\frac{N-M}{4}+\frac{\nu_{B}+\nu_{F}}{2},\frac{N-M}{4}+\frac{N_{B}+N_{F}}{2} \rangle
\end{equation}
The basis states are determined  considering the fact that fermionic part of the action of $\mathrm{S_{BF}^{+}} $is restricted by the Pauli Exclusion Principle and the action of $\mathrm{S_{BF}^{-}} $ terminates when a state of $\mathrm{\nu}$ unpaired particles is reached i.e
$\mathrm{S_{BF}^{-}|\nu_{B},\nu_{F} \rangle =0}$ \cite{15}.

The infinite dimensional generalized quasispin algebra that is generated by use of \cite{12}
\begin{equation}
\label{18}
S_{BF,n}^{\pm}=c_{s}^{2n+1} S_{B}^{\pm} (s)+c_{d}^{2n+1} S_{B}^{\pm} (d)+c_{f}^{2n+1} S_{F}^{\pm} (f)
\end{equation}
\begin{equation}
\label{19}
S_{BF,n}^{0}=c_{s}^{2n} S_{B}^{0} (s)+c_{d}^{2n} S_{B}^{0} (d)+c_{f}^{2n} S_{F}^{0} (f)
\end{equation}
Where $\mathrm{c_{s}}$ , $\mathrm{c_{d}}$ and $\mathrm{c_{f}}$ are real parameters and  n  can be
$\mathrm{0,\pm1,\pm2,....}$. These generators satisfy the commutation relations
\begin{equation}
\label{20}
[S_{BF,m}^{0}  ,S_{BF,n}^{\pm} ]=\pm S_{BF,m+n}^{\pm}
\quad\quad     ,     \quad\quad         [S_{BF,m}^{+},S_{BF,n}^{-}
]=-2S_{BF,m+n+1}^{0}
\end{equation}
Then, $\mathrm{{S_{BF,m}^{\mu},\mu=0,+,-; m=\pm1,\pm2,...} } $generate an
affine generalized quasispin algebra $ \mathrm{\widehat{GQA}} $ without central extension. For evaluating the eigenvalues, the eigenstates are considered as
\begin{equation}
\label{21}
|k;\nu_{s}\nu n_\Delta LM \rangle= \textit{N}
S_{BF}^{+}(x_{1})S_{BF}^{+}(x_{2})S_{BF}^{+}(x_{3})...S_{BF}^{+}(x_{k})|lw\rangle^{BF}
\end{equation}
\begin{equation}
\label{22}
S _{x_{i} }^{+}=\frac {c_{s}}{1-c_{s}^{2} x_{i} } S_{B}^{+} (s)+\frac
{c_{d}}{1-c_{d}^{2} x_{i} } S_{B}^{+} (d)+\frac
{c_{f}}{1-c_{f}^{2} x_{i} } S_{F}^{+} (f)
\end{equation}
The lowest weight state, $\mathrm{ |lw\rangle^{BF} }$, is defined as
\begin{align}
\label{23}
|lw\rangle^{BF}&=|N=N_{B}+N_{F},k_{d}=\frac{1}{2}
(\nu_{d}+\frac{5}{2}),\mu_{d}=\frac{1}{2}
(n_{d}+\frac{5}{2}),k_{s}=\frac{1}{2}
(\nu_{s}+\frac{1}{2}),\mu_{s}=\frac{1}{2} (n_{s}+\frac{1}{2}), k_{f}=\frac{1}{2}
\nonumber \\
&(\nu_{f}-\frac{2j+1}{2}),\mu_{f}=\frac{1}{2} (n_{f}-\frac{2j+1}{2}),J ,M \rangle
\end{align}
\begin{equation}
\label{24}
S_{n}^{0} |lw\rangle^{BF} =\Lambda_{n}^{0} |lw\rangle^{BF} \quad , \quad \Lambda_{n}^{0} = c_{s}^{2n}(\nu_{s}+\frac {1}{2})\frac
{1}{2}+c_{d}^{2n}(\nu_{d}+\frac {5}{2})\frac {1}{2}+c_{f}^{2n}(\nu_{f}-\frac {2j+1}{2})\frac {1}{2}
\end{equation}
In order to obtaining an algebraic solution for transitional region, we have used of dual algebraic structures. Such relations have often been used to effect simplifications of the calculations for two-level and multi-level systems. These relations were obtained  for both bosonic
and fermionic systems\cite{13,14}. We have established the duality relations for mixed boson-fermion system.
Because of duality relationships \cite{12,13,14}, It is known that
in even-even nuclei the base of $\mathrm{U(5)\supset SO(5)} $ and $ \mathrm{SO(6)\supset SO(5)} $ are
simultaneously the basis of $ \mathrm{SU^{d} (1,1)\supset U(1)} $ and $
\mathrm{SU^{sd} (1,1)\supset U(1)} $, respectively. By use of duality
relations \cite{12,13,14}, the Casimir operators of SO(5) and
SO(6) can also be expressed in terms of the Casimir operators
of $ \mathrm{SU^{d} (1,1)}$ and $ \mathrm{SU^{sd} (1,1)}$, respectively
\begin{equation}
\label{25}
  \hat{C } _{2}(SU^{d} (1,1))=\frac{5}{16}+\frac{1}{4} \hat{C } _{2}(SO^{B}(5) )
\end{equation}
\begin{equation}
\label{26}
  \hat{C } _{2}(SU^{sd} (1,1))=\frac{3}{4}+\frac{1}{4} \hat{C } _{2}(SO^{B}(6) )
\end{equation}
For a mixed boson-fermion system, the chain of subalgebras of unitary superalgebras U(6/M ) for j=1/2 and j=3/2 is shown in Fig.1 and Fig.2, respectively. The two-level pairing system has two dynamical symmetries defined with respect to the generalized quasispin algebras, corresponding to either the upper or lower subalgebra chains in Eq.(27).
\begin{equation}
GQA_{1}^{sf}\otimes GQA_{2}^{df} \supset \left \{
\begin{array}{cl}
GQA_{1,2}^{sdf} \\
U_{1}^{sf}(1)\otimes U_{2}^{df}(1)\\
\end{array}\right\}
\supset U_{1,2}^{sdf}
\end{equation}
The upper subalgebra chain is corresponding to strong-coupling dynamical symmetry limit while lower chain is weak-coupling limit.
For odd-A nuclei with j=1/2, we have obtained the relation between the Casimir operators SO(5) and $\mathrm{GQA^{df}}$ (generalized quasispin algebra of d bosons and single fermion with j=1/2) and the Casimir operators SO(6) and $\mathrm{GQA^{sdf}}$ (generalized quasispin algebra of s and d bosons and single fermion with j=1/2) according with Eqs.[25] and [26], respectively. The correspondence between the basis vectors in this case was shown in Ref.\cite{12}.For odd-A nuclei with j=3/2, we have obtained the relation between the Casimir operators $\mathrm{Spin^{BF}(5)}$ and $\mathrm{GQA^{df}}$ (generalized quasispin algebra of d bosons and single fermion with j=3/2) as

If $\mathrm{\tau_{1}=v_{d}-\frac{1}{2}} $ and $ \mathrm{\tau_{2}=\frac{1}{2}}$
\begin{equation}
\label{30}
\hat{C } _{2}(GQA^{df} )=\frac{1}{4} \hat{C } _{2}(Spin^{BF}(5) )-\frac{1}{4}(\tau_{1}+\frac{3}{4})
\end{equation}
If $\mathrm{\tau_{1}=v_{d}+\frac{1}{2}} $ and $ \mathrm{\tau_{2}=\frac{1}{2}}$
\begin{equation}
\label{31}
\hat{C } _{2}(GQA^{df} )=\frac{1}{4} \hat{C } _{2}(Spin^{BF}(5) )-\frac{1}{4}(3\tau_{1}+\frac{7}{4})
\end{equation}
By use of duality
relations, the correspondence between the basis vectors $\mathrm{Spin^{BF}(5)}$ and $\mathrm{GQA^{df}}$ is
\begin{equation}
\label{32}
|\mathcal{N};[N_{B}=N],{N_{F}=1},\nu_{d},(\tau_{1}=v_{d}-\frac{1}{2},\tau_{2}) ,n_\Delta JM \rangle = |\mathcal{N};k^{d}=\frac{1}{2}(\nu_{d}+\frac{5}{2}),k^{df}=\frac{\tau_{1}+2}{2},\mu^{df}=\frac{1}{4}+\frac{1}{2}(n_{d}+n_{f}) ,n_\Delta JM \rangle
\end{equation}
\begin{equation}
\label{33}
|\mathcal{N};[N_{B}=N],{N_{F}=1},\nu_{d},(\tau_{1}=v_{d}+\frac{1}{2},\tau_{2}) ,n_\Delta JM \rangle = |\mathcal{N};k^{d}=\frac{1}{2}(\nu_{d}+\frac{5}{2}),k^{df}=\frac{1+\tau_{1}}{2}, \mu^{df}=\frac{1}{4}+\frac{1}{2}(n_{d}+n_{f}) ,n_\Delta JM \rangle
\end{equation}
The Casimir operator
of $\mathrm{Spin^{BF}(6)}$ and $\mathrm{GQA^{sdf}}$ (generalized quasispin algebra of d and s bosons with fermion j=3/2 ) has the following correspondence
If $\mathrm{\sigma_{1}=\sigma-\frac{1}{2}} $ and $ \mathrm{\sigma_{2}=|\sigma_{3}|=\frac{1}{2}}$
\begin{equation}
\label{34}
\hat{C } _{2}(GQA^{sdf} )=\frac{1}{4} \hat{C } _{2}(Spin^{BF}(6) )-\frac{3}{4}(\sigma_{1}+\frac{3}{4})
\end{equation}
If $\mathrm{\sigma_{1}=\sigma+\frac{1}{2}} $ and $ \mathrm{\sigma_{2}= |\sigma_{3}|=\frac{1}{2}}$
\begin{equation}
\label{35}
\hat{C } _{2}(GQA^{sdf} )=\frac{1}{4} \hat{C } _{2}(Spin^{BF}(6) )-\frac{1}{4}(\sigma_{1}+\frac{3}{2})
\end{equation}
By use of duality
relations, the correspondence between the basis vectors $\mathrm{Spin^{BF}(6)}$ and $\mathrm{GQA^{sdf}}$ is
\begin{equation}
\label{36}
|\mathcal{N};[N_{B}=N],{N_{F}=1},\sigma,(\sigma_{1},\sigma_{2},\sigma_{3}),(\tau_{1},\tau_{2}) ,n_\Delta JM \rangle = |\mathcal{N};k^{sdf}=\frac{1}{2}(\sigma_{1}+\frac{3}{2}),\mu^{sdf}=\frac{1}{4}+\frac{1}{2}(n_{s}+n_{d}+n_{f}) ,n_\Delta JM \rangle
\end{equation}
\begin{equation}
\label{37}
|\mathcal{N};[N_{B}=N],{N_{F}=1},\sigma,(\sigma_{1},\sigma_{2},\sigma_{3}),(\tau_{1},\tau_{2}) ,n_\Delta JM \rangle = |\mathcal{N};k^{sdf}=\frac{1}{2}(\sigma_{1}+\frac{5}{2}),\mu^{sdf}=\frac{1}{4}+\frac{1}{2}(n_{s}+n_{d}+n_{f}) ,n_\Delta JM \rangle
\end{equation}
In this article, we investigate U(6/2) and U(6/4) supersymmetry. The detailed description of the U(6/2) and U(6/4) supersymmetry in U(5) and O(6) limits can be found in Refs. \cite{2,3,4}.
Here we restrict ourselves only to a brief discussion of the main ideas.
\subsection{U(6/2) Supersymmetry Model}
First we considered case that fermions occupy a single particle state with angular momentum j =1/2. Thus, the underlying graded algebra is U(6/2)\cite{2}.
In order to study the occurrence of supersymmetries, one must first identify the set of nuclei belonging to the representation $ \mathcal{N}$. Supermultiplets here contain 3 nuclei, if $\mathcal{N}\geq 2$. The three nuclei are characterized by\cite{2}
$$\mathrm{N_{B}}=\mathcal{N} ,\mathrm{N_{F}}=0$$
$$\mathrm{N_{B}}=\mathcal{N}-1 ,\mathrm{N_{F}}=1$$
$$\mathrm{N_{B}}=\mathcal{N}-2 ,\mathrm{N_{F}}=2$$
These nuclei are alternately even-even and odd-even nuclei and have an increasing number of unpaired fermions \cite{2}. States with $\mathrm{N_{F}}=0$ and $\mathrm{N_{F}}=1$ are the lowest states of the corresponding nuclei, while states with $\mathrm{N_{F}> 2}$ are at higher energies \cite{2}. The lattice of algebras in this model is shown in Fig.(1).

In the following we will explain how derive the even-even and odd-A Hamiltonians within the U(6/2) supersymmetry scheme.By employing the generators of $ \mathrm{\widehat{GQA} }$ and casimir operators of subalgebras, the following Hamiltonian for transitional region between U(5)-O(6) limits is prepared
\begin{equation}
\label{27}
\hat{H }=S_{BF,0}^{+} S_{BF,0}^{-}+\alpha S_{BF,1}^{0}+\beta \hat{C }_{2}  ̂(O^{B} (5) )+\delta\hat{C }_{2}  ̂(O^{B} (3) )+\gamma\hat{C }
_{2}(spin^{BF} (3) )
\end{equation}
$\mathrm{\alpha}$
, $\mathrm{\beta}$ ,$ \mathrm{\delta}$ ,$\mathrm{\gamma}$ are real parameters.
The eigenvalues of Hamiltonian Eq.(36) can then be expressed;
\begin{equation}
\label{28}
E^{(k) }=h^{(k) }+\alpha \Lambda_{1}^{0}+\beta\nu_{d}(\nu_{d}+3)+\delta L_{B} (L_{B}+1)+\gamma J(J+1)\quad  , \quad h^{(k) }=\sum_{i=1}^{k}{\frac {\alpha}{x_{i}}}
\end{equation}
\begin{equation}
\label{29}
\frac {\alpha}{x_{i}}=\frac{ c_{s}^{2} (\nu_{s}+\frac
{1}{2})}{1-c_{s}^{2} y_{i}}+\frac{ c_{d}^{2}(\nu_{d}+\frac
{5}{2})}{1- c_{d}^{2}x_{i}}+\frac{c_{f}^{2} (\nu_{f}-1)}{1-c_{f}^{2} x_{i}}-{\sum_{j\neq i}{\frac {2}{x_{i}-x_{j}}}}
\end{equation}
\subsection{U(6/4) Supersymmetry Model}
The case that fermions occupy a single particle state with angular momentum j =3/2 constitute the superalgebra U(6/4).
Supermultiplets here contain 5 nuclei, if $ \mathcal{N}\geq 4$.The five nuclei are characterized by\cite{2,3,4}
$$\mathrm{N_{B}}=\mathcal{N} , \mathrm{N_{F}}=0$$
$$\mathrm{N_{B}}=\mathcal{N}-1 , \mathrm{N_{F}}=1$$
$$\mathrm{N_{B}}=\mathcal{N}-2 , \mathrm{N_{F}}=2$$
$$\mathrm{N_{B}}=\mathcal{N}-3 , \mathrm{N_{F}}=3$$
$$\mathrm{N_{B}}=\mathcal{N}-4 , \mathrm{N_{F}}=4$$

In order to obtain Hamiltonian within the U(6/4) supersymmetry scheme, We have considered the lattice of algebras is shown in Fig.2.

So,following Hamiltonian for transitional region between U(5)-O(6) limits is prepared
\begin{equation}
\label{38}
\hat{H }=S_{BF,0}^{+} S_{BF,0}^{-}+\alpha S_{BF,1}^{0}+\beta'\hat{C }
_{2}(Spin^{BF}(5) )+\gamma\hat{C }
_{2}(spin^{BF} (3) )
\end{equation}
The eigenvalues of Hamiltonian Eq.(39) can then be expressed;
\begin{equation}
\label{39}
E^{(k) }=h^{(k) }+\alpha \Lambda_{1}^{0}+\beta'
(\tau _{1}(\tau _{1}+3)+\tau _{2}(\tau _{2}+1))+\gamma J(J+1)\quad  , \quad h^{(k) }=\sum_{i=1}^{k}{\frac {\alpha}{x_{i}}}
\end{equation}
\begin{equation}
\label{40}
\frac {\alpha}{x_{i}}=\frac{ c_{s}^{2} (\nu_{s}+\frac
{1}{2})}{1-c_{s}^{2} x_{i}}+\frac{ c_{d}^{2}(\nu_{d}+\frac
{5}{2})}{1- c_{d}^{2}x_{i}}+\frac{c_{f}^{2} (\nu_{f}-2)}{1-c_{f}^{2} x_{i}}-{\sum_{j\neq i}{\frac {2}{x_{i}-x_{j}}}}
\end{equation}
It can be shown that Hamiltonians Eq.(36) and Eq.(39) are equivalent to a boson Hamiltonian for the even-even nuclei if acting on the $ \mathrm{[N] \times [0]} $ representation of $ \mathrm{U^{B}(6)\times U^{F}(4)}$ and with boson-fermion Hamiltonian for odd-A nuclei if acting on the $ \mathrm{[N] \times [1]} $ representation of $ \mathrm{U^{B}(6)\times U^{F}(4)}$. By considering the neutron and proton degree of freedom in this model, we  can describe all nuclei belonging to the representation $[\mathcal{N}\}$. In odd-A nuclei, Hamiltonians Eq.(36) and Eq.(39) are equivalent to $ \mathrm{O^{BF}(6)}$
Hamiltonian when $\mathrm{c_{s}=c_{d}=c_{f}} $ and with $ \mathrm{U^{BF}(5)} $
Hamiltonian if $\mathrm{c_{s}=0} $ and $\mathrm{c_{d}\neq c_{f}\neq0} $. So, the $\mathrm{c_{s}\neq
c_{d} \neq c_{f} \neq 0}$  situation just corresponds to $\mathrm{U^{BF}
(5)\leftrightarrow O^{BF}(6)}$ transitional region.  Hamiltonians Eq.(36) and Eq.(39)  in even-even nuclei is equivalent to $ \mathrm{O(6)}$ Hamiltonian when $\mathrm{c_{s}=c_{d}}$ and with $ \mathrm{U(5)} $
Hamiltonian if $\mathrm{c_{s}=0 }$ and $\mathrm{c_{d}\neq0} $ and Hamiltonian in transitional region with $\mathrm{c_{s}\neq
c_{d} \neq 0}$.
 In our
calculation, we take  $\mathrm{c_{d}(=1)}$ constant value and $\mathrm{c_{s}}$
and  $\mathrm{c_{f}}$ change between 0 and $\mathrm{c_{d}}$

In order to obtain the numerical results for energy spectra
$\mathrm{(E^{(k) } )}$ of considered nuclei in two kinds supersymmetry approach, a set of non-linear
Bethe-Ansatz equations (BAE) with k- unknowns for k-pair
excitations must be solved  also constants of
Hamiltonian with least square fitting processes to experimental
data is obtained \cite{16}. To achieve this aim, we have changed variables as
$$ \mathrm{C_{s}=\frac  {c_{s}}{c_{d}} \leq 1    ,   C_{f}=\frac  {c_{f}}{c_{d}} \leq 1        ,    y_{i}=c_{d}^{2} x_{i}} $$
\begin{equation}
\label{41}
\frac {\alpha}{y_{i}}=\frac{ C_{s}^{2} (\nu_{s}+\frac
{1}{2})}{1-C_{s}^{2} y_{i}}+\frac{ (\nu_{d}+\frac
{5}{2})}{1- y_{i}}+\frac{C_{f}^{2} (\nu_{f}-\frac{2j+1}{2})}{1-C_{f}^{2} y_{i}}-{\sum_{j\neq i}{\frac {2}{y_{i}-y_{j}}}}
\end{equation}
The quantum number (k) is related to  $\mathcal{N}$
by
$ \mathcal{N}=\mathrm{2k+\nu_{s}+\nu_{d}+\nu_{f}}$. The quality of the fits is quantified by the values of $ \mathrm{\sigma = (\frac {1}{N_{tot}}\sum _{i,tot}{|E_{exp}(i)-E_{Cal}(i)|^2})^{\frac {1}{2}}}$ (keV)
and $ \mathrm{\phi = \frac{\sum\limits_{i}|E_{i}^{theor}-E_{i}^ {exp}| } {\sum\limits_{i} E_{i}^{exp}}} $ (\%)
($N_{tot}$ the number of energy levels where included in the
fitting processes) \cite{2,3,4}.  A measure
of the breaking of the supersymmetry scheme, the average absolute deviation
divided by the average excitation energy, give with $\phi$ parameter \cite{2,4}.
The method for optimizing the set of
parameters in the Hamiltonian $\mathrm{(\beta,\gamma,\delta)}$  includes
carrying out a least-square fit (LSF) of the excitation energies
of selected states \cite{16}.
\subsection{E2  transition probabilities}
The observables such as electric quadrupole transition
probabilities, $B(E2)$, as well as quadrupole moment ratios
within the low-lying state provide important information about
the nuclear structure and $QPTs$. In this section we discuss the calculation of E2
transition strengths for j=1/2 and j=3/2. Supersymmetry implies that all transitions in nuclei belonging to the same supermultiplet be explained by the same operator\cite{2}. The electric quadrupole transition operator $\hat
{T}^{(E2)}$ in odd-A nuclei consists of a bosonic and a fermionic
part\cite{2,17,18}. In the $j=1/2 $ case, the
E2 transitions for even-even and odd-A nuclei are completely determined by the bosonic part of
the E2 operator\cite{2} . The bosonic part  have the specific selection
rules, where for former term $ \Delta \nu_{d}=\pm 1 $, $ |\Delta
L |\leq 2 $ and for latter $ \Delta \nu_{d}=0,\pm 2 $, $ |\Delta
L |\leq 0,4 $. In the $j=3/2 $ case, we also consider the portion of fermionic term\cite{2,17,18}.
\begin{equation}
\hat {T}^{(E2)} =\hat {T} _{B}^{(E2)} +\hat {T} _{F}^{(E2)}
\end{equation}
With
\begin{equation}
\label{29}
{T} _{B, \mu}^{(E2)} =q_{2} [s^{+}\times \tilde{d} + d^{+} \times
\tilde{s}]_{\mu}^{(2) }+q'_{2} [d^{+} \times
\tilde{d}]_{\mu}^{(2) }=q_{B} Q_{B,\mu}
\end{equation}
\begin{equation}
\label{30}
Q_{B,\mu}=[s^{+}\times \tilde{d} + d^{+} \times
\tilde{s}]_{\mu}^{(2) }+\chi [d^{+} \times \tilde{d}]_{\mu}^{(2) }
\end{equation}
\begin{equation}
\label{31}
 {T} _{F}^{(E2)} =q_{f} \sum_{jj'}Q_{jj'}[a_{j}^{+}\times \tilde{a}_{j'} ]^{(2) }
\end{equation}
Where $ \mathrm{Q_{B}} $   and $ \mathrm{Q_{jj'}} $   are the boson and fermion
quadrupole operator and $\mathrm{q_{B}}$ and $\mathrm{q_{f}}$ are the effective
boson and fermion charges\cite{2,17,18}.

The reduced electric quadrupole transition rate
between the $ J_{i}\rightarrow J_{f} $ states is given by
\cite{1,2}
\begin{equation}
B(E2; \alpha _{i} J_{i} \rightarrow \alpha _{f} J_{f})=\frac
{|\langle \alpha _{f} J_{f} || T^{(E2)} || \alpha _{i}
J_{i}\rangle |^{2} }{2J_{i}+1}
\end{equation}
For evaluating B(E2), we have calculated the matrix elements of T(E2)  operators between the eigenstates of Eq.(21) that the
normalization factor obtain as:
\begin{equation}
N=\sqrt{{\frac {1}{ \prod_{p=1}^{k} \sum_{i=p}^{k}(\frac
{2C_{s}^{2}(k-p+\frac {1}{2}(\nu_s+\frac
{1}{2}))}{(1-C_{s}^{2}y_{k+1-p})(1-C_{s}^{2}y_{i})}+\frac {2(k-p+\frac
{1}{2}(\nu_d+\frac {5}{2}))}{(1-y_{k+1-p})(1-y_{i})}-\frac{2C_{f}^{2}(k-p+\frac {1}{2}(\nu_f-\frac
{2j+1}{2}))}{(1-C_{f}^{2}y_{k+1-p})(1-C_{f}^{2}y_{i})})}}}
\end{equation}
We calculate fermion part for j=3/2 by using Eq.(49)\cite{17,18}
\begin{eqnarray}
\langle \nu _{B}, \nu_{F}, (\tau_{1},\tau_{2}), L_{BF},J||T_{F}^{(\lambda,\mu)}||\nu' _{B}, \nu'_{F}, (\tau'_{1},\tau'_{2}), L'_{BF},J'\rangle=\sum_{\nu,L}\sum_{\nu',L'}\xi_{N+1/2,\tau_{1},J}^{N,\nu,L}\xi_{N+1/2,\tau'_{1},J'}^{N,\nu',L'}(-1)^{L+J'+3/2}(-1)^{J-J'}
\nonumber \\
\left \{
\begin{array}{ll}
3/2 \quad J   \quad L \\
J'  \quad 3/2 \quad 2\\
\end{array}\right\}\langle 3/2 ||T_{F}^{(E2)}||3/2\rangle
\end{eqnarray}

Where $\xi$ is isoscalar factor and  $ \langle 3/2 ||T_{F}^{(E2)}||3/2\rangle$ is a single-particle matrix element that is
obtained as\cite{2,17,18}
\begin{equation}
\langle j ||(a_{j_{1}}^{+}\times \widetilde{a}_{j_{2}})^{(\lambda)}||j'\rangle=-\sqrt{2\lambda+1}\delta_{j_{2},j'}\delta_{j_{1},j}
\end{equation}
A test of supersymmetry is to see the extent to which  the electric quadrupole transition
probabilities of even-even and odd-A nuclei can be describe with the same coefficients of Eq.(47). To determine boson effective charge, we have extracted these quantities from the empirical B(E2) values via least square technique. In the fitting process, empirical transition rates of both of even-even and odd-A nuclei were used.
\section{Quantal analysis}
This section presented the calculated phase transition observables such as level crossing,expectation values of the d-boson  and the fermion number operators.
\subsection{energy spectrum and level crossing}
In order to display how the energy levels change within the whole range of the $C_{s}$ and $C_{f}$ control parameters, the energy surfaces $ E(C_{s},C_{f} )$ with the other parameters fixed can be defined and calculated using procedure that explained in sec.($\prod$).
Fig.3 and Fig.4 show the energy surfaces of Hamiltonian of Eq.(36) and Eq.(39) for the neighboring even-even (left panel) and odd-A nuclei (right panel) with $ N=10 $, respectively. The calculation are performed by considering the same fit parameters for neighboring even-even and odd-A nuclei, where in Fig.3 the fixed parameters are $ \alpha=1000 $keV, $\beta=3.14 $keV, $ \delta=-5.26 $keV, $\gamma=0.0439$keV  and Fig.4 obtained
with $\alpha=1000 $keV, $\beta'=-1.29 $keV,  $\gamma=6.05 $keV. Figs show how the energy levels evolve from one phase to the
other as a function of $ C_{s}$and $C_{f} $ control parameters. These Figures clearly show how the transition from one dynamical symmetry to another occurs. It can be seen from Figs that numerous level crossings occur. The crossings are due to the fact that$\nu_{d}$, O(5) quantum number called seniority, is preserved along the whole path between O(6) and U(5)\cite{19,20}.
\subsection{expectation values of the d-boson number operator}
An appropriate quantal order parameter is:
$$ \langle \hat{n_{d}}\rangle=\frac {\langle \psi |\hat{n_{d}}|\psi\rangle}{N}$$
In order to obtain $\langle \hat{n_{d}}\rangle$, we act
$s_{m}^{0}$ on the eigenstate, $|k;\nu_{s}\nu n_\Delta LM \rangle$
\begin{equation}
\langle \hat{n_{d}}\rangle=\frac {2C_{s}^{2}C_{f}^{2}(\Lambda_{0}^{0}+k)-2(C_{s}^{2}+C_{f}^{2})(\Lambda_{1}^{0}+k y_{1}^{-1})+2(\Lambda_{2}^{0}+k y_{2}^{-2})}{N (1-C_{s}^{2})(1-C_{f}^{2})}-\frac {5}{2N}
\end{equation}
Fig.5 and Fig.6 show the expectation values of the d-boson number operator
for the lowest states even-even (left panel) and odd-A nuclei (right
panel) as a function of  control parameters for
 j=1/2 and j=3/2 with N=10 bosons, respectively. Fig.5 and Fig.6 (left panel) display that the expectation values of the
number of d bosons for each L, $n_{d}$, remain approximately
constant for $ C_{s}  <0.45 $ and only begin to change rapidly for $ C_{s}
>0.45$. The near constancy of $n_{d}$ for $C_{s} <0.45$, is a obvious
indication that $ U(5)$ dynamical symmetry preserves in this
region to a high degree and also the $n_{d}$ values change
rapidly with $ C_{s} $ over the range $0.65\leq C_{s} \leq 1$.

\subsection{expectation values of the fermion number operator}
The expectation values of the fermion number operator are obtained as
\begin{equation}
\langle \hat{n_{f}}\rangle=\frac {\langle \psi |\hat{n_{f}}|\psi\rangle}{N}=\frac{2(1+C_{s}^{2})(\Lambda_{0}^{0}-\Lambda_{1}^{0}+k(1-y_{1}^{-1}))-2(\Lambda_{0}^{0}-\Lambda_{2}^{0}+k(1-y_{2}^{-2}))}{N(1-C_{f}^{2})(C_{s}^{2}-C_{f}^{2})}+\frac {2j+1}{2N}
\end{equation}
Fig.7 and Fig.8 show also $\langle n_{f}\rangle$ as a function of the $ C_{s} $ and $ C_{f}$ control parameters.

It can bee seen from figs that combination of multi-symmetry in the point of phase transition cause dissimilar behavior with neighboring points that  representing the transition of u(5) limit to O(6) limit.
\section{Experimental evidence}
This section presented the calculated results of low-lying states of Rh-Ru Supermultiplets and Zn - Cu Supermultiplets. The results include energy levels and the B(E2)values and Key observables of E(5) and E(5/2j+1) symmetry.
\subsection{The Rh-Ru Supermultiplets}
Nuclei in the mass region around $\mathrm{A \sim 100}$ have transitional characteristics intermediate between spherical and  gamma-unstable shapes \cite{21,22}. Stachel et al.\cite{23} have been studied the Ru isotopes and are found that these isotopes have U(5)-O(6) transition features. The odd-A Rh isotopes were studied in the framework IBFM by Vervier and Janssens \cite{24} and Refs.\cite{16,21,22,25}.  A. Frank and collaborators have studied the Rh-Ru Supermultiplets by using supersymmetry approach \cite{6}. They have studied successfully a combination of $\mathrm{U^{BF}(5)}$ and $\mathrm{SO^{BF}(6)}$ symmetry by using $\mathrm{U(6/12)}$ supersymmetry for the Ru and Rh isotopes. We have also analyzed the negative parity states of the odd-proton nuclei,$ \mathrm{_{45}^{101-109}Rh}$ and positive parity states of the even-even nuclei,$ \mathrm{_{44}^{100-108}Ru}$.The negative parity states in the odd-even nuclei Rh are built mainly on the $ \mathrm{2p_{\frac{1}{2}}}$ shell model orbit\cite{21}. In order to obtain energy spectrum and realistic calculation for these nuclei, we need to specify Hamiltonian parameters Eq.(36). According to the supersymmetry between bosons and fermions, the fermion is transformed into a boson,  the system is described by the same set of parameters thus to achieve a better fit, states of both of even-even and odd-A nuclei were used. Eigenvalues of these systems are obtained by solving Bethe-Ansatz equations with least square fitting processes to experimental data to obtain constants of Hamiltonian. The best fits for Hamiltonian\textquoteright s parameters, namely $\mathrm{\alpha}$, $ \mathrm{\beta}$, $ \mathrm{\delta}$ and $\mathrm{ \gamma} $, used in the present work are shown in table 1. Fig.9 and Fig.10 show a comparison between the available experimental levels and the predictions of our results for the $ \mathrm{_{45}^{101-109}Rh}$ and $\mathrm{ _{44}^{100-108}Ru}$ isotopes in the low-lying region of spectra. The quantum numbers of different states of each isotope and a comparison between theoretical prediction and experimental values for considered isotopes are presented in Tables.2(a,b...e). An acceptable degree of agreement is obvious between them.

The most important successful nuclear model characteristic is a good description  of electromagnetic properties of
the nucleus in addition to  its  energy spectrum. So,
we have calculated B(E2) transition rates for Ru-Rh Supermultiplets. In the fitting process, because there is no empirical data for Rh isotopes except $ ^{103}Rh $ we have used only of the even-mass Ru experimental data for
extracting of effective charge parameters. In
$ ^{102}Ru-^{103}Rh$ supermultiplet, we have used of experimental
transition rates of both nuclei. Extracted values for effective charge parameters have presented in Table 3. Tables.4(a,b,c,d) show
experimental and calculated values of B(E2) for negative parity
states of Rh  and positive parity states of  Ru.
\subsection{The Zn-Cu Supermultiplets}
In order to test the predictions of the U(6/4) scheme  in dynamical symmetry limits and transition region, we  choose the negative parity states of the odd-proton nuclei,$ ^{61-69}Cu$ and positive parity states of the even-even nuclei,$ ^{62-70}Zn$. These nuclei have not considered in other theoretical studies completely but for some of them which we have found similar counterparts.
The Cu isotopes  across the N=40 shell closure and have a single proton outside the Z=28 closed shell. The ground state spin/parity of the odd-A copper isotopes( Between  $ ^{57}Cu$ and   $ ^{69}Cu$) is $(\frac{3}{2})^{-}$.
The low-lying levels of $\mathrm{^{63,65} Cu}$ were studied by Bijker and Kota in 1984 \cite{37}. They showed that the Spin(5) symmetry could occur in the odd-proton nuclei in the Cu region with the odd-proton occupying the  $\mathrm{2p_{3/2}}$ orbit.
Since the low-lying negative parity states in $\mathrm{^{63,65} Cu}$ are built mainly on the $\mathrm{2p_{3/2}}$ shell model orbit and the adjacent even-even nuclei  $\mathrm{^{64,66} Zn}$ show a vibrational type
of spectrum, it has been suggested \cite{37,38}, that the Cu-Zn mass-region provides
experimental evidence for the existence of the Spin(5) spinor symmetry and the
U(6/4) supersymmetry \cite{37}.  The odd-even nuclei $\mathrm{^{63,65} Cu}$ alone provide a test of the Spin(5) symmetry, while the pairs of nuclei Zn-Cu provide a test of the U(6/4) supersymmetry \cite{37}. Since vibrational nuclei are known to exhibit other excitation modes such as quasi particle modes  or intruder states due to the excitation of two
protons into the next major shell in  Zn \cite{39}, great care must
be taken in comparing with experiment.
We have used
a similar procedure as has done for Ru-Rh supermultiplets calculations to extract the parameters of
Hamiltonian Eq.(39) for Zn-Cu supermultiplets.
In the fitting process, in addition to the states in Cu, the states of Zn isotopes were considered.
The best fits for Hamiltonian\textquoteright s parameters, namely $\alpha$, $ \beta'$ and $ \gamma $, used in the present work are shown in table 5. Figs.11 and 12 show the comparison of the  experimental  and the theoretical level scheme. Experimental data of $ ^{61-69}Cu$ and $ ^{62-70}Zn$ isotopes was taken from Refs.[40-49]. The quantum numbers of different states of each isotope and a comparison between theoretical prediction and experimental values for considered isotopes are presented in Tables.6(a,b,..,e).

we have also calculated B(E2) transition rates for Zn-Cu Supermultiplets. To determine boson effective charge, we have extracted these quantities from the empirical B(E2) values via least square technique. In the fitting process, we have used of experimental
transition rates of both nuclei in a Supermultiplet.
Extracted values for effective charge parameters have presented in Tables 7. Tables.8(a,b,.,e) show
experimental and calculated values of B(E2) for negative parity
states of Cu  and positive parity states of  Zn.
In Zn-Cu Supermultiplet, improved results my be obtained by considering neutron and proton degree of freedom separately.

In the following, we  concentrated upon those distinguishing observables which  vary along the
U(5)- SO(6) transition. The observables such as the energy ratios, the B(E2) values that are most sensitive to the U(5)-O(6) structural transition.
M.A.Caprio and F.Iachello \cite{18} obtained analytic descriptions for transitional nuclei near the critical point. The solutions provided baselines for experimental studies of even- even [E(5)] and odd-mass E(5/4) nuclei near the critical point of the spherical to gamma-unstable phase transition. Their results have provided benchmarks for nuclei near the critical point of the U(5)-O(6) phase transition and can be used as a basis for comparison with experiment. So, we have also  calculated  these  quantities   for  the Ru-Rh   and   Zn-Cu supermultiplets by using
their method and performed an analysis for these isotopes.
One of the most basic structural predictions of $\mathrm{ U
(5)-O(6)} $ transition is a $ R_{\frac{4}{2} }=\frac{E(4_{1}^{+})}{E(2_{1}^{+})}$ value. The ratio equal to  2.2-2.3
indicates the spectrum of transitional nuclei \cite{18,50,51}. Thus, we
calculated this quantity for even-even Ru and Zn Isotopes. Fig.\ref{fig:13}(a) and Fig.\ref{fig:14}(a) show
 $ R_{\frac{4}{2}}$ prediction  along with the experimental values for Ru and Zn
isotopes, respectively. We have also calculated observables such as $ \frac{E(0_{2}^{+})}{E(2_{1}^{+})}$
,the vibrational excitation measure, values for control parameter, namely the ratio of shape phase transition, and  ratios
$B(\frac{E2;4_{1}^{+} \longrightarrow 2_{1}^{+}}{E2;2_{1}^{+} \longrightarrow 0_{1}^{+}})$ and $B(\frac{E2;0_{2}^{+}\longrightarrow 2_{1}^{+}}{E2;2_{1}^{+}\longrightarrow 0_{1}^{+}})$.
The
comparison with experiment for these quantities in Ru and Zn isotopes is
summarized in Table.9 and Table.10, respectively.

Because of the occurrence
of  single particle orbitals that can considerably perturb the
spectrum and electromagnetic transition strengths, comparison of the critical point description with experimental data in odd-mass nuclei is more
difficult than even-even nuclei\cite{18}.
For Cu and Rh isotopes,  we  calculated  only the energy ratio quantity. Fig.\ref{fig:13}(b) and Fig.\ref{fig:14}(b) show
$\mathrm{\frac {E(\nu_{d}=2)}{E(\nu_{d}=1)})}$ prediction  along with the experimental values for these isotopes.

From these figures and tables, one can conclude, the calculated energy spectra in this approach are approximately in good agreements with the experimental data. It means that our
suggestion to use this transitional Hamiltonian for the
description of the Rh-Ru and Zn-Cu supermultiplets would not have any
contradiction with other theoretical studies done with special
hypotheses about mixing of intruder and normal configurations. On
the other hand, predictions of our model for the control
parameters of considered supermultiplet, $C_{s}$ and $C_{f}$ , describe the vibrational,
i.e.$C_{s}=0$, or rotational, namely $C_{s}=1$, and also $C_{f}\neq0$ for odd-A nuclei and $C_{f}=0$ for even-even nuclei, confirm this mixing of
both vibrating and rotating structures in these nuclei when
$C_{s}\sim 0.5\rightarrow 0.65$.
The values of the control parameter, $C_{s}$,
suggest structural changes in nuclear deformation and shape-phase
transitions in  Rh-Ru and Zn-Cu supermultiplets.  we
proposed $C_{s}\sim 0.5\rightarrow 0.65$ as critical point.
We conclude
from the values of control parameter which has been obtained, observables such as the energy ratios and the B(E2) values
that $^{104}Ru- ^{105}Rh$ and $ ^{64}Zn-^{63}Cu$ supermultiplets  are as
the best candidates for $ U(5)-O(6)$ transition. Also, theoretical  B(E2)transition probabilities of the Rh-Ru and Zn-Cu supermultiplets, which have obtained by using the model perspectives, exhibit nice agreement with experimental ones.
\subsection{two-neutron separation energies}
Shape phase transitions in nuclei can be studied experimentally by considering
the behavior of the ground state energies of a series of isotopes, or, more conveniently, the behavior of the two-neutron separation energies, $S_{2n}$ \cite{1,52}. On the other hand, the ground-state two-neutron separation energies,$S_{2n}$, are observables very sensitive to the details of the nuclear structure and indirect test of the supersymmetry scheme. The occurrence of continuities in the
behavior of two-neutron separation energies describe a
second-order shape-phase transition between spherical and
$\mathrm{\gamma- unstable}$  rotor limits \cite{1,52}.In due to, we have
investigated the evolution of two-neutron separation energies
along the Ru,Rh,Zn and Cu isotopic chains by both experimental and
theoretical values, which have been presented in Fig.\ref{fig:15} and Fig.\ref{fig:16}.

Two-neutron separation energies for even-even and odd-even nuclei defined by\cite{2,4}
 \begin{equation}
S_{2} (\mathcal{N},N_{f}=0)=E_{B} (\mathcal{N}+1,N_{f}=0)-E_{B} (\mathcal{N},N_{f}=0)
 \end{equation}
 \begin{equation}
 S_{2} (\mathcal{N},N_{f}=1)=E_{B} (\mathcal{N}+1,N_{f}=1)-E_{B} (\mathcal{N},N_{f}=1)
  \end{equation}
Where $E_{B}$ denotes the binding energy. It can be shown that if supersymmetry applies, the separation energies in even-even and odd-even nuclei should be linear functions of $\mathcal{N}$ as\cite{2,4}
\begin{equation}
S_{2} (\mathcal{N},N_{f}=0)=S_{2} (\mathcal{N},N_{f}=1)=D_{1}+C_{1} \mathcal{N}
\end{equation}
Using the $\mathrm{S_{2n}}$ empirical values for these isotopic chains
 \cite{36} we have extracted $\mathrm{D_{1}=2.28\times 10^{{4}} , 3.0814\times 10^{{4}}}
 $ MeV and $\mathrm{C_{1}=-898.5 , -2.5 \times 10^{{3}}    }$ MeV for Ru-Rh and Zn-Cu supermultiplets,
respectively. Then, we obtained the two-neutron separation
energies, which are shown in Fig.\ref{fig:15} and Fig.\ref{fig:16}, together with the
experimental values.

It can be seen from Fig.\ref{fig:15} and Fig.\ref{fig:16} that exist continuities (linear variation) in the behavior of two-neutron separation energies thus the phase transition for Ru, Rh, Zn and Cu isotopic chains is of second order and also approximately equal slopes for Ru(Zn) even-even and Rh(Cu) odd-mass nuclei is a result of supersymmetry scheme.
Our result Confirmed the predictions of done in refs. [1,2,4,52].

 \section{Conclusions}
In this paper, we have proposed exactly-solvable supersymmetry
Richardson-Gaudin (R-G) model for transitional region.
We apply supersymmetry ideas
to the description of  even and odd nuclei and display that several states in
many nuclei can be explained by schemes based on the U(6/2) and U(6/4) supergroups. We have employed
the nuclear supersymmetry approach for description of
the transitional region between spherical and gamma -unstable phase shape in addition to dynamical symmetry
limits in one chain isotopic. Key
observables of phase transition such as level crossing,
expectation values of the d-boson and fermion number   operator were  calculated.
New experimental data on the even-mass ruthenium and the odd-mass rhodium isotopes and Zn-Cu Supermultiplets were used to test the predictions of the new Supersymmetry scheme in dynamical symmetry limits and transition region
and performed an analysis for these isotopes via
a GQA-based Hamiltonian. The results indicate that the energy
spectra of the Rh-Ru and Zn-Cu Supermultiplets can be reproduced approximately well. The observables such as the energy ratios, the B(E2) values that are most sensitive to the U(5)-O(6) structural transition were calculated and compared
with the available experimental data. Our results
show that the Rh-Ru and Zn-Cu Supermultiplets have gamma-unstable rotor features but the
vibrational character is dominant and also
$^{104}Ru- ^{105}Rh$ and $ ^{64}Zn-^{63}Cu$ Supermultiplets are as the best
candidates for $U (5)-O (6)$ transition.

\clearpage
\begin{table}
\begin{tabular}{p{9.3cm}}

\footnotesize Table 1.Parameters of Hamiltonian
Eq.(36) used in the calculation of the Ru-Rh Supermultiplets.
\footnotesize All parameters  are given in keV.\\
\end{tabular}

\begin{tabular}{cccccccccc}
\hline

Nucleus      &$ \mathcal{N} $    & $C_{s}$ & $C_{f}$ & $\alpha$ & $\beta$ & $\delta$ & $\gamma$ & $\sigma$ & $\phi$    \\
\hline
\hline
\quad\\
$ ^{100}Ru-^{101}Rh$   & 6& 0.38 & 0.99 & 279.2 & 3.974 & 7.89 & -2.217 & 164.53 & 11.23\%  \\
\quad\\
$ ^{102}Ru-^{103}Rh$ & 7 & 0.23 & 0.8 & 170.32 & 0.8854 & -1.207 & 6.874 & 104.74 & 6.38 \%   \\
\quad\\
$^{104}Ru- ^{105}Rh$ & 8 & 0.54 & 0.99 & 70.61 & 0.1915 & 46.69 & -15.123 & 137.8 & 13.2\%  \\
\quad\\
$ ^{106}Ru-^{107}Rh$ & 9 & 0.34 & 0.7 & 197.52 & 0.329 & 51.14& -35.72 & 128.73& 11.79 \%  \\
\quad\\
$ ^{108}Ru-^{109}Rh$ & 10 & 0.58 & 0.99& 248.2 & 4.464 & 1.21 & -7.863 & 128.54 & 12.8 \% \\

\hline
\end{tabular}
\end{table}

\begin{table}
\begin{center}
\begin{tabular}{p{9.8cm}}

\footnotesize Table 2a.Energy spectra for $ _{44}^{100}Ru $ - $ _{45}^{101}Rh $ isotope.The experimental
data for $ _{44}^{100}Ru $ - $ _{45}^{101}Rh $
supermultiplet are taken from \cite{26,27,36}.\\
\end{tabular}

\begin{tabular}{cccccc}

\hline
 Nuclei &$ J^{\pi}$ & \quad $ K $ & \quad $\nu_{d}$  & \quad $E_{exp} $(keV) & \quad $ E_{cal} $(keV)   \\
\hline
\quad\\
$ _{44}^{100}Ru $  &    $0_{1}^{+} $  &      \quad 3&\quad 0& \quad 0& \quad 0 \\
 \quad &    $2_{1}^{+} $      &  \quad 2&\quad 1 &\quad 539.5 &\quad 503 \\
 \quad&    $0_{2}^{+} $   &     \quad 3&\quad 2 &\quad 1130.3 &\quad 1284.9  \\
 \quad &    $2_{2}^{+} $     &   \quad 2&\quad 2& \quad 1362.16 &\quad 1219.1  \\
\quad  &    $4_{1}^{+} $  &      \quad 2&\quad 2& \quad 1226.48& \quad 1126.72 \\
\quad  &    $0_{3}^{+} $      &  \quad 1&\quad 3 &\quad 1740.98 &\quad 1938.6 \\
 \quad  &    $3_{1}^{+} $    &    \quad 1&\quad 3& \quad 1881.04& \quad 2018.3 \\
 \quad &    $2_{3}^{+} $     &   \quad 2&\quad 1 &\quad 1865.1 &\quad 2018.3\\
 \quad  &    $4_{2}^{+} $  &      \quad 1&\quad 3& \quad 2062.51 &\quad 2086.5 \\
  \quad &    $6_{1}^{+} $      &  \quad 1&\quad 3 &\quad 2076.1 &\quad 2256.9 \\
  \quad&    $0_{4}^{+} $   &     \quad 3&\quad 0 &\quad 2051.65 &\quad 2290.8  \\
  \quad &    $2_{4}^{+} $     &   \quad 2&\quad 2& \quad 2099.1 &\quad 2207.7  \\
$ _{45}^{101}Rh $& $(1/2)_{1}^{-} $  &  \quad       2&\quad  0 &\quad  0 &\quad  0 \\
  \quad & $(3/2)_{1}^{-}  $&  \quad 2&\quad 1 &\quad 305.5& \quad 315.4\\
  \quad &$(5/2)_{1}^{-}  $ & \quad 2&\quad 1 &\quad 305.5 &\quad 292.3\\
  \quad &$(3/2)_{2}^{-} $ & \quad 1&\quad 2 &\quad 355.3& \quad 536.9\\
  \quad &$(5/2)_{2}^{-} $ & \quad 1&\quad 2& \quad 355.3& \quad 425\\
  \quad &$(7/2)_{1}^{-} $ & \quad 1&\quad 2& \quad 851.4& \quad 860.4\\
  \quad &$(9/2)_{1}^{-} $  &\quad 1&\quad 2& \quad 851.4& \quad 826.7\\
  \quad &$(9/2)_{2}^{-} $ & \quad 1&\quad 2& \quad 899.3& \quad 1104.2\\
  \quad &$(5/2)_{3}^{-} $ & \quad 2&\quad 1& \quad 996.4& \quad 914.4\\
  \quad &$(3/2)_{4}^{-} $&  \quad 1&\quad 2& \quad 1058& \quad 895.7\\

\hline
\end{tabular}
\end{center}
\end{table}

\begin{table}
\begin{center}
\begin{tabular}{p{9.8cm}}

\footnotesize Table 2b.Energy spectra for$ _{44}^{102}Ru $ - $ _{45}^{103}Rh $ isotope.The experimental
data for $ _{44}^{102}Ru $ - $ _{45}^{103}Rh $
supermultiplet are taken from \cite{28,29,36}.\\
\end{tabular}

\begin{tabular}{cccccc}

\hline
Nuclei  &$ J^{\pi}$ & \quad $ K $ & \quad $\nu_{d}$  & \quad $E_{exp} $(keV) & \quad $ E_{cal} $(keV)   \\
\hline
\quad\\
$ _{44}^{102}Ru $   &    $0_{1}^{+} $  &      \quad 3&\quad 0& \quad 0& \quad 0 \\
 \quad &    $2_{1}^{+} $      &  \quad 3&\quad 1 &\quad 475.08 &\quad 510.8 \\
 \quad&    $0_{2}^{+} $   &     \quad 3&\quad 0 &\quad 943.69 &\quad 828.9  \\
 \quad &    $2_{2}^{+} $     &   \quad 2&\quad 2& \quad 1103.15 &\quad 1267.4  \\
\quad  &    $4_{1}^{+} $  &      \quad 2&\quad 2& \quad 1106.36& \quad 1215 \\
\quad  &    $0_{3}^{+} $      &  \quad 2&\quad 3 &\quad 1837.10 &\quad 1702.6 \\
 \quad  &    $3_{1}^{+} $    &    \quad 2&\quad 3& \quad 1521.67& \quad 1621.4 \\
 \quad &    $2_{3}^{+} $     &   \quad 3&\quad 1 &\quad 1580.56 &\quad 1577.1\\
 \quad  &    $4_{2}^{+} $  &      \quad 2&\quad 3& \quad 1602.9 &\quad 1671.4 \\
  \quad &    $6_{1}^{+} $      &  \quad 2&\quad 3 &\quad 1873.23 &\quad 1809 \\
  \quad&    $0_{4}^{+} $   &     \quad 3&\quad 0 &\quad 1968.66 &\quad 2131  \\
  \quad &    $2_{4}^{+} $     &   \quad 2&\quad 2& \quad 2036.9 &\quad 2360.3  \\
$ _{45}^{103}Rh $  &   $(1/2)_{1}^{-} $     &   \quad 3 &\quad 0 & \quad 0 & \quad 0  \\
  \quad &$(3/2)_{1}^{-} $       & \quad 2&\quad 1& \quad 294.984& \quad 276.8\\
  \quad  & $(5/2)_{1}^{-} $     &   \quad 2&\quad 1& \quad 357.408 &\quad 325.5 \\
   \quad&   $(1/2)_{2}^{-} $   &     \quad 3&\quad 0 &\quad 803.07& \quad 966.5 \\
   \quad& $(3/2)_{2}^{-} $      &  \quad 2&\quad 2 &\quad 803.07& \quad 647.1 \\
   \quad& $(7/2)_{1}^{-} $     &   \quad 2&\quad 2 &\quad 847.58& \quad 717.6 \\
  \quad& $(5/2)_{2}^{-} $      &  \quad 2&\quad 2& \quad 880.47& \quad 689.9 \\
  \quad & $(9/2)_{1}^{-} $      &  \quad 2&\quad 2 &\quad 920.1 &\quad 794.5 \\
  \quad & $(3/2)_{3}^{-} $      &  \quad 2&\quad 2& \quad 1277.04 &\quad 1247.9 \\
  \quad & $(13/2)_{1}^{-} $     &   \quad 2&\quad 3 &\quad 1637.64 &\quad 1551.1 \\
   \quad& $(15/2)_{1}^{-} $      &  \quad 1&\quad 4 &\quad 2221.2 &\quad 2255.5 \\
  \quad & $(17/2)_{1}^{-} $      &  \quad 1&\quad 4 &\quad 2345.35 &\quad 2400.9 \\

\hline
\end{tabular}
\end{center}
\end{table}
\begin{table}
\begin{center}
\begin{tabular}{p{9.8cm}}

\footnotesize Table 2c.Energy spectra for$ _{44}^{104}Ru $ - $ _{45}^{105}Rh $  isotope.The experimental
data for $ _{44}^{104}Ru $ - $ _{45}^{105}Rh $
supermultiplet are taken from \cite{30,31,36}.\\
\end{tabular}

\begin{tabular}{cccccc}

\hline
Nuclei  &$ J^{\pi}$ & \quad $ K $ & \quad $\nu_{d}$  & \quad $E_{exp} $(keV) & \quad $ E_{cal} $(keV)   \\
\hline
\quad\\
$ _{44}^{104}Ru $   &    $0_{1}^{+} $  &      \quad 4&\quad 0& \quad 0& \quad 0 \\
 \quad &    $2_{1}^{+} $      &  \quad 3&\quad 1 &\quad 358.02 &\quad 406.3 \\
 \quad&    $0_{2}^{+} $   &     \quad 4&\quad 0 &\quad 988.3 &\quad 1294.3  \\
 \quad &    $2_{2}^{+} $     &   \quad 3&\quad 2& \quad 893.1 &\quad 1139.4  \\
\quad  &    $4_{1}^{+} $  &      \quad 3&\quad 2& \quad 888.5& \quad 944.9 \\
\quad  &    $3_{1}^{+} $    &    \quad 2&\quad 3& \quad 1242.4& \quad 1233.5 \\
\quad &    $2_{3}^{+} $     &   \quad 3&\quad 1 &\quad 2285 &\quad 2103.8\\
$ _{45}^{105}Rh $ &$(1/2)_{1}^{-} $   &     \quad 3&\quad 0 &\quad 129.781& \quad 175.9\\
\quad &$(3/2)_{1}^{-}  $ & \quad 3&\quad 1& \quad 392.65& \quad 447.4 \\
\quad &$(5/2)_{1}^{-}  $ & \quad 3&\quad 1 &\quad 455.61& \quad 371.8\\
\quad &$(3/2)_{2}^{-} $  &\quad 2&\quad 2 &\quad 762.11& \quad 642.9\\
\quad &$(3/2)_{3}^{-} $  &\quad 2&\quad 1& \quad 783 &\quad 642.9\\
\quad &$(5/2)_{2}^{-} $  &\quad 2&\quad 2 &\quad 817 &\quad 1006.7\\
\quad &$(7/2)_{1}^{-} $  &\quad 2&\quad 2& \quad 817 &\quad 900.8\\
\quad &$(5/2)_{3}^{-} $  &\quad 2&\quad 3& \quad 866& \quad 1112.6\\
\quad &$(7/2)_{2}^{-} $  &\quad 2&\quad 3& \quad 898& \quad 1115.1\\
\quad &$(7/2)_{3}^{-} $  &\quad 2&\quad 2 &\quad 976& \quad 1009.2\\
\quad &$(9/2)_{1}^{-} $  &\quad 2&\quad 2 &\quad 976& \quad 802.5 \\
\quad &$(3/2)_{4}^{-} $  &\quad 2&\quad 3 &\quad 1147& \quad 1174.2\\
\quad &$(5/2)_{4}^{-} $  &\quad 3&\quad 1 &\quad 1147& \quad 987.7 \\
\hline
\end{tabular}
\end{center}
\end{table}
\begin{table}
\begin{center}
\begin{tabular}{p{9.8cm}}

\footnotesize Table 2d.Energy spectra for$ _{44}^{106}Ru $ - $ _{45}^{107}Rh $  isotope.The experimental
data for $ _{44}^{106}Ru $ - $ _{45}^{107}Rh $
supermultiplet are taken from \cite{32,33,36}.\\
\end{tabular}

\begin{tabular}{cccccc}

\hline
Nuclei   &$ J^{\pi}$ & \quad $ K $ & \quad $\nu_{d}$  & \quad $E_{exp} $(keV) & \quad $ E_{cal} $(keV)   \\
\hline
\quad\\
$ _{44}^{106}Ru $  &    $0_{1}^{+} $  &      \quad 4&\quad 0& \quad 0& \quad 0 \\
 \quad &    $2_{1}^{+} $      &  \quad 4&\quad 1 &\quad 270.07 &\quad 321.1 \\
 \quad&    $0_{2}^{+} $   &     \quad 4&\quad 0 &\quad 990.62 &\quad 1086.5  \\
 \quad &    $2_{2}^{+} $     &   \quad 3&\quad 2& \quad 792.31 &\quad 613.1  \\
\quad  &    $4_{1}^{+} $  &      \quad 3&\quad 2& \quad 714.69& \quad 811.3 \\
 \quad  &    $3_{1}^{+} $    &    \quad 3&\quad 3& \quad 1091.55& \quad 974 \\
 \quad &    $2_{3}^{+} $     &   \quad 4&\quad 1 &\quad 1392.21 &\quad 1173.6\\
 \quad  &    $4_{2}^{+} $  &      \quad 3&\quad 3& \quad 2367 &\quad 2321 \\
  \quad &    $6_{1}^{+} $      &  \quad 3&\quad 3 &\quad 1295.8 &\quad 1232.6 \\
$ _{45}^{107}Rh $ &$(1/2)_{1}^{-} $&        \quad 4&\quad 0 &\quad 268.36 &\quad 320 \\
\quad &$(3/2)_{1}^{-} $  &      \quad 3&\quad 1 &\quad 485.66 &\quad 613.4  \\
\quad &$(5/2)_{1}^{-} $  &      \quad 3&\quad 1 &\quad 543.84 &\quad 451.3\\
\quad & $(9/2)_{1}^{-} $ &       \quad 3&\quad 2 &\quad 559.97 &\quad 662.7 \\
\quad & $(3/2)_{2}^{-} $  &      \quad 3&\quad 2& \quad 752.55 &\quad 711.7\\
\quad &$(5/2)_{2}^{-} $   &     \quad 3&\quad 2 &\quad 877.75& \quad 1139.8 \\
\quad &$(3/2)_{3}^{-} $   &     \quad3&\quad 2& \quad 974.44& \quad 1146.5\\
\quad &$(5/2)_{3}^{-} $    &    \quad 2&\quad 3 &\quad 974.44& \quad 967.9\\
\quad &$(7/2)_{1}^{-} $ &       \quad 3&\quad 2 &\quad 974.44& \quad 994.3 \\
\quad  &$(3/2)_{4}^{-} $    &    \quad 2&\quad 3 &\quad 1009.76& \quad 1146.5\\
\hline
\end{tabular}
\end{center}
\end{table}
\begin{table}
\begin{center}
\begin{tabular}{p{9.8cm}}

\footnotesize Table 2e.Energy spectra for$ _{44}^{108}Ru $ - $ _{45}^{109}Rh $ isotope.The experimental
data for $ _{44}^{108}Ru $ - $ _{45}^{109}Rh $
supermultiplet are taken from \cite{34,35,36}.\\
\end{tabular}

\begin{tabular}{cccccc}

\hline
Nuclei   &$ J^{\pi}$ & \quad $ K $ & \quad $\nu_{d}$  & \quad $E_{exp} $(keV) & \quad $ E_{cal} $(keV)   \\
\hline
\quad\\
$ _{44}^{108}Ru $ &    $0_{1}^{+} $  &      \quad 5&\quad 0& \quad 0& \quad 0 \\
 \quad &    $2_{1}^{+} $      &  \quad 4&\quad 1 &\quad 242.24 &\quad 308.9 \\
 \quad&    $0_{2}^{+} $   &     \quad 5&\quad 0 &\quad 975.96 &\quad 1051.8  \\
 \quad &    $2_{2}^{+} $     &   \quad 4&\quad 2& \quad 707.82 &\quad 825.3 \\
\quad  &    $4_{1}^{+} $  &      \quad 4&\quad 2& \quad 665.2& \quad 752.3 \\
 \quad  &    $3_{1}^{+} $    &    \quad 3&\quad 3& \quad 974.8& \quad 1110.2 \\
 \quad &    $2_{3}^{+} $     &   \quad 4&\quad 1 &\quad 1249.19 &\quad 1126.9\\
 \quad  &    $4_{2}^{+} $  &      \quad 3&\quad 3& \quad 1183.03 &\quad 1254.2 \\
  \quad &    $6_{1}^{+} $      &  \quad 3&\quad 3 &\quad 1240 &\quad 1139.3 \\
  \quad &    $0_{3}^{+} $     &   \quad 3&\quad 3& \quad 1218.8 &\quad 1358.5  \\
$ _{45}^{109}Rh $&    $(1/2)_{1}^{-} $ &       \quad 4&\quad 0 &\quad 374.1& \quad 557.7  \\
 \quad&  $(3/2)_{1}^{-} $    &    \quad 4&\quad 1 &\quad 568.2& \quad 634.7\\
  \quad  & $(5/2)_{1}^{-} $  &      \quad 4&\quad 1 &\quad 623.2& \quad 806.1 \\
  \quad &  $(3/2)_{2}^{-} $  &      \quad 3&\quad 2& \quad 704.9& \quad 914.3 \\
 \quad & $(5/2)_{2}^{-} $    &    \quad 3&\quad 2 &\quad 856.1 &\quad 914.5 \\
 \quad& $(5/2)_{3}^{-} $     &   \quad 4&\quad 1& \quad 926.9& \quad 1154.7 \\
 \quad & $(3/2)_{3}^{-} $    &    \quad 4&\quad 1 &\quad 1162.3 &\quad 1040.4 \\
 \quad & $(3/2)_{4}^{-} $    &    \quad 3&\quad 2 &\quad 1214.3& \quad 1332.7 \\
 \quad& $(5/2)_{4}^{-} $     &   \quad 3&\quad 2 &\quad 1283.9 &\quad 1332.9 \\
 \quad & $(1/2)_{2}^{-} $    &    \quad 4&\quad 0& \quad 1631& \quad 1503.8\\

\hline
\end{tabular}
\end{center}
\end{table}

\begin{table}
\begin{center}
\begin{tabular}{p{7cm}}

\footnotesize Table 3.The coefficients of $T
(E_{2} ) $used in the present work
for the Ru-Rh Supermultiplets.Experimental values are taken form Refs. [26-36] .\\
\end{tabular}

\begin{tabular}{ccc}

\hline
Nucleus      &$ q_{B} $    &$ q_{f} $    \\
\hline
\quad\\

$ ^{100}Ru-^{101}Rh$  & 4.27       &              0  \\
$ ^{102}Ru-^{103}Rh$  & 1.5221      &           0\\
$ ^{104}Ru-^{105}Rh$  & 2.2782      &            0\\
$ ^{108}Ru-^{109}Rh$  & 2.2602      &           0\\
\hline
\end{tabular}
\end{center}
\end{table}
\begin{table}
\begin{center}
\begin{tabular}{p{9.2cm}}
\footnotesize Table 4a.B(E2)values for $
_{44}^{100}Ru$ and $ _{45}^{101}Rh$ isotopes.Experimental values are taken form Refs.[26,27] and are presented in Weisskopf units(W.u.).\\
\end{tabular}

\begin{tabular}{ccc}

\hline
Nucleus      &$ J_{i}^{\pi}\longrightarrow J_{j}^{\pi}  $    &$ \frac {B(E_{2} (W.u.)}{exp. \quad\quad\quad calc.} $    \\
\hline
\quad\\
$ _{44}^{100}Rh$   &   $2_{1}^{+} \longrightarrow 0_{1}^{+}$        &    35.6      \quad\quad     35.6  \\
 \quad\quad &  $0_{2}^{+} \longrightarrow 2_{1}^{+}$  &      35      \quad\quad     33.39\\
\quad\quad &  $4_{1}^{+} \longrightarrow 2_{1}^{+}$  & 51              \quad\quad      55.26 \\
\quad\quad &  $2_{2}^{+} \longrightarrow 2_{1}^{+}$  & 30.9               \quad\quad     15.71 \\
\quad\quad &  $2_{2}^{+} \longrightarrow 0_{1}^{+}$  & 1.9    \quad\quad    1.9\\

$ _{45}^{101}Rh$   &   $(3/2)_{1}^{-} \longrightarrow (1/2)_{1}^{-}$        &                      \quad\quad    2.2323   \\
 \quad\quad &  $(5/2)_{1}^{-} \longrightarrow (1/2)_{1}^{-}$  &                   \quad\quad     0 \\
\quad\quad &  $(5/2)_{2}^{-} \longrightarrow (1/2)_{1}^{-}$  &                              \quad\quad      0 \\
\quad\quad &  $(5/2)_{2}^{-} \longrightarrow (3/2)_{1}^{-}$  &                            \quad\quad     0.9945 \\
\quad\quad &  $(5/2)_{2}^{-} \longrightarrow (5/2)_{1}^{-}$  &                                 \quad\quad  0.0043 \\
\quad\quad &  $(7/2)_{1}^{-} \longrightarrow (3/2)_{1}^{-}$  &                                     \quad\quad    5.3894\\
\quad\quad &  $(9/2)_{1}^{-} \longrightarrow (5/2)_{1}^{-}$  &                                     \quad\quad    0.9934\\

\hline
\end{tabular}
\end{center}
\end{table}
\begin{table}
\begin{center}
\begin{tabular}{p{9.2cm}}
\footnotesize Table 4b.B(E2)values for $
_{44}^{102}Ru$ and $ _{45}^{103}Rh$ isotopes.Experimental values are taken form Refs.[28,29] and are presented in Weisskopf units(W.u.).\\
\end{tabular}

\begin{tabular}{ccc}

\hline
Nucleus      &$ J_{i}^{\pi}\longrightarrow J_{j}^{\pi}  $    &$ \frac {B(E_{2} (W.u.)}{exp. \quad\quad\quad calc.} $    \\
\hline
\quad\\
$ _{44}^{102}Ru$   &   $2_{1}^{+} \longrightarrow 0_{1}^{+}$        &    45.1      \quad\quad     32.1487  \\
 \quad\quad &  $0_{2}^{+} \longrightarrow 2_{1}^{+}$  &      35      \quad\quad    39.46\\
\quad\quad &  $4_{1}^{+} \longrightarrow 2_{1}^{+}$  & 66              \quad\quad      56.34 \\
\quad\quad &  $2_{2}^{+} \longrightarrow 2_{1}^{+}$  & 32               \quad\quad     8.47 \\
\quad\quad &  $2_{2}^{+} \longrightarrow 0_{1}^{+}$  & 1.14    \quad\quad    0.11\\

$ _{45}^{103}Rh$   &   $(3/2)_{1}^{-} \longrightarrow (1/2)_{1}^{-}$        &  36                    \quad\quad    55.98   \\
 \quad\quad &  $(5/2)_{1}^{-} \longrightarrow (1/2)_{1}^{-}$  &     44              \quad\quad     29.3311 \\
\quad\quad &  $(5/2)_{2}^{-} \longrightarrow (1/2)_{1}^{-}$  &       1.4                       \quad\quad      1.4002 \\
\quad\quad &  $(5/2)_{2}^{-} \longrightarrow (3/2)_{1}^{-}$  &        2.7                    \quad\quad     0 \\
\quad\quad &  $(5/2)_{2}^{-} \longrightarrow (5/2)_{1}^{-}$  &         4                        \quad\quad  1.34 \\
\quad\quad &  $(7/2)_{1}^{-} \longrightarrow (3/2)_{1}^{-}$  &         34                            \quad\quad    37.03\\
\quad\quad &  $(9/2)_{1}^{-} \longrightarrow (5/2)_{1}^{-}$  &         46                            \quad\quad    29.86\\

\hline
\end{tabular}
\end{center}
\end{table}
\begin{table}
\begin{center}
\begin{tabular}{p{9.2cm}}
\footnotesize Table 4c.B(E2)values for $
_{44}^{104}Ru$ and $ _{45}^{105}Rh$ isotopes.Experimental values are taken form Refs.[30,31] and are presented in Weisskopf units(W.u.).\\
\end{tabular}

\begin{tabular}{ccc}

\hline
Nucleus      &$ J_{i}^{\pi}\longrightarrow J_{j}^{\pi}  $    &$ \frac {B(E_{2} (W.u.)}{exp. \quad\quad\quad calc.} $    \\
\hline
\quad\\
$ _{44}^{104}Ru$   &   $2_{1}^{+} \longrightarrow 0_{1}^{+}$        &    57.9      \quad\quad     35.8  \\
 \quad\quad &  $0_{2}^{+} \longrightarrow 2_{1}^{+}$  &      25      \quad\quad    15.98\\
\quad\quad &  $4_{1}^{+} \longrightarrow 2_{1}^{+}$  & 83             \quad\quad      53.36 \\
\quad\quad &  $2_{2}^{+} \longrightarrow 2_{1}^{+}$  & 0.6               \quad\quad     4.94 \\
\quad\quad &  $2_{2}^{+} \longrightarrow 0_{1}^{+}$  & 2.8    \quad\quad    2.8 \\

$ _{45}^{105}Rh$   &   $(3/2)_{1}^{-} \longrightarrow (1/2)_{1}^{-}$        &                      \quad\quad    1.37   \\
 \quad\quad &  $(5/2)_{1}^{-} \longrightarrow (1/2)_{1}^{-}$  &                   \quad\quad     0.16 \\
\quad\quad &  $(5/2)_{2}^{-} \longrightarrow (1/2)_{1}^{-}$  &                              \quad\quad     0.006 \\
\quad\quad &  $(5/2)_{2}^{-} \longrightarrow (3/2)_{1}^{-}$  &                            \quad\quad     0.048 \\
\quad\quad &  $(5/2)_{2}^{-} \longrightarrow (5/2)_{1}^{-}$  &                                \quad\quad  1.32 \\
\quad\quad &  $(7/2)_{1}^{-} \longrightarrow (3/2)_{1}^{-}$  &                                     \quad\quad    2.6\\
\quad\quad &  $(9/2)_{1}^{-} \longrightarrow (5/2)_{1}^{-}$  &                                     \quad\quad    0.82\\

\hline
\end{tabular}
\end{center}
\end{table}

\begin{table}
\begin{tabular}{p{9.2cm}}
\footnotesize Table 4d.B(E2)values for $
_{44}^{108}Ru$ and $ _{45}^{109}Rh$ isotopes.Experimental values are taken form Refs.[34,35] and are presented in Weisskopf units(W.u.).\\
\end{tabular}

\begin{tabular}{ccc}

\hline
Nucleus      &$ J_{i}^{\pi}\longrightarrow J_{j}^{\pi}  $    &$ \frac {B(E_{2} (W.u.)}{exp. \quad\quad\quad calc.} $    \\
\hline
\quad\\
$ _{44}^{108}Ru$   &   $2_{1}^{+} \longrightarrow 0_{1}^{+}$        &    58      \quad\quad    58.03  \\
 \quad\quad &  $4_{1}^{+} \longrightarrow 2_{1}^{+}$  &   102         \quad\quad    95.3138\\
\quad\quad &  $2_{2}^{+} \longrightarrow 0_{1}^{+}$  &  0.5            \quad\quad      0.5 \\
\quad\quad &  $6_{1}^{+} \longrightarrow 2_{2}^{+}$  &   0.08             \quad\quad     0 \\
\quad\quad &  $6_{1}^{+} \longrightarrow 0_{1}^{+}$  & 0.004    \quad\quad   0 \\

$ _{45}^{109}Rh$   &   $(3/2)_{1}^{-} \longrightarrow (1/2)_{1}^{-}$        &                      \quad\quad    2.8083   \\
 \quad\quad &  $(5/2)_{1}^{-} \longrightarrow (1/2)_{1}^{-}$  &                   \quad\quad     0.3349 \\
\quad\quad &  $(5/2)_{2}^{-} \longrightarrow (1/2)_{1}^{-}$  &                              \quad\quad    2.1831 \\
\quad\quad &  $(5/2)_{2}^{-} \longrightarrow (3/2)_{1}^{-}$  &                            \quad\quad     0.0026 \\
\quad\quad &  $(5/2)_{2}^{-} \longrightarrow (5/2)_{1}^{-}$  &                                \quad\quad  0.0203 \\
\quad\quad &  $(7/2)_{1}^{-} \longrightarrow (3/2)_{1}^{-}$  &                                     \quad\quad    0.5573\\
\quad\quad &  $(9/2)_{1}^{-} \longrightarrow (5/2)_{1}^{-}$  &                                     \quad\quad    1.0983\\

\hline
\end{tabular}
\end{table}

\begin{table}
\begin{tabular}{p{9.3cm}}

\footnotesize Table 5. \footnotesize Parameters of Hamiltonian
(30) used in the calculation of the Zn-Cu Supermultiplets.
All parameters  are given in keV.\\
\end{tabular}

\begin{tabular}{ccccccccc}
\hline
Nucleus      &$ \mathcal{N} $    & $C_{s}$ & $C_{f}$ & $\alpha$ & $\beta'$  & $\gamma$ & $\sigma$ & $\phi$    \\
\hline
\hline
\quad\\
$ ^{62}Zn-^{61}Cu$   & 3& 0.4 & 0.87 & 889.63  & 0.077 & 15.067 & 217.2& 10.86\%  \\
\quad\\
$ ^{64}Zn-^{63}Cu$ & 4 & 0.51 & 1 & 946.2 & 0.2498 & 8.3402  & 189 & 8.92 \%   \\
\quad\\
$ ^{66}Zn-^{65}Cu$ & 5 & 0.65 & 1 & 739.6 & -0.136 & 5.947 &  181.92 & 7.05\%  \\
\quad\\
$ ^{68}Zn-^{67}Cu$ & 6 & 0.03 & 0.9 & 988.7 & 2.367 &  -43.11 & 202.43 & 6.9 \%  \\
\quad\\
$ ^{70}Zn-^{69}Cu$ & 6 & 0.01 & 0.4& 914.99 & 1.537  & -0.0468 & 228.18 & 10.23 \% \\

\hline
\end{tabular}
\end{table}
\begin{table}
\begin{center}
\begin{tabular}{p{8.8cm}}

\footnotesize Table 6a.Energy spectra for$ _{30}^{62}Zn $- $ _{29}^{61}Cu $  isotope. Experimental values are taken form Refs.[40,41]
\end{tabular}

\begin{tabular}{cccccc}

\hline
Nuclei   &$ J^{\pi}$ & \quad $ K $ & \quad $\nu_{d}$  & \quad $E_{exp} $(keV) & \quad $ E_{cal} $(keV)   \\
\hline
\quad\\
 $ _{30}^{62}Zn $ &    $0_{1}^{+} $  &      \quad 1&\quad 0& \quad 0& \quad 0 \\
 \quad &    $2_{1}^{+} $      &  \quad 1&\quad 1 &\quad 954 &\quad 785.4 \\
 \quad &    $2_{2}^{+} $     &   \quad 0&\quad 2& \quad 1804.7 &\quad 2266.4  \\
\quad  &    $4_{1}^{+} $  &      \quad 0&\quad 2& \quad 2186.1& \quad 1773.3 \\
\quad  &    $0_{2}^{+} $  &      \quad 1&\quad 0& \quad 2330& \quad 2425.7 \\
 \quad  &    $3_{1}^{+} $    &    \quad 0&\quad 3& \quad 2384.5& \quad 2686.5 \\
\quad &    $2_{3}^{+} $     &   \quad 1&\quad 1 &\quad 2810 &\quad 2826.6\\
 \quad  &    $4_{2}^{+} $  &      \quad 0&\quad 3& \quad 2743.5 &\quad 2807.1 \\
  \quad &    $2_{4}^{+} $      &  \quad 0&\quad 2 &\quad 2890 &\quad 3156 \\
  \quad&    $2_{5}^{+} $   &     \quad 1&\quad 1 &\quad 3470 &\quad 3387.5  \\
  \quad &    $2_{6}^{+} $     &   \quad 0&\quad 2& \quad 3640 &\quad 3484.7  \\
$ _{29}^{61}Cu $ &$(3/2)_{1}^{-} $&        \quad 1&\quad 0 &\quad 0 &\quad 0 \\
  \quad &$(1/2)_{1}^{-} $  &      \quad 0&\quad 1 &\quad 475.1 &\quad 524.3  \\
  \quad &$(5/2)_{1}^{-} $  &      \quad 0&\quad 1 &\quad 970.06 &\quad 726.1\\
  \quad & $(7/2)_{1}^{-} $ &       \quad 0&\quad 1 &\quad 1310.55 &\quad 1431.6 \\
  \quad & $(5/2)_{2}^{-} $  &      \quad 0&\quad 2& \quad 1394.2 &\quad 1855.7\\
  \quad &$(7/2)_{2}^{-} $   &     \quad 0&\quad 2 &\quad 1732.61& \quad 1961.3 \\
  \quad &$(5/2)_{3}^{-} $   &     \quad 0&\quad 2& \quad 1904.18& \quad 1855.7\\
  \quad &$(7/2)_{3}^{-} $    &    \quad 0&\quad 2 &\quad 1942.49& \quad 1961.3\\
  \quad &$(1/2)_{2}^{-} $ &       \quad 0&\quad 2 &\quad 2088.86& \quad 1734.7 \\
  \quad  &$(5/2)_{4}^{-} $    &    \quad 0&\quad 1 &\quad 2203.4& \quad 2415.81\\
  \quad  &$(9/2)_{1}^{-} $    &    \quad 0&\quad 2 &\quad 2295.09& \quad 2096.9\\
\hline
\end{tabular}
\end{center}
\end{table}
\begin{table}
\begin{center}
\begin{tabular}{p{8.8cm}}

\footnotesize Table 6b.Energy spectra for$ _{30}^{64}Zn $- $ _{29}^{63}Cu $  isotope.Experimental values are taken form Refs.[42,43]\\
\end{tabular}

\begin{tabular}{cccccc}

\hline
Nuclei   &$ J^{\pi}$ & \quad $ K $ & \quad $\nu_{d}$  & \quad $E_{exp} $(keV) & \quad $ E_{cal} $(keV)   \\
\hline
\quad\\
 $ _{30}^{64}Zn $ &    $0_{1}^{+} $  &      \quad 2&\quad 0& \quad 0& \quad 0 \\
 \quad &    $2_{1}^{+} $      &  \quad 1&\quad 1 &\quad 991.54 &\quad 1040.878 \\
 \quad &    $2_{2}^{+} $     &   \quad 1&\quad 2& \quad 1799.41 &\quad 1406.2  \\
\quad  &    $4_{1}^{+} $  &      \quad 1&\quad 2& \quad 2306.68& \quad 2487.7 \\
 \quad  &    $3_{1}^{+} $    &    \quad 1&\quad 1& \quad 2979.81& \quad 2820.4 \\
\quad &    $2_{3}^{+} $     &   \quad 1&\quad 1 &\quad 2793.7 &\quad 2770.4\\
 \quad  &    $4_{2}^{+} $  &      \quad 0&\quad 3& \quad 2736.6 &\quad 2887.2 \\
  \quad &    $2_{4}^{+} $      &  \quad 1&\quad 2 &\quad 3005.72 &\quad 3170.3 \\
  \quad&    $2_{5}^{+} $   &     \quad 0&\quad 4 &\quad 3094.6 &\quad 2820.4  \\
  \quad &    $0_{2}^{+} $     &   \quad 2&\quad 0& \quad 1910.3 &\quad 1725.6  \\
   \quad &    $0_{3}^{+} $     &   \quad 2&\quad 0& \quad 2609.45 &\quad 2324.5  \\
   \quad &    $4_{3}^{+} $     &   \quad 0&\quad 3& \quad 3078.43 &\quad 2887.2  \\
      \quad &    $3_{2}^{+} $     &   \quad 0&\quad 3& \quad 3094.6 &\quad 2820.4  \\
$ _{29}^{63}Cu $ &$(3/2)_{1}^{-} $&        \quad 1&\quad 0 &\quad 0 &\quad 0 \\
  \quad &$(1/2)_{1}^{-} $  &      \quad 1 &\quad 1 &\quad 669.67 &\quad 724.4  \\
  \quad &$(5/2)_{1}^{-} $  &      \quad 1 &\quad 1 &\quad 962.1 &\quad 1295.6\\
  \quad & $(7/2)_{1}^{-} $ &       \quad 1&\quad 1 &\quad 1327.01 &\quad 1105.4 \\
  \quad & $(5/2)_{2}^{-} $  &      \quad 0&\quad 2& \quad 1412.05 &\quad 1844.4\\
  \quad &$(3/2)_{1}^{-} $   &     \quad 1&\quad 1 &\quad 1547.05& \quad 1404.1 \\
  \quad &$(7/2)_{2}^{-} $   &     \quad 0&\quad 2& \quad 1904.5& \quad 1861.61 \\
  \quad &$(7/2)_{3}^{-} $    &    \quad 0&\quad 2 &\quad 2092.6& \quad 2204.5\\
  \quad &$(3/2)_{2}^{-} $ &       \quad 1&\quad 0 &\quad 2011.25& \quad 1802 \\
  \quad  &$(1/2)_{2}^{-} $    &    \quad 0&\quad 2 &\quad 2062.23& \quad 1777.7\\
  \quad  &$(9/2)_{1}^{-} $    &    \quad 0&\quad 2 &\quad 2207.88& \quad 2179.6\\
    \quad  &$(5/2)_{3}^{-} $    &    \quad 0&\quad 2 &\quad 2336.58& \quad 2595.9\\
\hline
\end{tabular}
\end{center}
\end{table}
\begin{table}
\begin{center}
\begin{tabular}{p{8.8cm}}

\footnotesize Table 6c.Energy spectra for$ _{30}^{66}Zn $- $ _{29}^{65}Cu $  isotope.Experimental values are taken form Refs.[44,45]
\end{tabular}

\begin{tabular}{cccccc}

\hline
Nuclei   &$ J^{\pi}$ & \quad $ K $ & \quad $\nu_{d}$  & \quad $E_{exp} $(keV) & \quad $ E_{cal} $(keV)   \\
\hline
\quad\\
 $ _{30}^{66}Zn $ &    $0_{1}^{+} $  &      \quad 2&\quad 0& \quad 0& \quad 0 \\
 \quad &    $2_{1}^{+} $      &  \quad 2&\quad 1 &\quad 1039.39 &\quad 1069.4 \\
 \quad &    $2_{2}^{+} $     &   \quad 1&\quad 2& \quad 1872.94 &\quad 2245.9  \\
\quad  &    $4_{1}^{+} $  &      \quad 1&\quad 2& \quad 2451.13& \quad 2660.1 \\
\quad &    $2_{3}^{+} $     &   \quad 2&\quad 1 &\quad 2780.55 &\quad 2772.5\\
 \quad  &    $4_{2}^{+} $  &      \quad 1&\quad 3& \quad 2765.69 &\quad 2913.6 \\
  \quad &    $2_{4}^{+} $      &  \quad 1&\quad 2 &\quad 2938.45 &\quad 2985.5 \\
  \quad &    $0_{2}^{+} $     &   \quad 2&\quad 0& \quad 2372.53 &\quad 2524.7  \\
   \quad &    $0_{3}^{+} $     &   \quad 2&\quad 0& \quad 3030 &\quad 3262.3  \\
   \quad &    $4_{3}^{+} $     &   \quad 1&\quad 3& \quad 3077.93 &\quad 3280.8  \\

$ _{29}^{65}Cu $ &$(3/2)_{1}^{-} $&        \quad 2&\quad 0 &\quad 0 &\quad 0 \\
  \quad &$(1/2)_{1}^{-} $  &      \quad 1 &\quad 1 &\quad 770.64 &\quad 794.4  \\
  \quad &$(5/2)_{1}^{-} $  &      \quad 1 &\quad 1 &\quad 1115.556 &\quad 1523\\
  \quad & $(7/2)_{1}^{-} $ &       \quad 1&\quad 1 &\quad 1481.83 &\quad 1564.6 \\
  \quad & $(5/2)_{2}^{-} $  &      \quad 1&\quad 2& \quad 1623.43 &\quad 1736.1\\
  \quad &$(3/2)_{2}^{-} $   &     \quad 1&\quad 1 &\quad 1725& \quad 1922.9 \\
  \quad &$(7/2)_{2}^{-} $   &     \quad 2&\quad 1& \quad 2094.34& \quad 2208.1 \\
  \quad &$(7/2)_{3}^{-} $    &    \quad 0&\quad 2 &\quad 2092.6& \quad 2204.5\\
  \quad &$(3/2)_{3}^{-} $ &       \quad 2&\quad 0 &\quad 2329.05& \quad 2449.2 \\
  \quad  &$(1/2)_{2}^{-} $    &    \quad 1&\quad 2 &\quad 2212.84& \quad 2118.9\\
  \quad  &$(5/2)_{3}^{-} $    &    \quad 1&\quad 2 &\quad 2107.44& \quad 2365.5\\
    \quad  &$(5/2)_{4}^{-} $    &    \quad 1&\quad 1 &\quad 2532.04& \quad 2691.6\\
\hline
\end{tabular}
\end{center}
\end{table}
\begin{table}
\begin{center}
\begin{tabular}{p{8.8cm}}

\footnotesize Table 6d.Energy spectra for$ _{30}^{68}Zn $- $ _{29}^{67}Cu $  isotope.Experimental values are taken form Refs.[46,47]
\end{tabular}

\begin{tabular}{cccccc}

\hline
Nuclei   &$ J^{\pi}$ & \quad $ K $ & \quad $\nu_{d}$  & \quad $E_{exp} $(keV) & \quad $ E_{cal} $(keV)   \\
\hline
\quad\\
 $ _{30}^{68}Zn $ &    $0_{1}^{+} $  &      \quad 3&\quad 0& \quad 0& \quad 0 \\
 \quad &    $2_{1}^{+} $      &  \quad 2&\quad 1 &\quad 1077.37 &\quad 1183.3 \\
 \quad &    $2_{2}^{+} $     &   \quad 2&\quad 2& \quad 1883.14 &\quad 1989.8  \\
\quad  &    $4_{1}^{+} $  &      \quad 2&\quad 2& \quad 2417.44& \quad 2379.2 \\
\quad &    $2_{3}^{+} $     &   \quad 2&\quad 1 &\quad 2338.29 &\quad 2470\\
 \quad  &    $4_{2}^{+} $  &      \quad 1&\quad 3& \quad 2955.9 &\quad 2893.5 \\
  \quad &    $2_{4}^{+} $      &  \quad 2&\quad 2 &\quad 2821.58 &\quad 2978.5 \\
  \quad &    $0_{2}^{+} $     &   \quad 3&\quad 0& \quad 1655.94 &\quad 1627.5  \\
   \quad &    $0_{3}^{+} $     &   \quad 3&\quad 0& \quad 3102.45 &\quad 3213.6  \\
   \quad &    $3_{1}^{+} $     &   \quad 1&\quad 3& \quad 3009.2 &\quad 3238.4  \\

$ _{29}^{67}Cu $ &$(3/2)_{1}^{-} $&        \quad 2&\quad 0 &\quad 0 &\quad 0 \\
  \quad &$(1/2)_{1}^{-} $  &      \quad 2 &\quad 1 &\quad 2272 &\quad 2212.6  \\
  \quad & $(3/2)_{2}^{-} $  &      \quad 2&\quad 1& \quad 1937.1 &\quad 2070\\
  \quad &$(3/2)_{2}^{-} $   &     \quad 1&\quad 1 &\quad 1725& \quad 1922.9 \\
  \quad &$(3/2)_{3}^{-} $   &     \quad 2&\quad 0& \quad 2272& \quad 2564.8 \\
  \quad &$(1/2)_{2}^{-} $    &    \quad 1&\quad 2 &\quad 2623.1& \quad 2707.9\\
  \quad &$(3/2)_{4}^{-} $ &       \quad 1&\quad 2 &\quad 2623.1& \quad 2595.1 \\
  \quad  &$(3/2)_{5}^{-} $    &    \quad 2&\quad 1 &\quad 2680.1& \quad 3058.7\\
  \quad  &$(1/2)_{3}^{-} $    &    \quad 2&\quad 1 &\quad 2841.1& \quad 3201.3\\
    \quad  &$(3/2)_{6}^{-} $    &    \quad 1&\quad 3 &\quad 2841.1& \quad 3090.4\\
\hline
\end{tabular}
\end{center}
\end{table}
\begin{table}
\begin{center}
\begin{tabular}{p{8.8cm}}

\footnotesize Table 6e.Energy spectra for$ _{30}^{70}Zn $- $ _{29}^{69}Cu $  isotope.Experimental values are taken form Refs.[48,49]
\end{tabular}

\begin{tabular}{cccccc}

\hline
Nuclei   &$ J^{\pi}$ & \quad $ K $ & \quad $\nu_{d}$  & \quad $E_{exp} $(keV) & \quad $ E_{cal} $(keV)   \\
\hline
\quad\\
 $ _{30}^{70}Zn $ &    $0_{1}^{+} $  &      \quad 3&\quad 0& \quad 0& \quad 0 \\
 \quad &    $2_{1}^{+} $      &  \quad 2&\quad 1 &\quad 884.8 &\quad 952.3 \\
 \quad &    $2_{2}^{+} $     &   \quad 2&\quad 2& \quad 1759.1 &\quad 2074  \\
\quad  &    $4_{1}^{+} $  &      \quad 2&\quad 2& \quad 1786.5& \quad 1923.1 \\
\quad &    $2_{3}^{+} $     &   \quad 2&\quad 1 &\quad 1957.7 &\quad 2389.6\\
 \quad  &    $4_{2}^{+} $  &      \quad 1&\quad 3& \quad 2693.6 &\quad 2458.6 \\
  \quad &    $2_{4}^{+} $      &  \quad 2&\quad 2 &\quad 2538 &\quad 2464.3 \\
  \quad &    $0_{2}^{+} $     &   \quad 3&\quad 0& \quad 1068.3 &\quad 1163.7  \\
   \quad &    $0_{3}^{+} $     &   \quad 3&\quad 0& \quad 2140.4 &\quad 2016.9  \\
   \quad &    $3_{1}^{+} $     &   \quad 1&\quad 3& \quad 2949.6 &\quad 2838.1  \\
    \quad &    $2_{6}^{+} $      &  \quad 2&\quad 1 &\quad 3635 &\quad 3438.1 \\
    \quad &    $0_{4}^{+} $      &  \quad 1&\quad 3 &\quad 3680 &\quad 3465.2 \\
   \quad &    $2_{7}^{+} $      &  \quad 2&\quad 2 &\quad 3710.6 &\quad 3904.6 \\

$ _{29}^{69}Cu $ &$(3/2)_{1}^{-} $&        \quad 2&\quad 0 &\quad 0 &\quad 0 \\
  \quad &$(1/2)_{1}^{-} $  &      \quad 2 &\quad 1 &\quad 1096 &\quad 1540.4  \\
  \quad &$(7/2)_{1}^{-} $  &      \quad 2 &\quad 1 &\quad 1710.8 &\quad 1539.7\\
  \quad & $(7/2)_{2}^{-} $ &       \quad 1&\quad 2 &\quad 1870 &\quad 2007.7 \\

\hline
\end{tabular}
\end{center}
\end{table}

\clearpage

\begin{table}
\begin{center}
\begin{tabular}{p{7cm}}

\footnotesize Table 7. The coefficients of $T
(E_{2} ) $used in the present work
for the Ru-Rh Supermultiplets..Experimental values are taken form Refs. [40-49] .\\\\
\end{tabular}

\begin{tabular}{ccc}

\hline
Nucleus      &$ q_{B} $    &$ q_{f} $    \\
\hline
\quad\\

$ ^{61}Cu-^{62}Zn$  & 2.4197       &              -0.9731  \\
$ ^{63}Cu-^{64}Zn$   & 2.1278      &           -0.28\\
$ ^{65}Cu-^{66}Zn$   & 8.775      &            0\\
$ ^{67}Cu-^{68}Zn$   & 4.9527      &           0\\
$ ^{69}Cu-^{70}Zn$   & 7.7603      &           0\\
\hline
\end{tabular}
\end{center}
\end{table}

\begin{table}
\begin{center}
\begin{tabular}{p{9.2cm}}
\footnotesize Table 8a. \footnotesize B(E2)values for
$ ^{61}Cu$ and $ ^{62}Zn$ isotopes. Experimental values are taken form Refs.[40,41]and are presented in Weisskopf units(W.u.).\\
\end{tabular}

\begin{tabular}{ccc}

\hline
Nucleus      &$ J_{i}^{\pi}\longrightarrow J_{j}^{\pi}  $    &$ \frac {B(E_{2} (W.u.)}{exp. \quad\quad\quad calc.} $    \\
\hline
\quad\\
$ ^{61}Cu$   &   $2_{1}^{+} \longrightarrow 0_{1}^{+}$        &    16.8 \quad       15.3989  \\
  &  $2_{2}^{+} \longrightarrow 2_{1}^{+}$  &      18      \quad          17.2585\\
 &  $4_{1}^{+} \longrightarrow 2_{1}^{+}$  & 26       \quad           25.7274 \\
 &  $2_{2}^{+} \longrightarrow 0_{1}^{+}$  & 0.32  \quad             0.14 \\
 &  $3_{1}^{+} \longrightarrow 2_{2}^{+}$  & 110                \quad            45.21\\
 &  $3_{1}^{+} \longrightarrow 2_{1}^{+}$  & 0.5             \quad           0.5466\\

$ ^{62}Zn$    &   $(5/2)_{1}^{-} \longrightarrow (1/2)_{1}^{-}$        &               17                        \quad           9.006   \\
 \quad\quad &  $(5/2)_{1}^{-} \longrightarrow (3/2)_{1}^{-}$  &  7.2 \quad     1.4164 \\
\quad\quad &  $(7/2)_{2}^{-} \longrightarrow (5/2)_{1}^{-}$  &     0.03  \quad      0 \\
\quad\quad &  $(7/2)_{2}^{-} \longrightarrow (3/2)_{1}^{-}$  &                   18            \quad     20.4437 \\
\quad\quad &  $(3/2)_{2}^{-} \longrightarrow (3/2)_{1}^{-}$  &                    1.5          \quad  1.1757 \\

\hline
\end{tabular}
\end{center}
\end{table}

\begin{table}
\begin{center}
\begin{tabular}{p{9.2cm}}
\footnotesize Table 8b. \footnotesize B(E2)values for $ ^{64}Zn$ and $ ^{63}Cu$  isotopes. Experimental values are taken form Refs.[42,43]and are presented in Weisskopf units(W.u.).\\
\end{tabular}

\begin{tabular}{ccc}

\hline
Nucleus      &$ J_{i}^{\pi}\longrightarrow J_{j}^{\pi}  $    &$ \frac {B(E_{2} (W.u.)}{exp. \quad\quad\quad calc.} $    \\
\hline
\quad\\
$ ^{64}Zn$    &   $2_{1}^{+} \longrightarrow 0_{1}^{+}$        &    20      \quad\quad     28.9867  \\
 \quad\quad &  $0_{2}^{+} \longrightarrow 2_{1}^{+}$  &      0.057      \quad\quad   3.2783\\
\quad\quad &  $4_{1}^{+} \longrightarrow 2_{1}^{+}$  &        12.2       \quad\quad      9.7 \\
\quad\quad &  $2_{2}^{+} \longrightarrow 2_{1}^{+}$  &         39       \quad\quad    35.186 \\
\quad\quad &  $2_{2}^{+} \longrightarrow 0_{1}^{+}$  & 0.24    \quad\quad    0.92\\
\quad\quad &  $0_{3}^{+} \longrightarrow 2_{2}^{+}$  & 1.3    \quad\quad    0.00\\

 $ ^{63}Cu$   &   $(1/2)_{1}^{-} \longrightarrow (3/2)_{1}^{-}$        &  15.2                    \quad\quad    15.7863   \\
 \quad\quad &  $(5/2)_{1}^{-} \longrightarrow (3/2)_{1}^{-}$  &     15.7              \quad\quad     12.7599 \\
\quad\quad &  $(7/2)_{1}^{-} \longrightarrow (3/2)_{1}^{-}$  &       12.7                       \quad\quad      11.05 \\
\quad\quad &  $(5/2)_{2}^{-} \longrightarrow (1/2)_{1}^{-}$  &        6              \quad\quad     5.3356 \\
\quad\quad &  $(5/2)_{2}^{-} \longrightarrow (3/2)_{1}^{-}$  &        1                       \quad\quad  0.002 \\
\quad\quad &  $(3/2)_{2}^{-} \longrightarrow (3/2)_{1}^{-}$  &         3.7                            \quad\quad    0.00\\
\quad\quad &  $(7/2)_{2}^{-} \longrightarrow (3/2)_{1}^{-}$  &         1.4                           \quad\quad    0.94 \\

\hline
\end{tabular}
\end{center}
\end{table}
\begin{table}
\begin{center}
\begin{tabular}{p{9.2cm}}
\footnotesize Table 8c. \footnotesize B(E2)values for $ ^{66}Zn $ and $ ^{65}Cu$ isotopes. Experimental values are taken form Refs.[44,45]and are presented in Weisskopf units(W.u.).\\
\end{tabular}

\begin{tabular}{ccc}

\hline
Nucleus      &$ J_{i}^{\pi}\longrightarrow J_{j}^{\pi}  $    &$ \frac {B(E_{2} (W.u.)}{exp. \quad\quad\quad calc.} $    \\
\hline
\quad\\
$ ^{66}Zn $   &   $2_{1}^{+} \longrightarrow 0_{1}^{+}$        &    17.5      \quad\quad     8.0597  \\
 \quad\quad &  $2_{2}^{+} \longrightarrow 2_{1}^{+}$  &      330      \quad\quad    330.3182\\
\quad\quad &  $2_{2}^{+} \longrightarrow 0_{1}^{+}$  &      0.032      \quad\quad    0.032\\
\quad\quad &  $4_{1}^{+} \longrightarrow 2_{1}^{+}$  & 18             \quad\quad      14.7453 \\
\quad\quad &  $2_{3}^{+} \longrightarrow 2_{1}^{+}$  & 0.13              \quad\quad     0 \\
\quad\quad &  $2_{3}^{+} \longrightarrow 0_{1}^{+}$  & 0.54    \quad\quad   4.6768 \\

$ ^{65}Cu$   &   $(1/2)_{1}^{-} \longrightarrow (3/2)_{1}^{-}$        &                  \quad\quad    0.003  \\
 \quad\quad &  $(5/2)_{1}^{-} \longrightarrow (3/2)_{1}^{-}$  &                  \quad\quad     31.5248 \\
\quad\quad &  $(7/2)_{1}^{-} \longrightarrow (5/2)_{1}^{-}$  &                              \quad\quad      0.001 \\
\quad\quad &  $(5/2)_{2}^{-} \longrightarrow (1/2)_{1}^{-}$  &                      \quad\quad     0.3517 \\
\quad\quad &  $(5/2)_{2}^{-} \longrightarrow (3/2)_{1}^{-}$  &                               \quad\quad  5.5851 \\

\hline
\end{tabular}
\end{center}
\end{table}

\begin{table}
\begin{tabular}{p{9.2cm}}
\footnotesize Table 8e. \footnotesize B(E2)values for $ ^{68}Zn $ and $ ^{67}Cu$ isotopes. Experimental values are taken form Refs.[46,47]and are presented in Weisskopf units(W.u.).\\
\end{tabular}

\begin{tabular}{ccc}

\hline
Nucleus      &$ J_{i}^{\pi}\longrightarrow J_{j}^{\pi}  $    &$ \frac {B(E_{2} (W.u.)}{exp. \quad\quad\quad calc.} $    \\
\hline
\quad\\
$ ^{68}Zn $  &   $2_{1}^{+} \longrightarrow 0_{1}^{+}$        &    14.69      \quad\quad    9.61  \\
 \quad\quad &  $4_{1}^{+} \longrightarrow 2_{1}^{+}$  &   10.8         \quad\quad    7.7922\\
\quad\quad &  $2_{2}^{+} \longrightarrow 0_{1}^{+}$  &  16           \quad\quad      8.425 \\
\quad\quad &  $0_{2}^{+} \longrightarrow 2_{1}^{+}$  &   5.5            \quad\quad     3.366 \\
\quad\quad &  $2_{2}^{+} \longrightarrow 0_{2}^{+}$  & 16   \quad\quad   8.425 \\
\quad\quad &  $2_{2}^{+} \longrightarrow 2_{1}^{+}$  & 28.6   \quad\quad   29.683 \\

$ ^{67}Cu$   &   $(1/2)_{1}^{-} \longrightarrow (3/2)_{1}^{-}$        &                  \quad\quad    0.058  \\
 \quad\quad &  $(5/2)_{1}^{-} \longrightarrow (3/2)_{1}^{-}$  &                  \quad\quad     0.3646 \\
\quad\quad &  $(7/2)_{1}^{-} \longrightarrow (5/2)_{1}^{-}$  &                              \quad\quad      0.00 \\
\quad\quad &  $(5/2)_{2}^{-} \longrightarrow (1/2)_{1}^{-}$  &                      \quad\quad     0.6664 \\
\quad\quad &  $(5/2)_{2}^{-} \longrightarrow (3/2)_{1}^{-}$  &                               \quad\quad  3.2829 \\

\hline
\end{tabular}
\end{table}

\begin{table}
\begin{tabular}{p{9.2cm}}
\footnotesize Table 8d. B(E2)values for $ ^{70}Zn $ and $ ^{69}Cu$ isotopes. Experimental values are taken form Refs.[48,49]and are presented in Weisskopf units(W.u.).\\
\end{tabular}

\begin{tabular}{ccc}

\hline
Nucleus      &$ J_{i}^{\pi}\longrightarrow J_{j}^{\pi}  $    &$ \frac {B(E_{2} (W.u.)}{exp. \quad\quad\quad calc.} $    \\
\hline
\quad\\
$ ^{70}Zn $   &   $2_{1}^{+} \longrightarrow 0_{1}^{+}$        &    16.5      \quad\quad    12.3661  \\
 \quad\quad &  $0_{2}^{+} \longrightarrow 2_{1}^{+}$  &   37.3         \quad\quad    37.1785\\
\quad\quad &  $2_{2}^{+} \longrightarrow 2_{1}^{+}$  &  69           \quad\quad      42.23 \\
\quad\quad &  $2_{2}^{+} \longrightarrow 0_{1}^{+}$  &   2.4             \quad\quad     1.275 \\
\quad\quad &  $2_{4}^{+} \longrightarrow 0_{1}^{+}$  & 0.15    \quad\quad   0.8133 \\

 $ ^{69}Cu$   &     $(1/2)_{1}^{-} \longrightarrow (3/2)_{1}^{-}$        &                  \quad\quad    0.145  \\
  \quad\quad &  $(5/2)_{1}^{-} \longrightarrow (3/2)_{1}^{-}$  &                  \quad\quad     1.002 \\
 \quad\quad &  $(7/2)_{1}^{-} \longrightarrow (5/2)_{1}^{-}$  &                              \quad\quad      0.002 \\
 \quad\quad &  $(5/2)_{2}^{-} \longrightarrow (1/2)_{1}^{-}$  &                      \quad\quad     0.328 \\
 \quad\quad &  $(5/2)_{2}^{-} \longrightarrow (3/2)_{1}^{-}$  &                               \quad\quad  0.418 \\

\hline
\end{tabular}
\end{table}
\clearpage

\begin{table}
\begin{tabular}{p{10.3cm}}
\footnotesize Table 9.Comparison of key observable in Ru isotopes with the E(5) symmetry.
\end{tabular}
\begin{tabular}{cccccc}
\hline
Nucleus      & $ R_{\frac{4}{2}} $    & $C_{s}$ & $\frac{E(0_{2}^{+})}{E(2_{1}^{+})}$ & $B(\frac{E2;4_{1}^{+} \longrightarrow 2_{1}^{+}}{E2;2_{1}^{+} \longrightarrow 0_{1}^{+}})$ & $B(\frac{E2;0_{2}^{+}\longrightarrow 2_{1}^{+}}{E2;2_{1}^{+}\longrightarrow 0_{1}^{+}})$    \\
\hline
\quad\\
E(5) & 2.19 & - & 3.03 & 1.68 & 0.68 \\

$ ^{100}Ru$  & 2.24 & 0.38 & 2.5544 & 1.55 & 0.9379\\
\quad\\
$ ^{102}Ru$ & 2.37 & 0.23 & 1.6227 & 1.7524 & 1.2274   \\
\quad\\
$^{104}Ru$ & 2.32 & 0.54 & 3.18 & 1.49 & 0.446  \\
\quad\\
$ ^{106}Ru$ & 2.526 & 0.34 & 3.38 & - & -  \\
\quad\\
$ ^{108}Ru$ & 2.435 & 0.65 & 3.4 & 1.64 & - \\
\hline
\end{tabular}
\end{table}
\begin{table}
\begin{tabular}{p{10.3cm}}
\footnotesize Table 10.Comparison of key observable in Zn isotopes with the E(5) symmetry.\\
\end{tabular}
\begin{tabular}{cccccc}
\hline
Nucleus      & $ R_{\frac{4}{2}} $    & $C_{s}$ & $\frac{E(0_{2}^{+})}{E(2_{1}^{+})}$ & $B(\frac{E2;4_{1}^{+} \longrightarrow 2_{1}^{+}}{E2;2_{1}^{+} \longrightarrow 0_{1}^{+}})$ & $B(\frac{E2;0_{2}^{+}\longrightarrow 2_{1}^{+}}{E2;2_{1}^{+}\longrightarrow 0_{1}^{+}})$    \\
\hline
E(5) & 2.19 & - & 3.03 & 1.68 & 0.68 \\
$ ^{62}Zn$  & 2.257 & 0.51 & 3.088 & 1.55 & 0.9379\\
\quad\\
$  ^{64}Zn$ & 2.391 & 0.4 & 1.657 & 1.7524 & 1.2274   \\
\quad\\
$ ^{66}Zn$ & 2.48 & 0.65 & 2.36 & 1.49 & 0.446  \\
\quad\\
$ ^{68}Zn$ & 2.014 & 0.03 & 1.375 & - & -  \\
\quad\\
$  ^{70}Zn$ & 2.0194 & 0.01 & 1.22 & 1.64 & - \\
\hline
\end{tabular}
\end{table}

\clearpage

\begin{figure}
\begin{center}
\includegraphics[height=8cm]{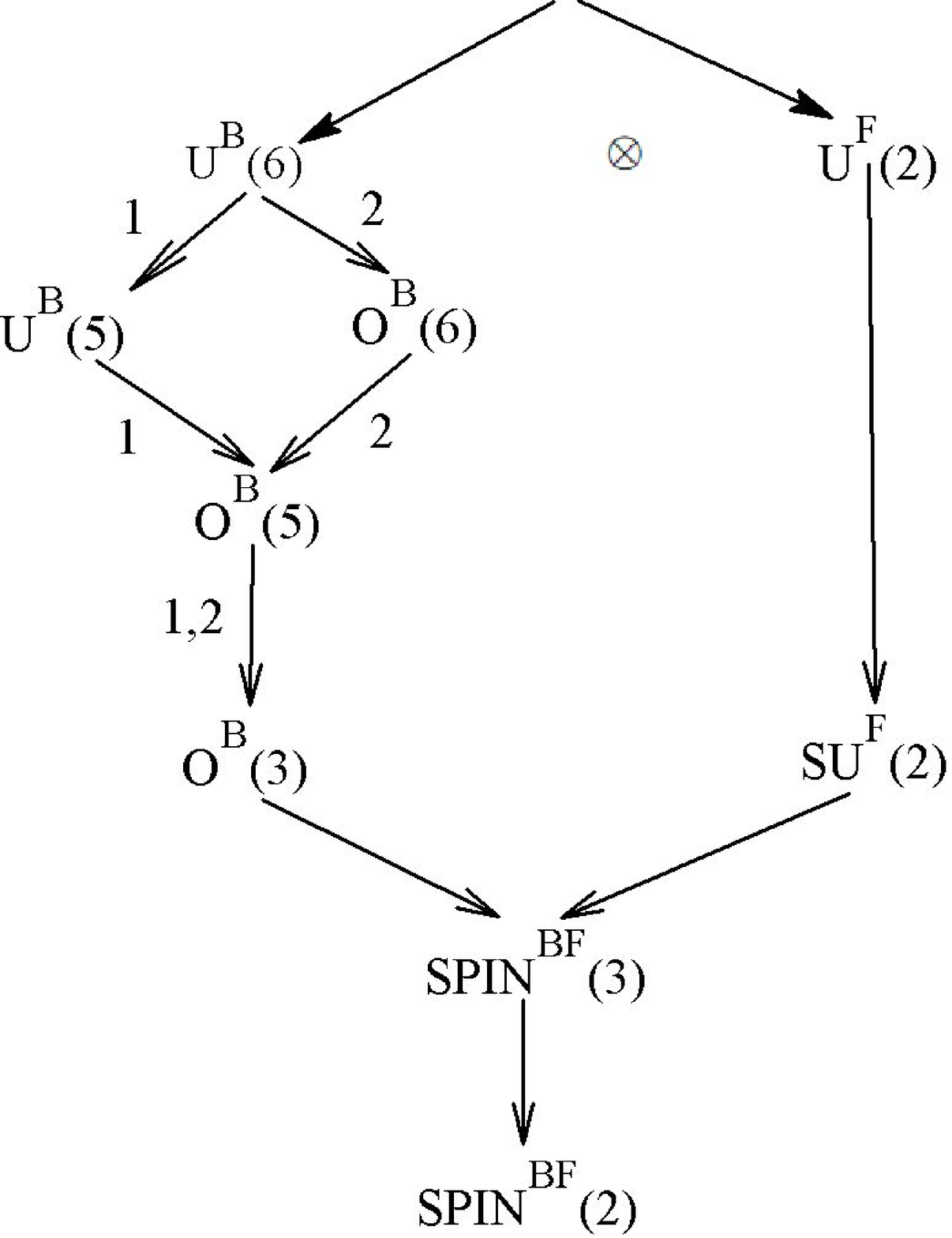}
\caption{The lattice of algebras in the U(6/2) supersymmetry scheme.}
\label{fig:1}
\end{center}
\end{figure}
\begin{figure}[htb]
\begin{center}
\includegraphics[height=8cm]{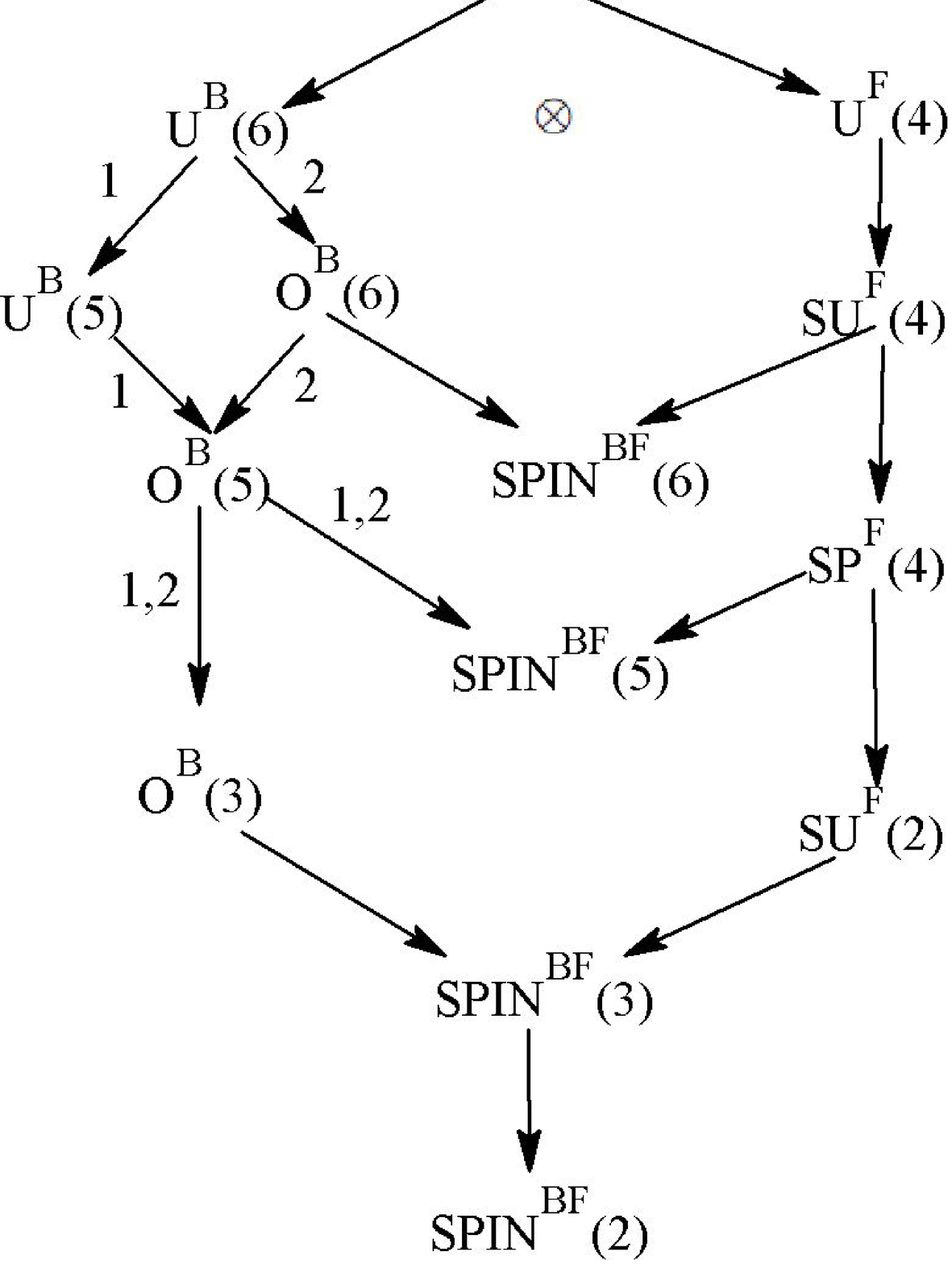}
\caption{The Lattice of algebras in the U(6/4) supersymmetry scheme.\label{fig:2}}
\end{center}
\end{figure}
\begin{figure}
\includegraphics[height=4.5cm]{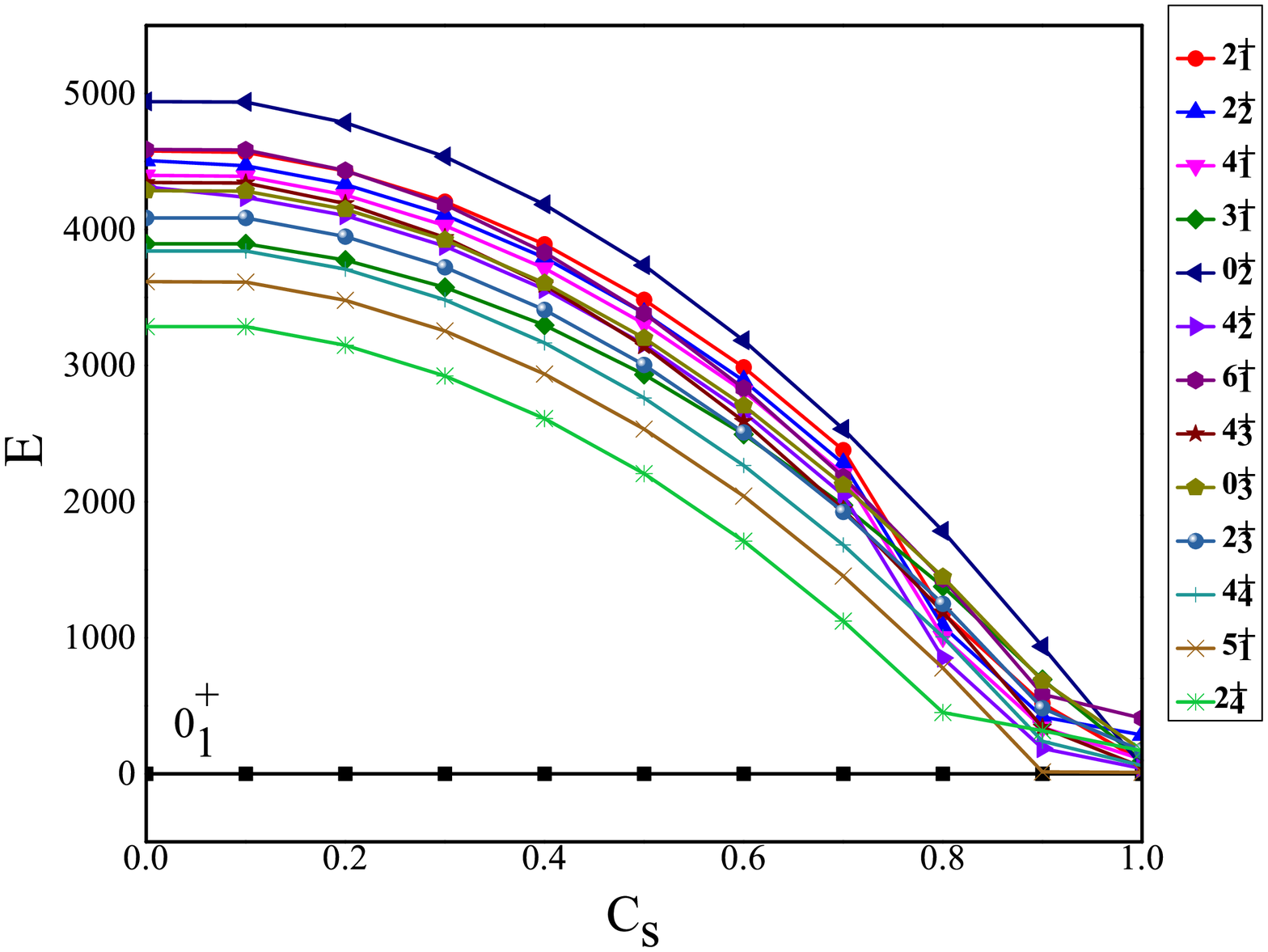}
\includegraphics[height=4cm]{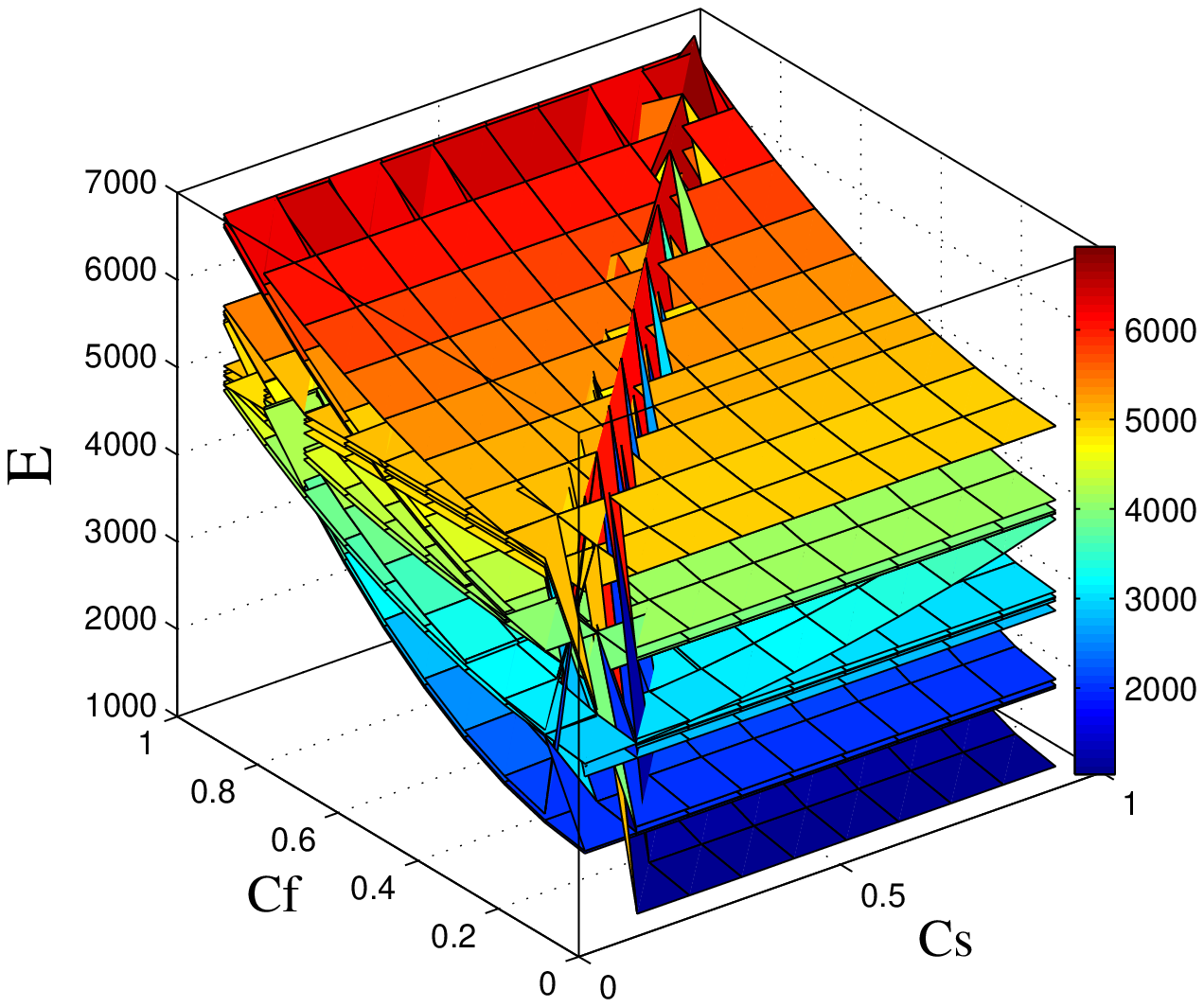}
\caption{Energy levels as a function of  $ C_{s} $ control parameter for a even-even nuclei (left panel) and for odd-A nuclei as a function of the $ C_{s} $ and $ C_{f}$ control parameters(right
panel)
in the Hamiltonian (27) for $N=10$ bosons
with$ \alpha=1000 $, $\beta=3.14 $, $ \delta=-5.26 $, $\gamma=0.0439$.\label{fig:3}}
\end{figure}
\begin{figure}
\includegraphics[height=4.5cm]{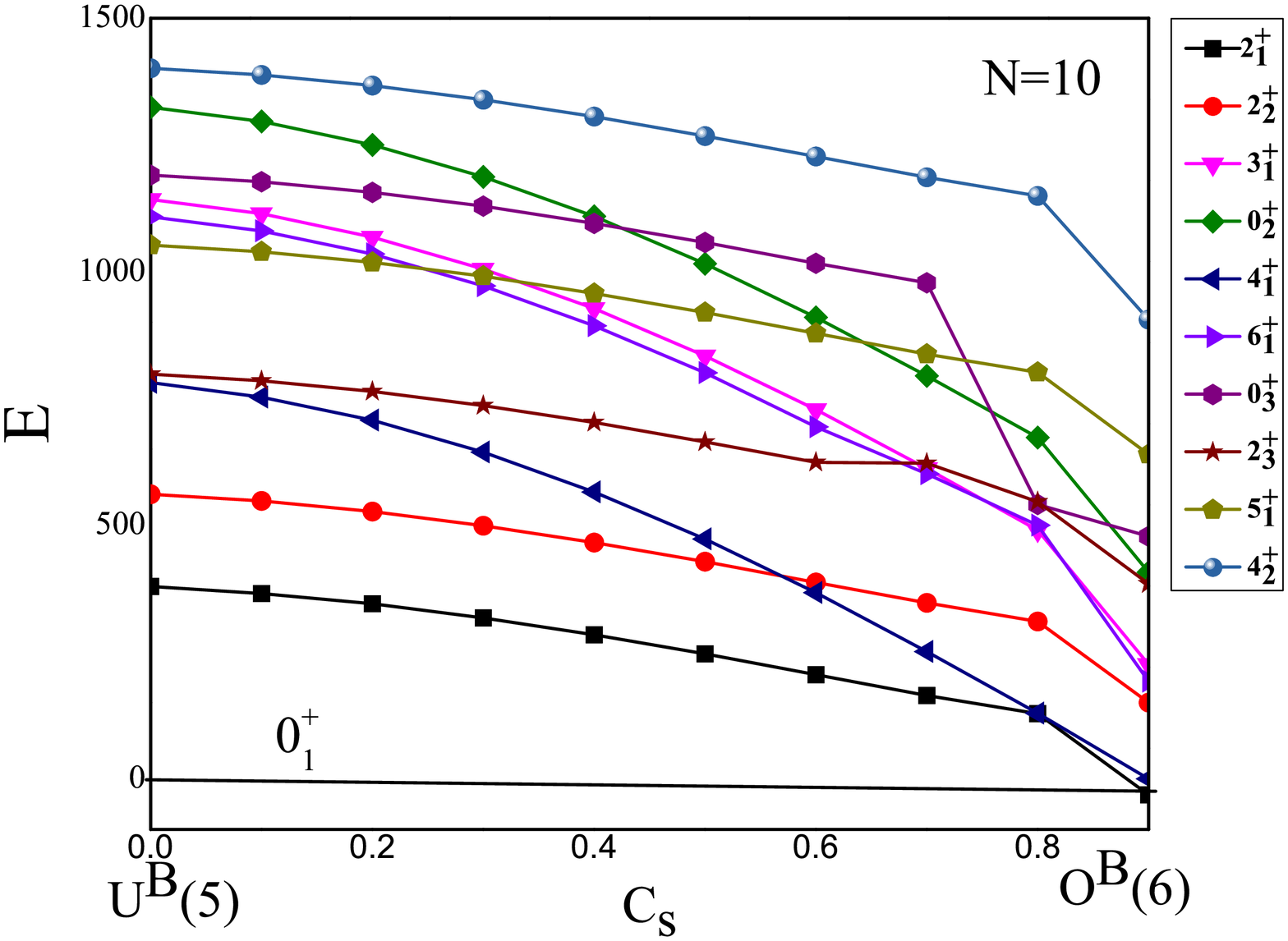}
\includegraphics[height=4cm]{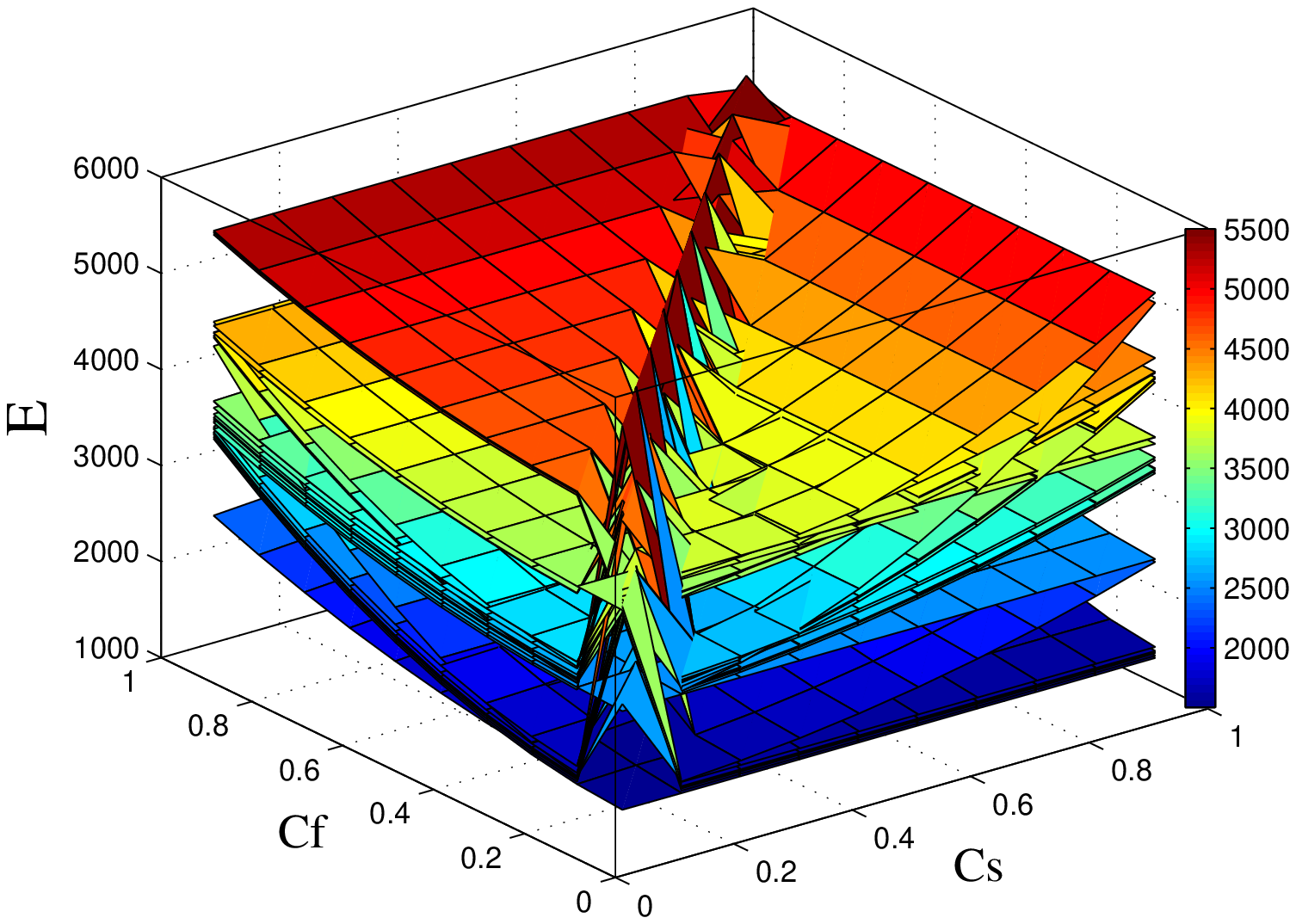}
\caption{Energy levels as a function of  $ C_{s} $ control parameter for a even-even nuclei (left panel) and for odd-A nuclei as a function of the $ C_{s} $ and $ C_{f}$ control parameters(right
panel)
in the Hamiltonian (30) for $N=10$ bosons
with $\alpha=1000,\beta'=-1.29,\gamma=6.05$.\label{fig:4}}
\end{figure}
\begin{figure}
\includegraphics[height=4.5cm]{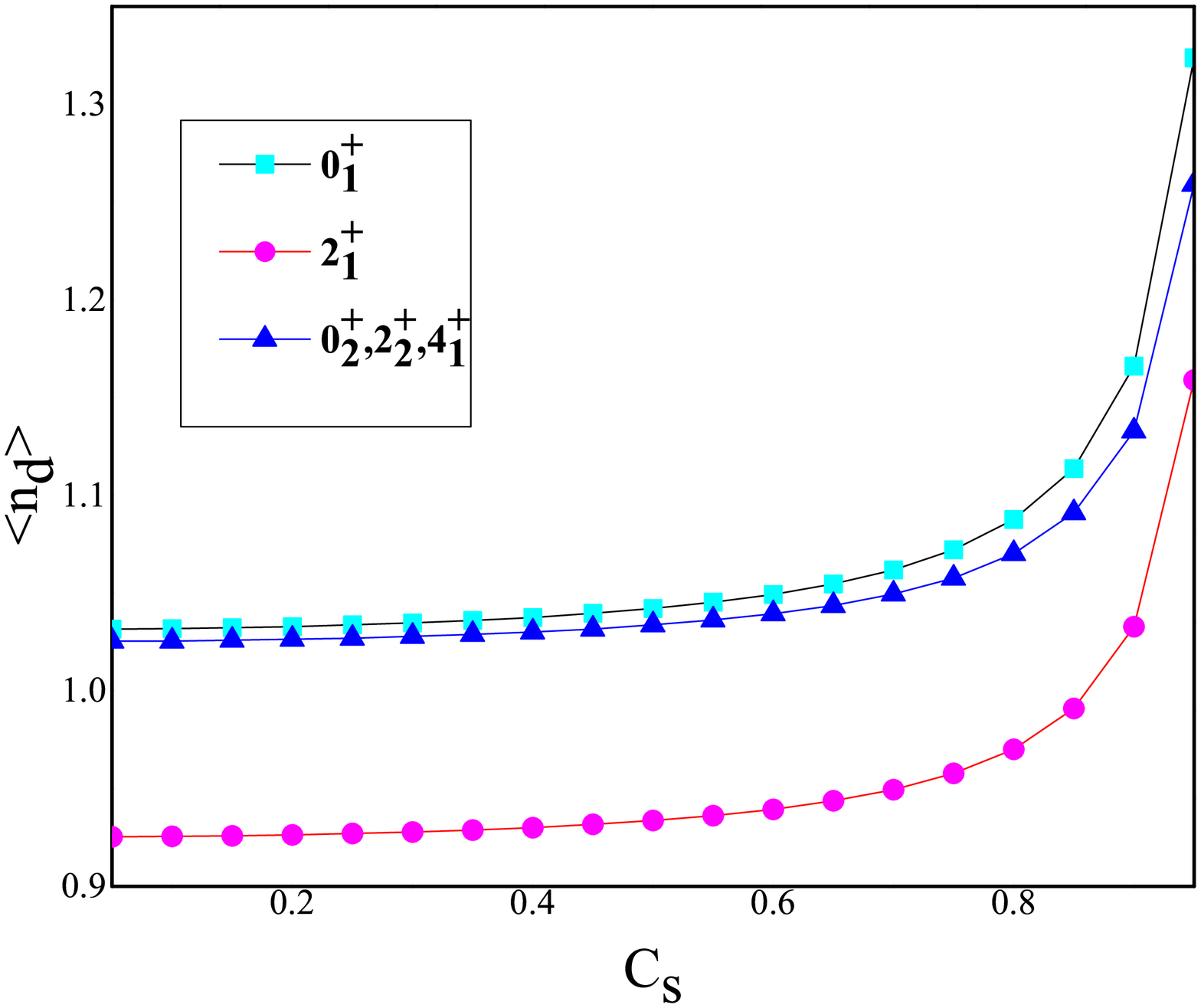}
\includegraphics[height=4cm]{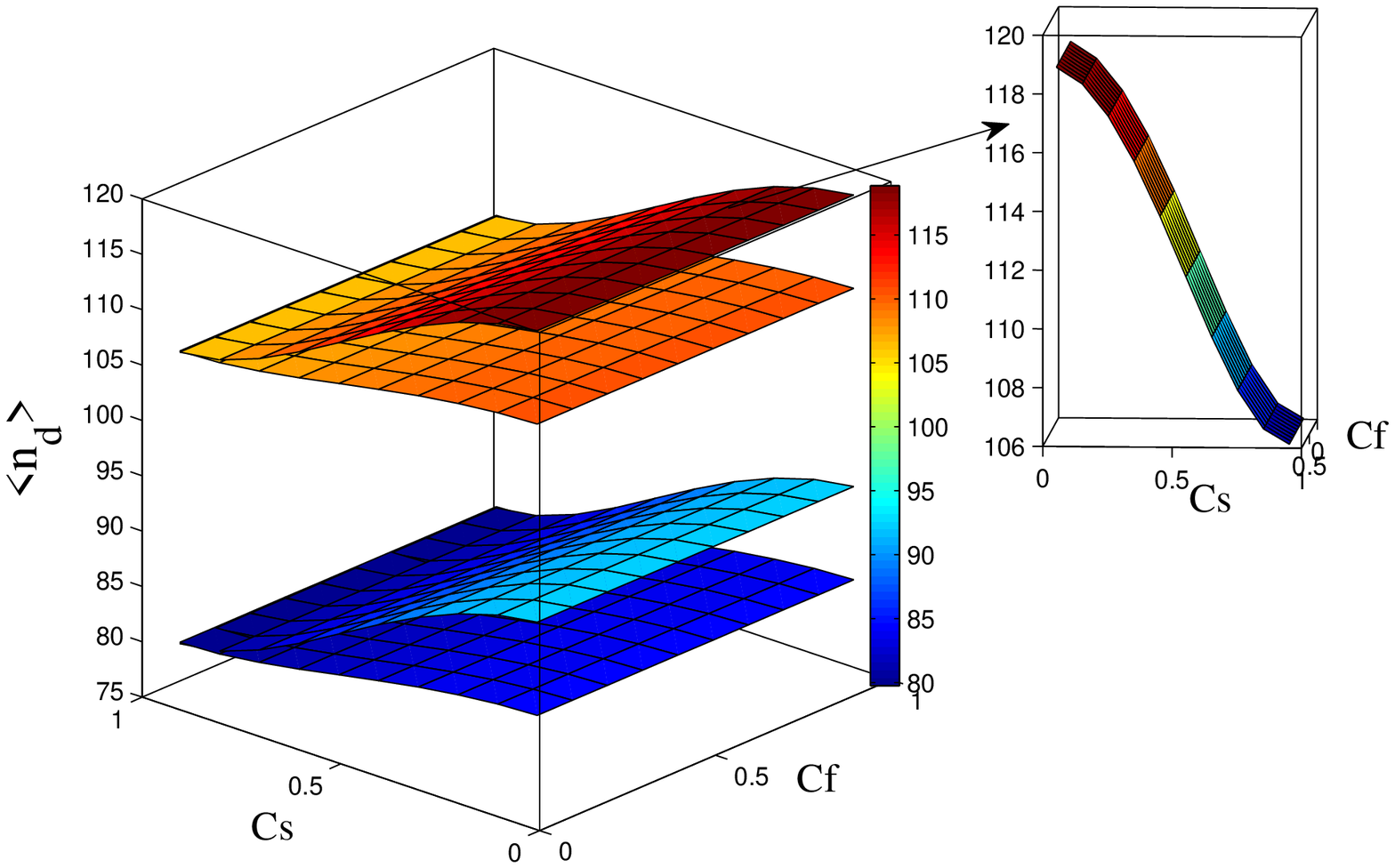}
\caption{The expectation values of the $d$-boson number operator
for the lowest states as a function of  $ C_{s} $ control parameter for an even-even nuclei (left panel) and for odd-A nuclei as a function of the $ C_{s} $ and $ C_{f}$ control parameters(right
panel) for j=1/2.\label{fig:5}}
\end{figure}
\begin{figure}
\includegraphics[height=4.5cm]{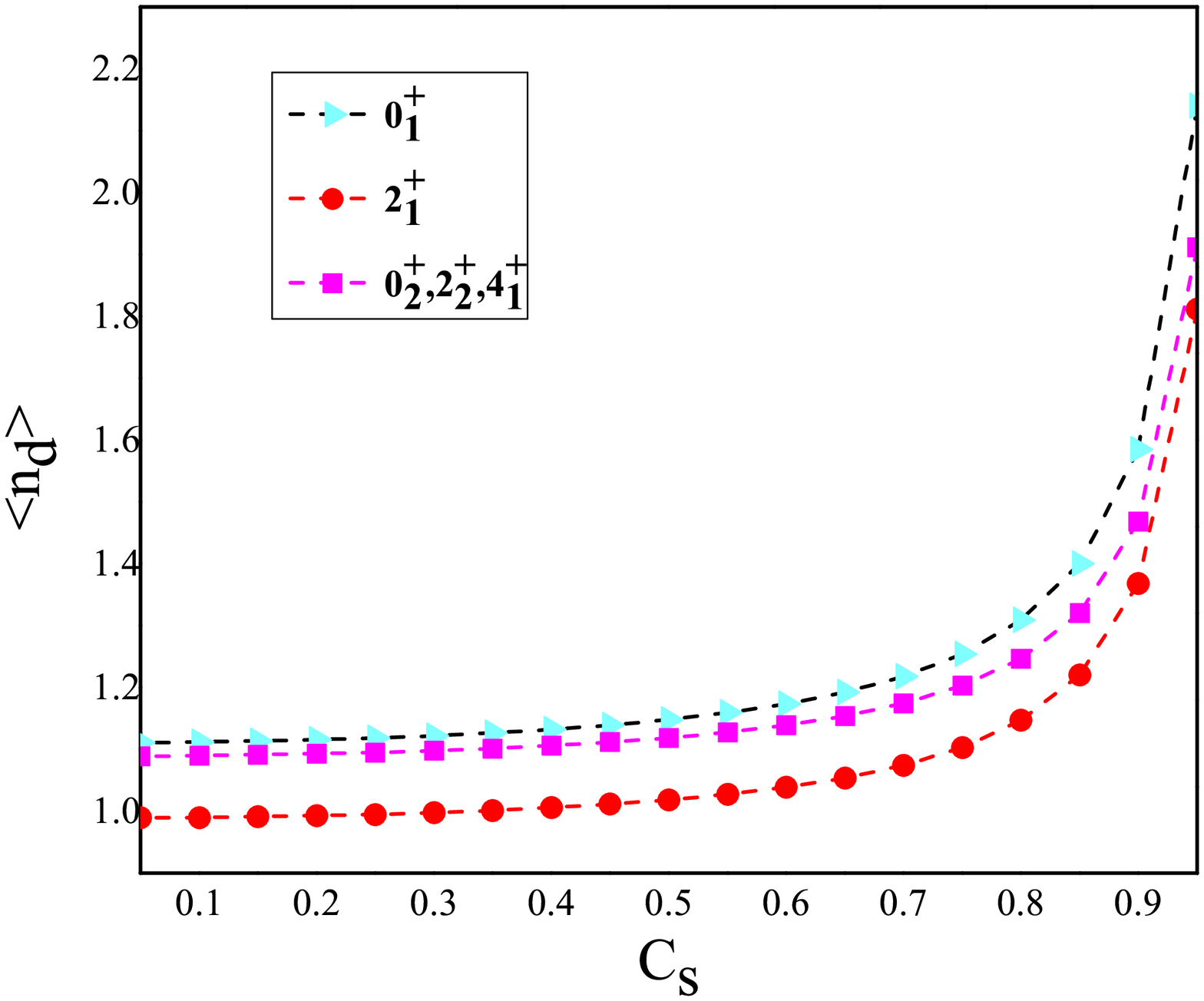}
\includegraphics[height=4cm]{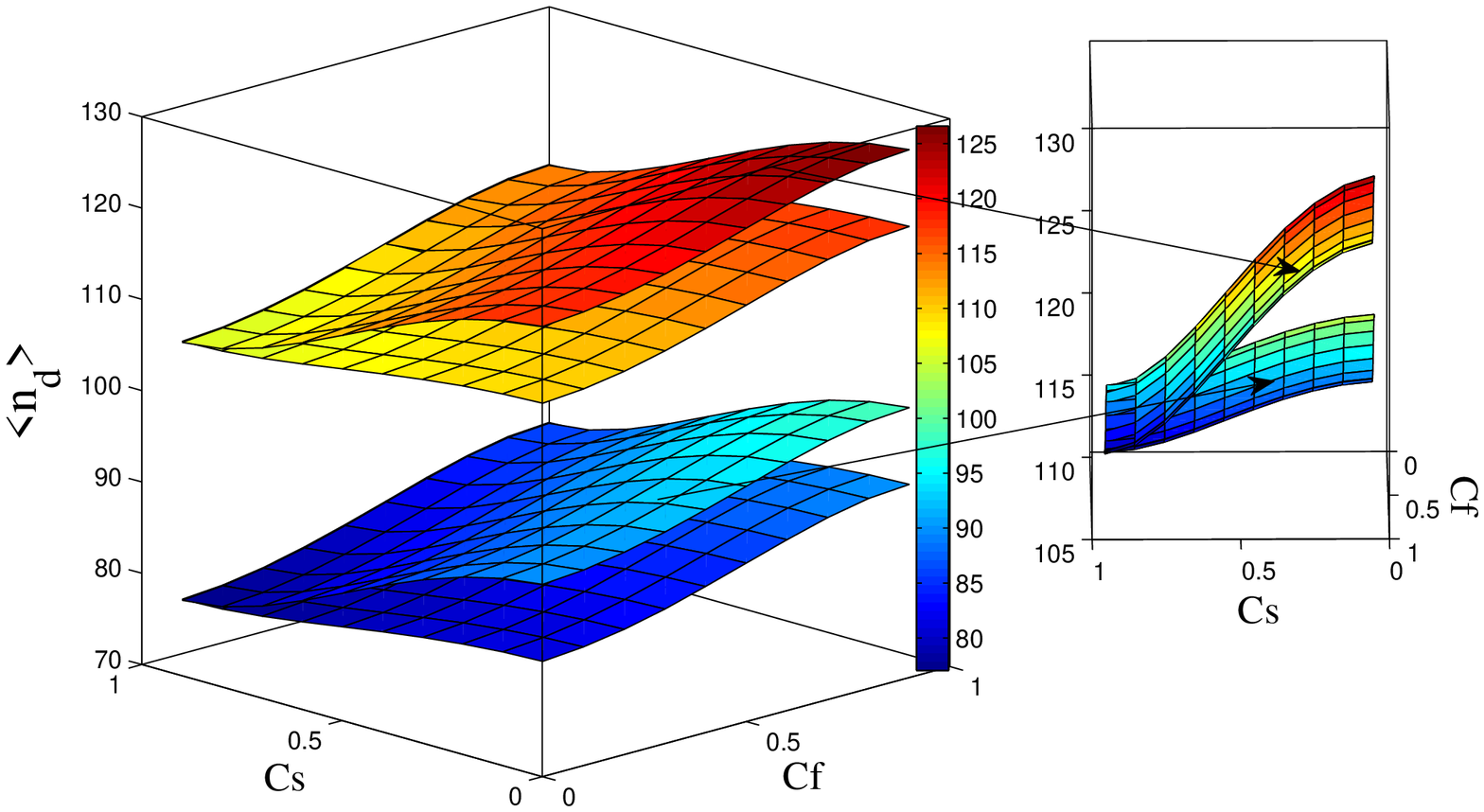}
\caption{The expectation values of the $d$-boson number operator
for the lowest states as a function of  $ C_{s} $ control parameter for an even-even nuclei (left panel) and for odd-A nuclei as a function of the $ C_{s} $ and $ C_{f}$ control parameters(right
panel) for j=3/2.\label{fig:6}}
\end{figure}
\begin{figure}
\includegraphics[height=6cm]{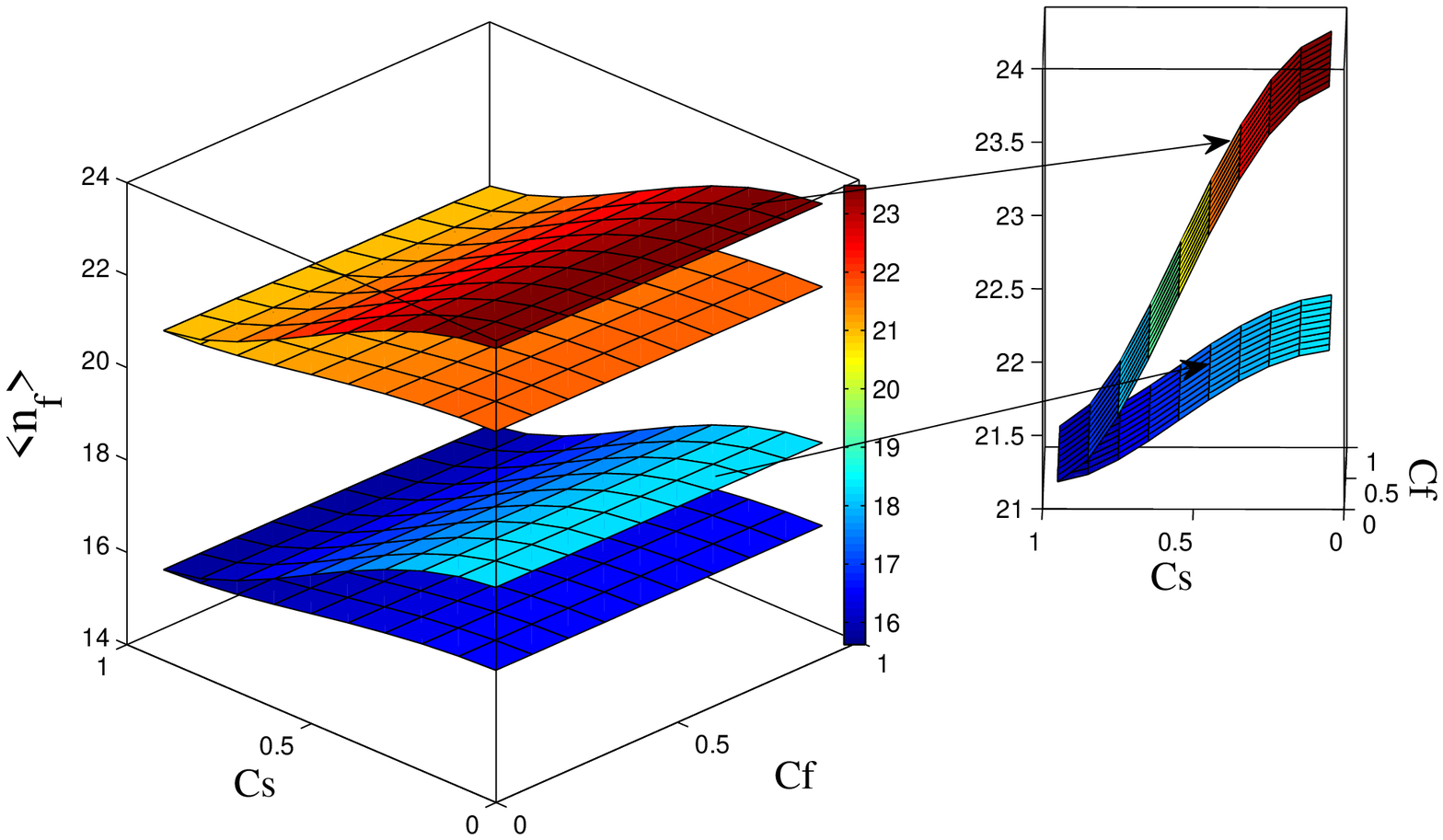}
\caption{The expectation values of the fermion number operator for odd-A nuclei for the lowest states as a function of  $ C_{s} $ and $ C_{f}$ control parameters.\label{fig:7}}
\end{figure}
\begin{figure}[htb]
\includegraphics[height=6cm]{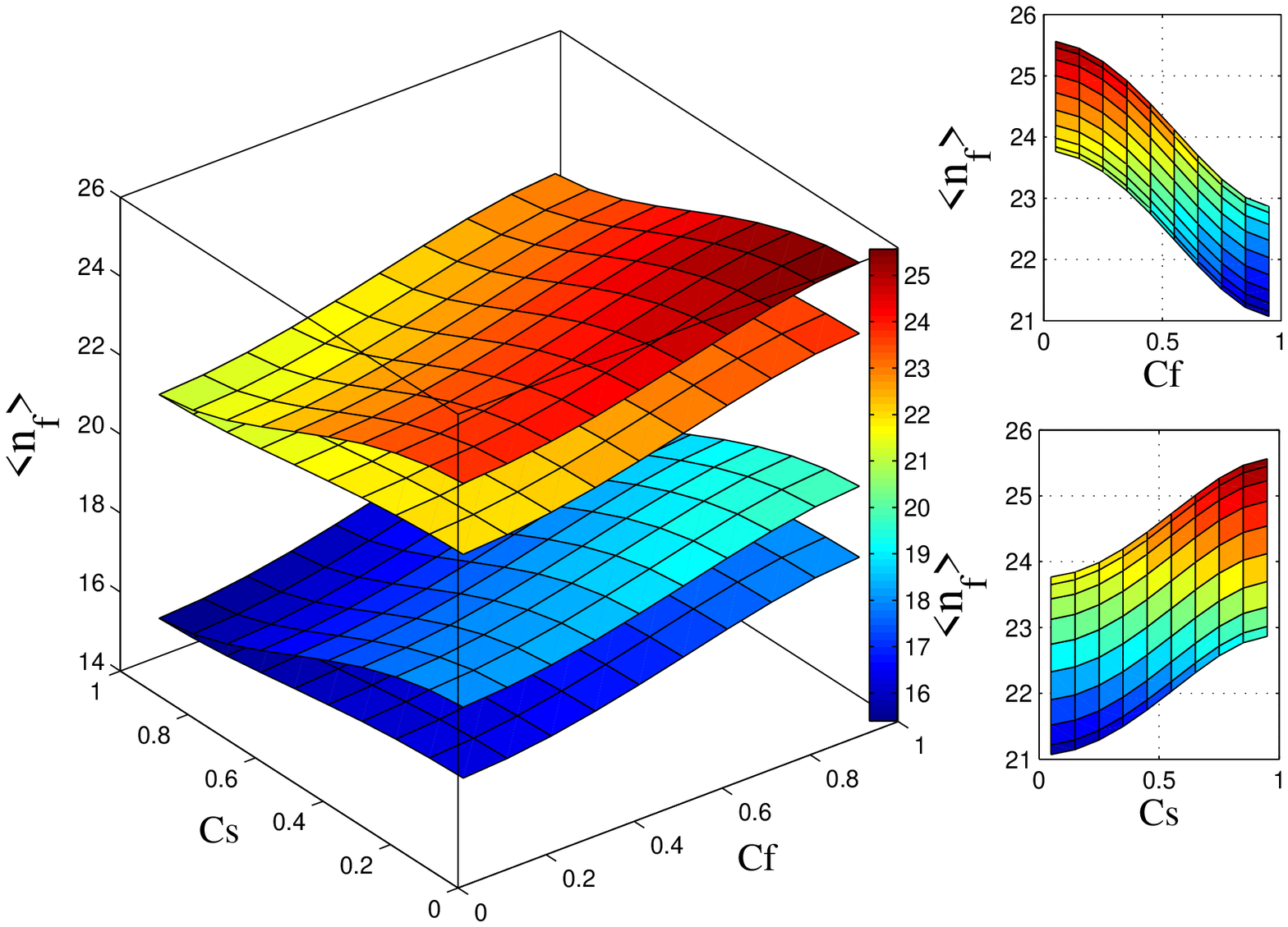}
\caption{The expectation values of the fermion number operator for odd-A nuclei for the lowest states as a function of  $ C_{s} $ and $ C_{f}$ control parameters.\label{fig:8}}
\end{figure}
\begin{figure}
\begin{center}
\includegraphics[height=6cm]{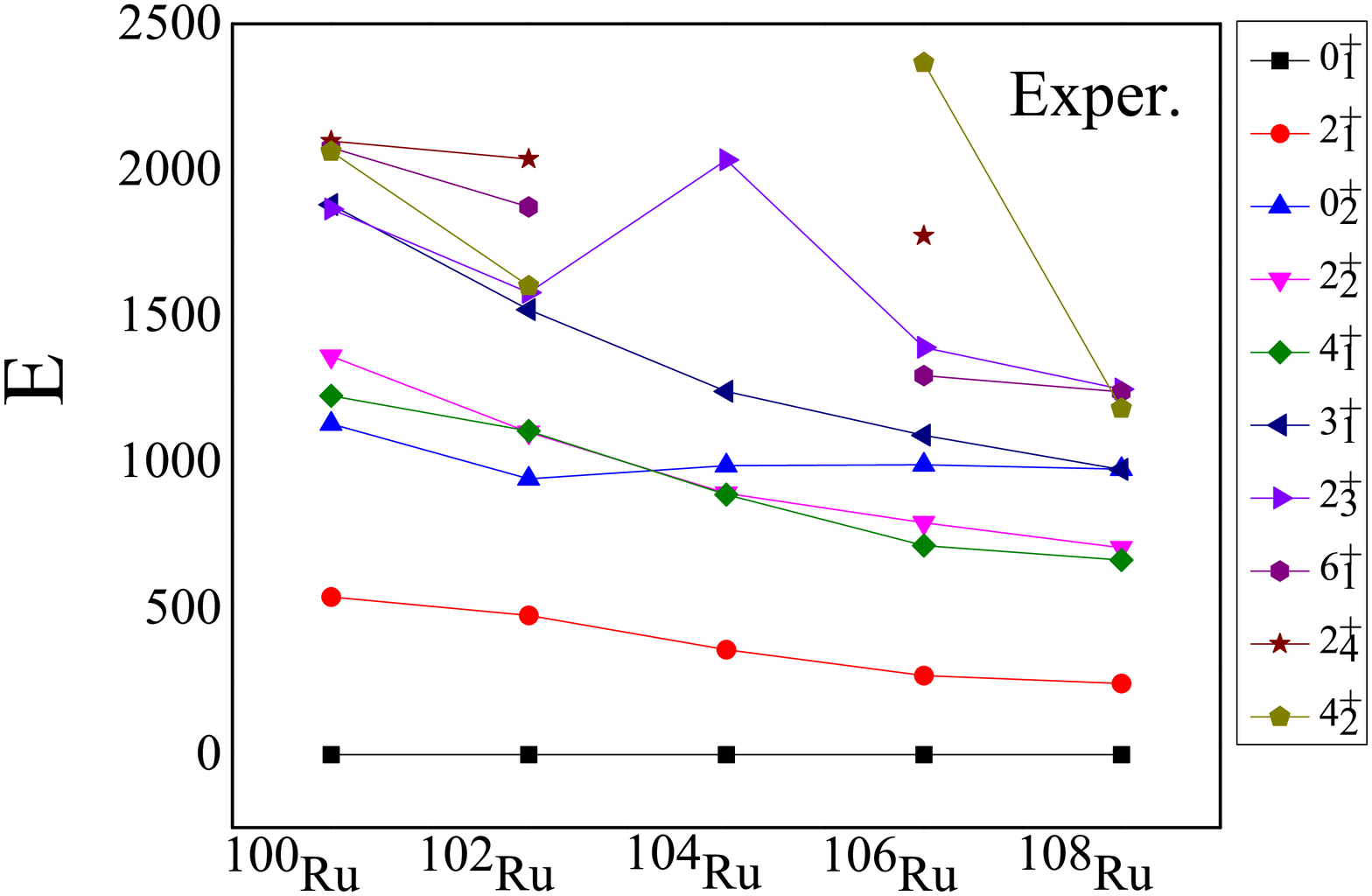}
\includegraphics[height=6cm]{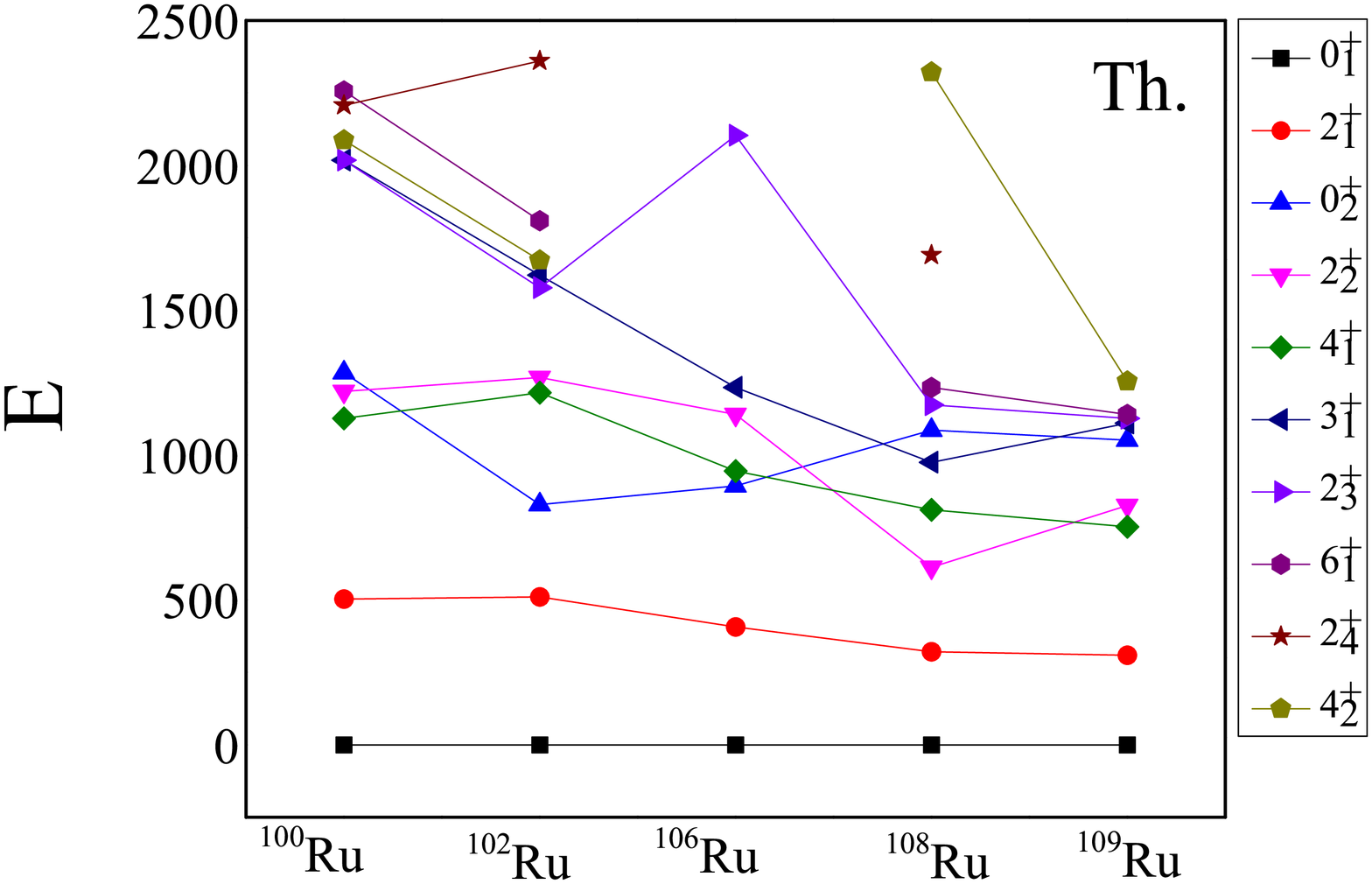}
\caption{Comparison between calculated and experimental spectra
of positive parity states in Ru Isotopes. The parameters of the
calculation are given in Tables 2. In the experimental spectra,
taken from \cite{26,28,30,32,34,36}.\label{fig:9}}
\end{center}
\end{figure}
\begin{figure}
\begin{center}
\includegraphics[height=6cm]{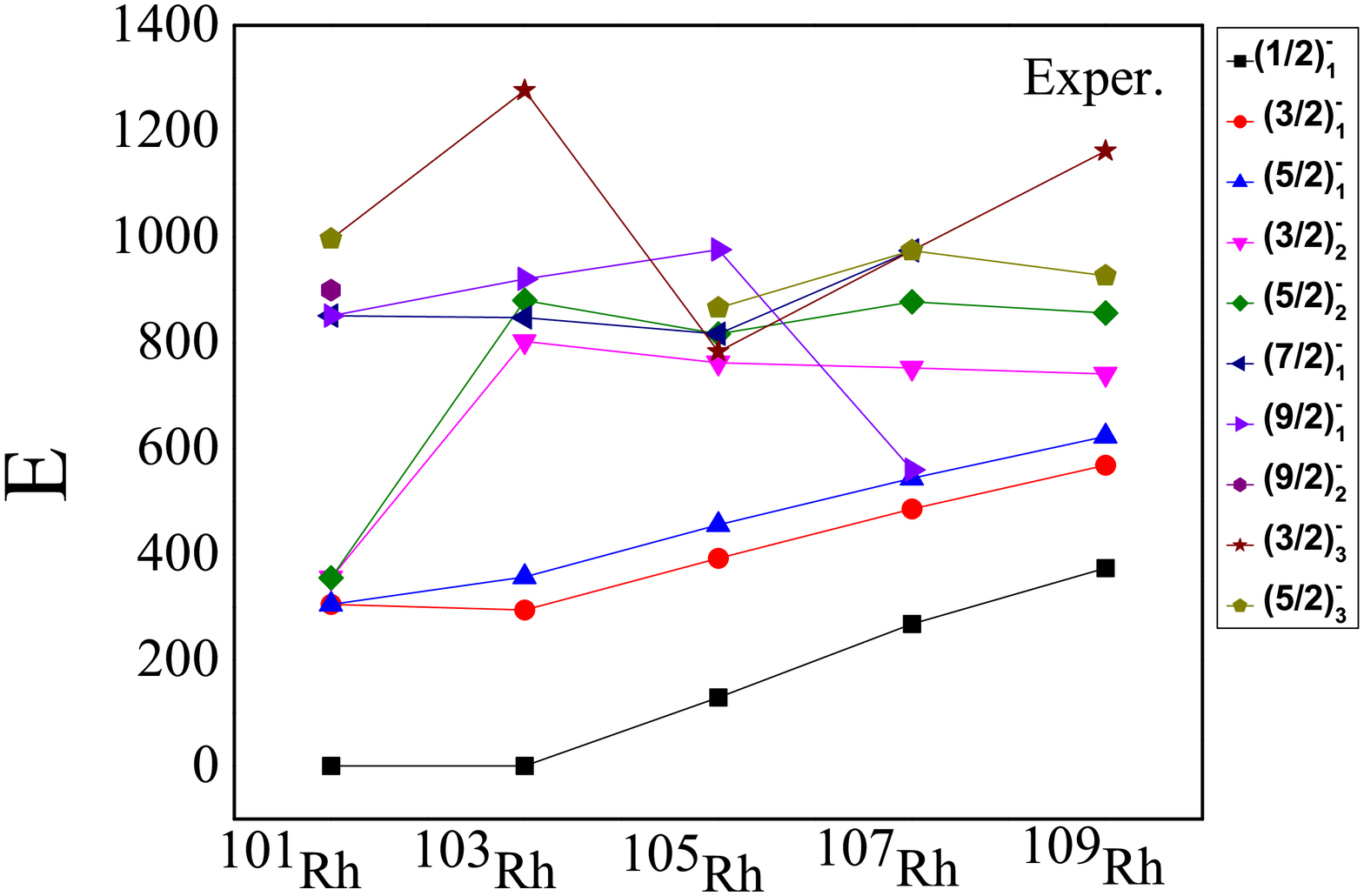}
\includegraphics[height=6cm]{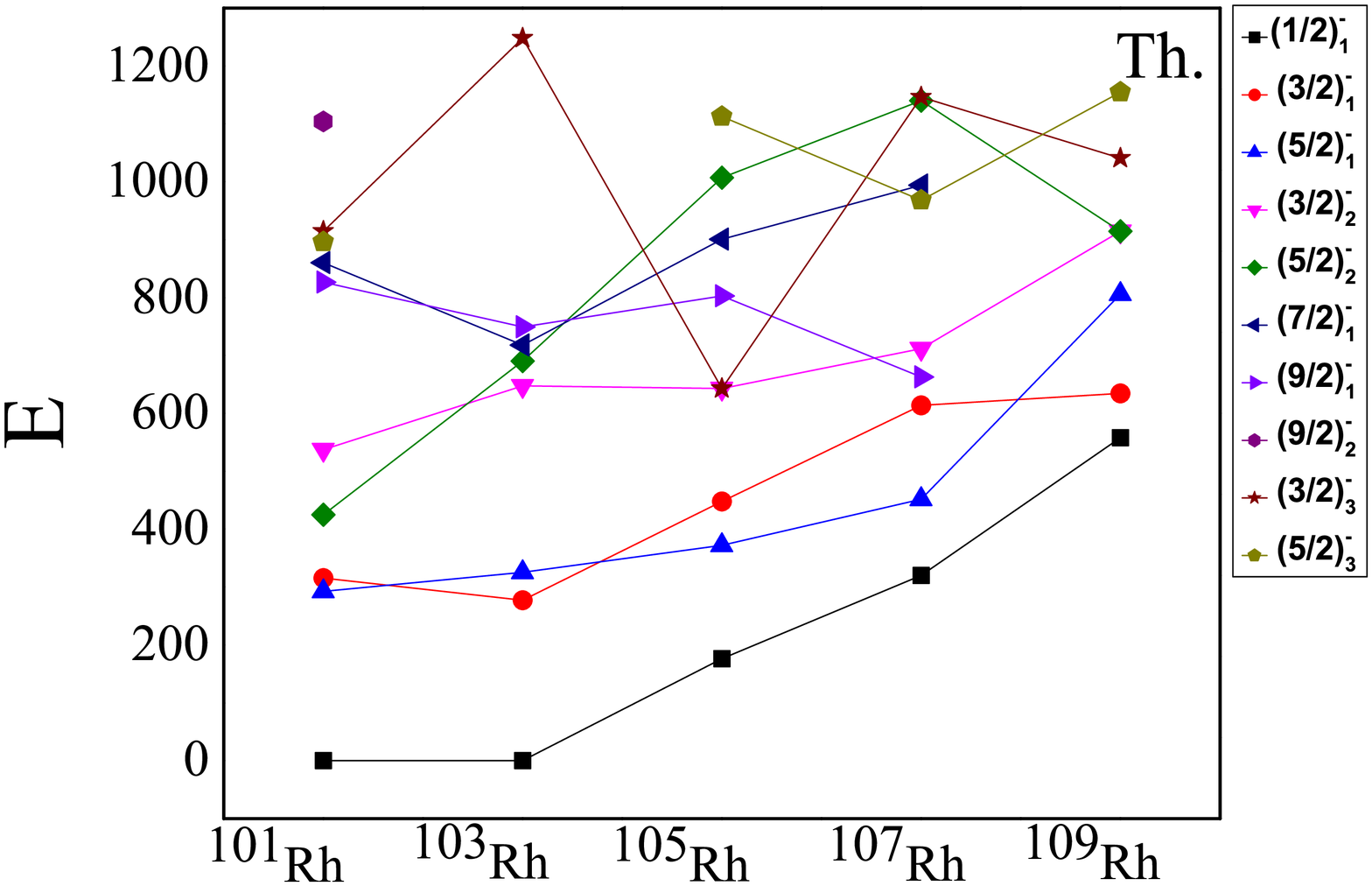}
\caption{Comparison between calculated and experimental spectra
of negative parity states in Rh Isotopes. The parameters of the
calculation are given in Tables 1. In the experimental spectra,
taken from \cite{27,29,31,33,35,36}.\label{fig:10}}
\end{center}
\end{figure}
\begin{figure}
\begin{center}
\includegraphics[height=6cm]{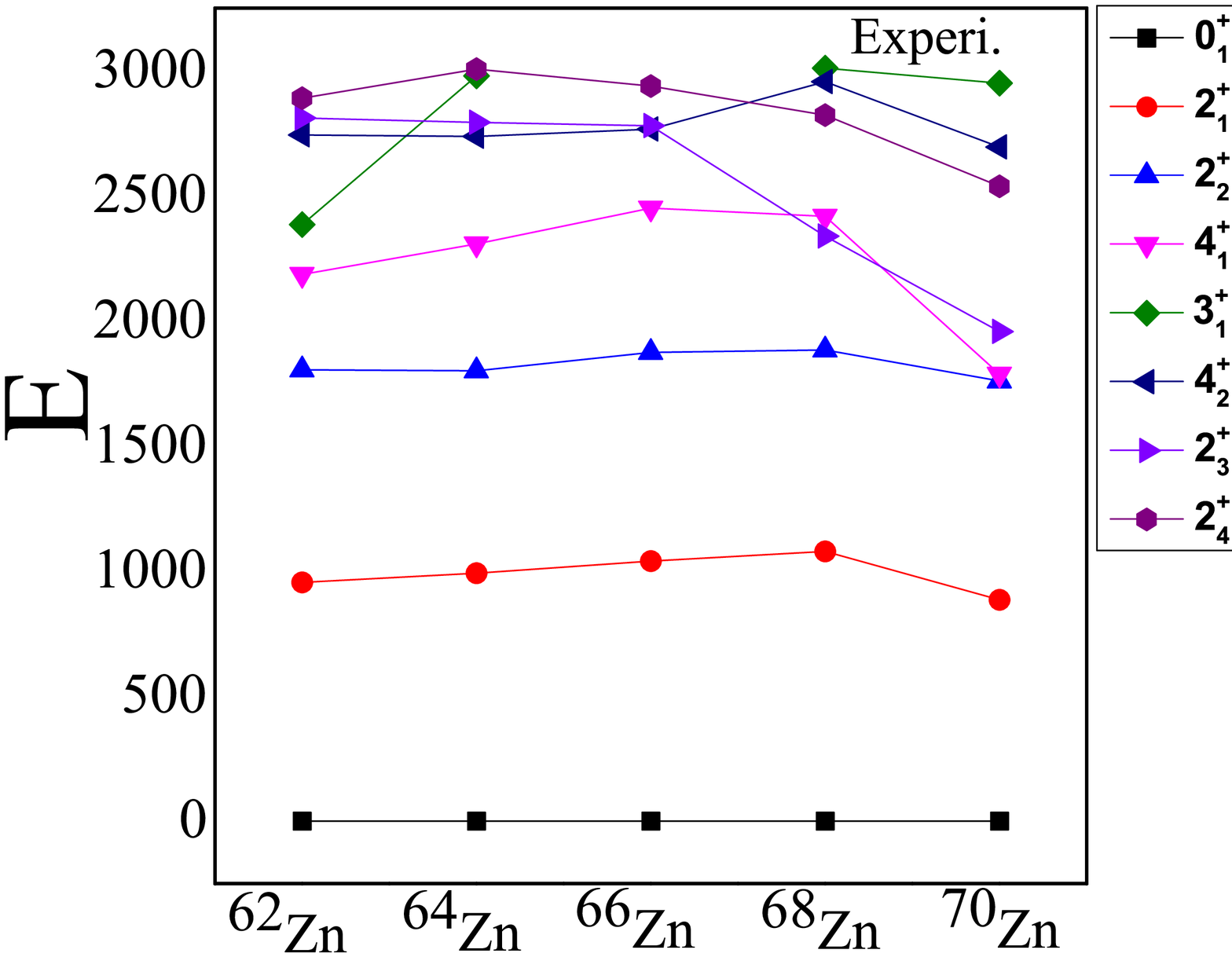}
\includegraphics[height=6cm]{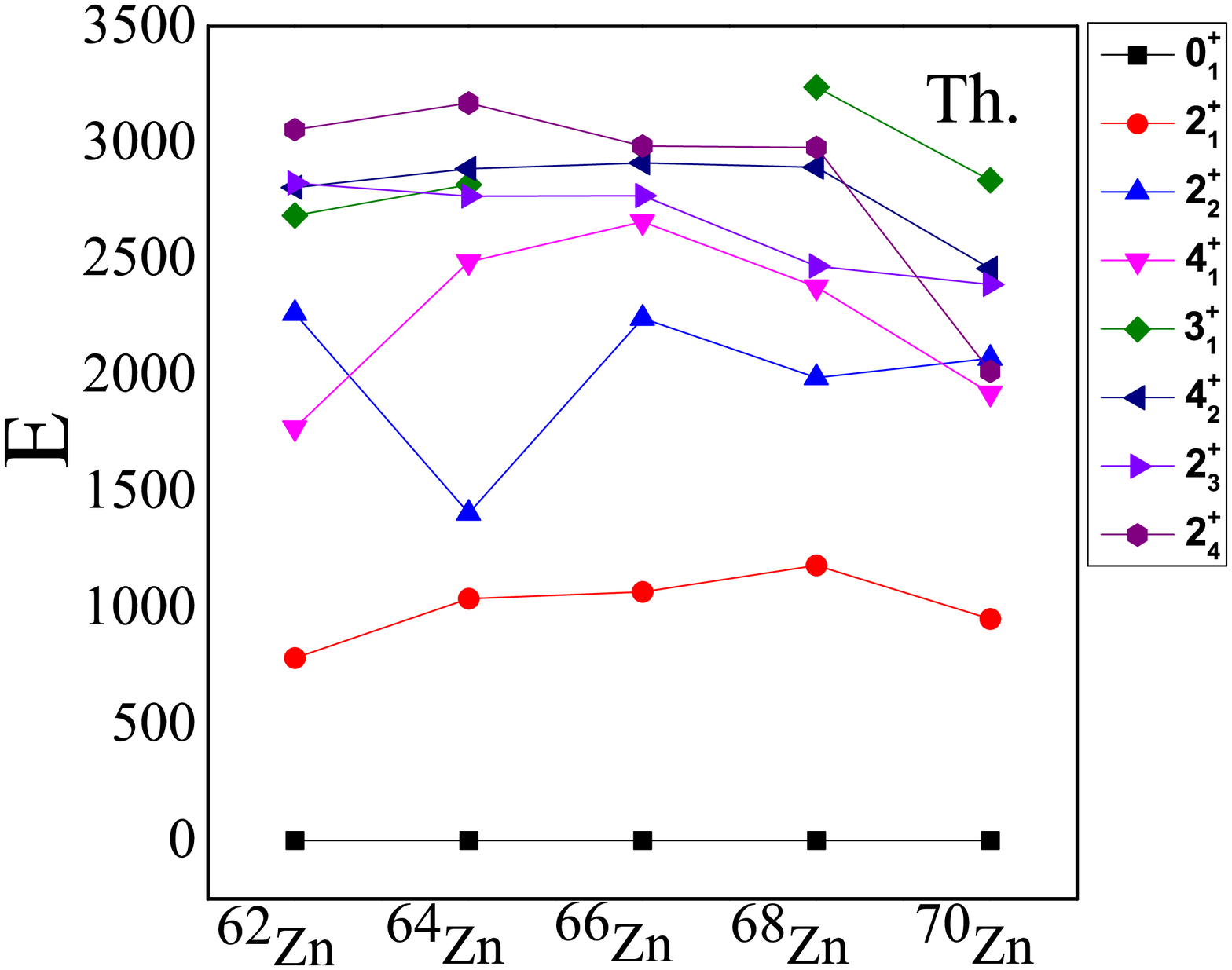}
\caption{Comparison between calculated and experimental spectra
of positive parity states in Zn Isotopes. The parameters of the
calculation are given in Tables 2. In the experimental spectra,
taken from \cite{41,43,45,47,49,36}.\label{fig:11}}
\end{center}
\end{figure}
\begin{figure}
\begin{center}
\includegraphics[height=6cm]{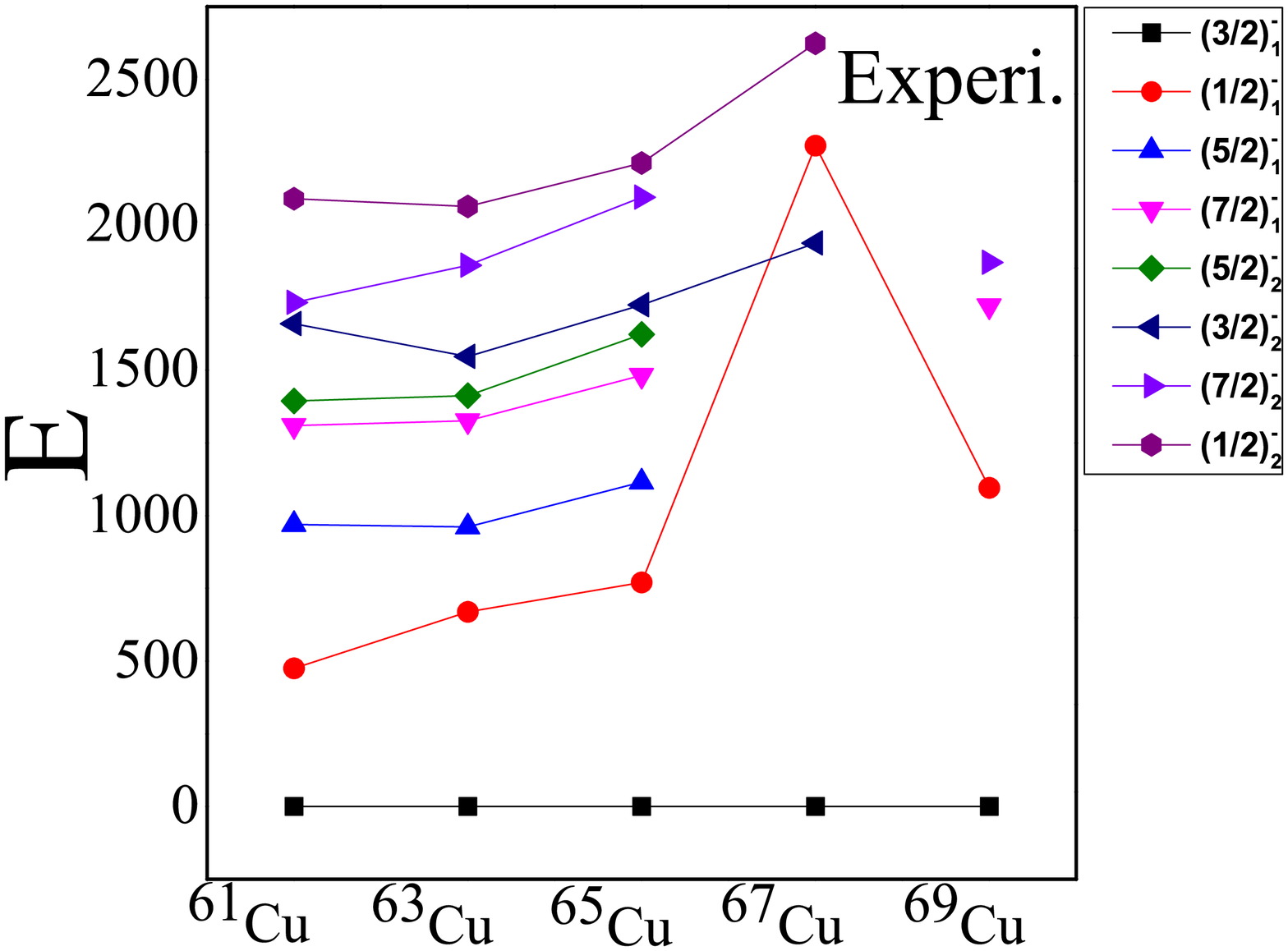}
\includegraphics[height=6cm]{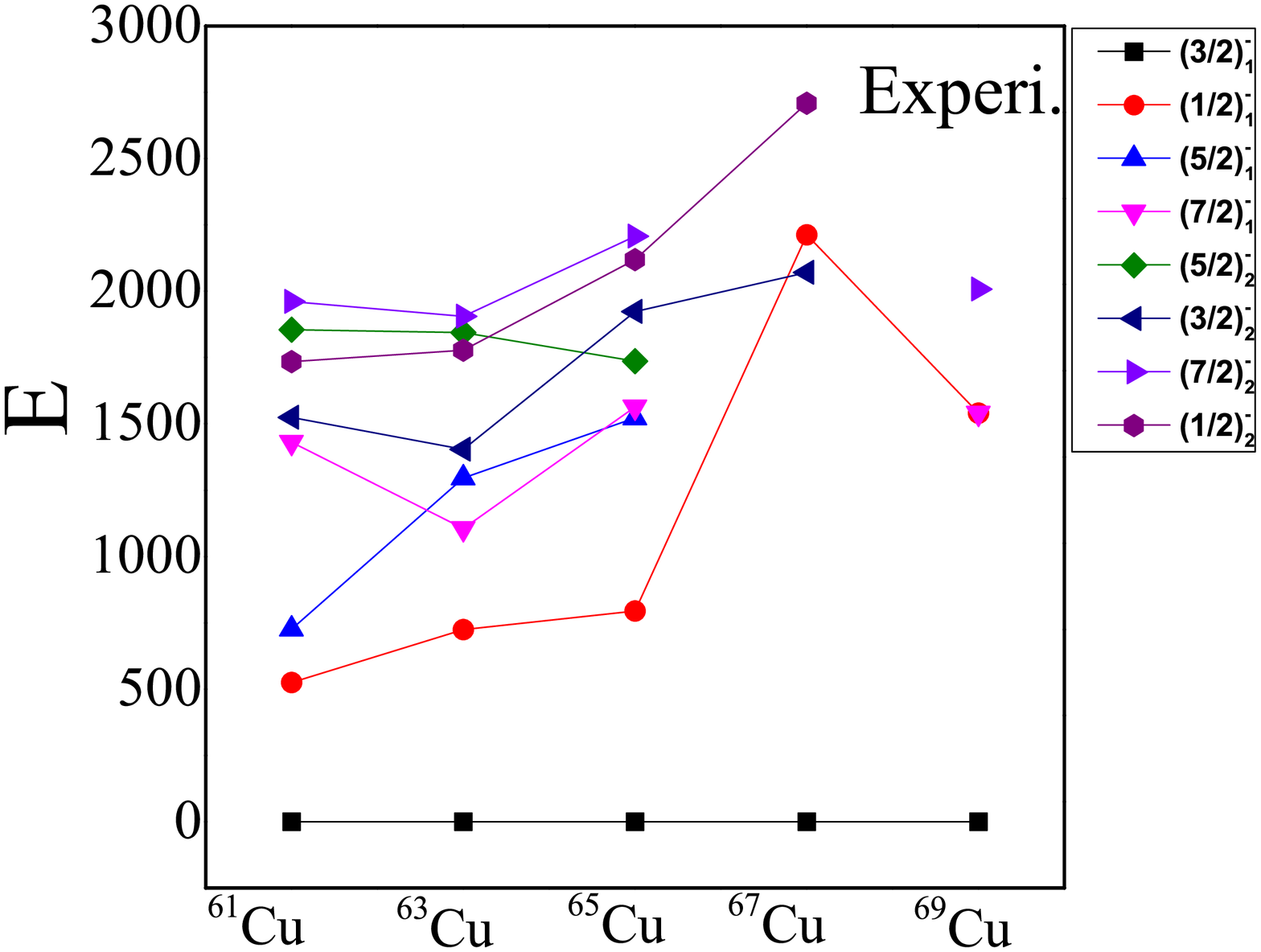}
\caption{Comparison between calculated and experimental spectra
of negative parity states in Cu Isotopes. The parameters of the
calculation are given in Tables 1. In the experimental spectra,
taken from \cite{40,42,44,46,48,36}.\label{fig:12}}
\end{center}
\end{figure}
\begin{figure}
\begin{center}
\includegraphics[height=6cm]{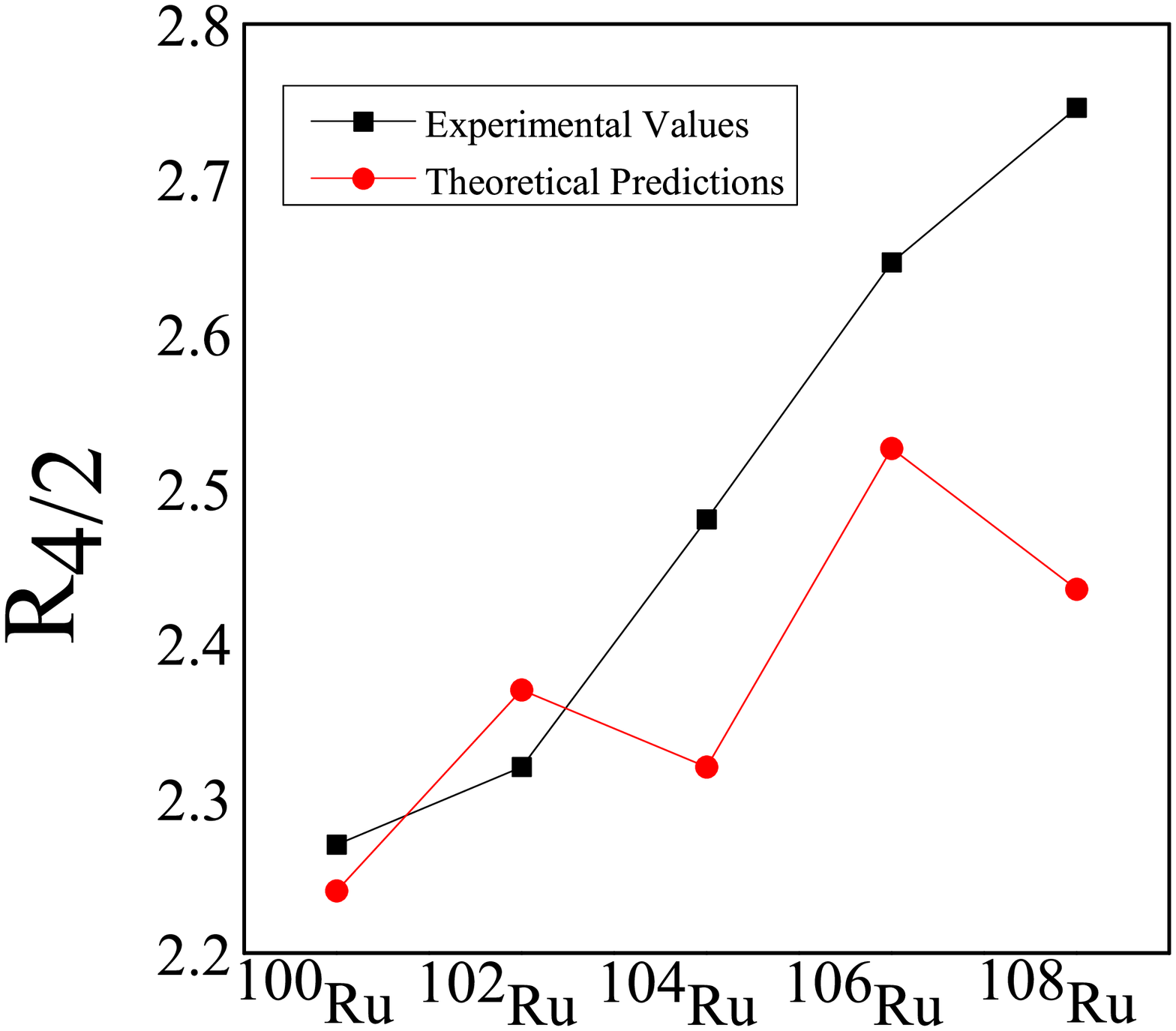}
\includegraphics[height=6cm]{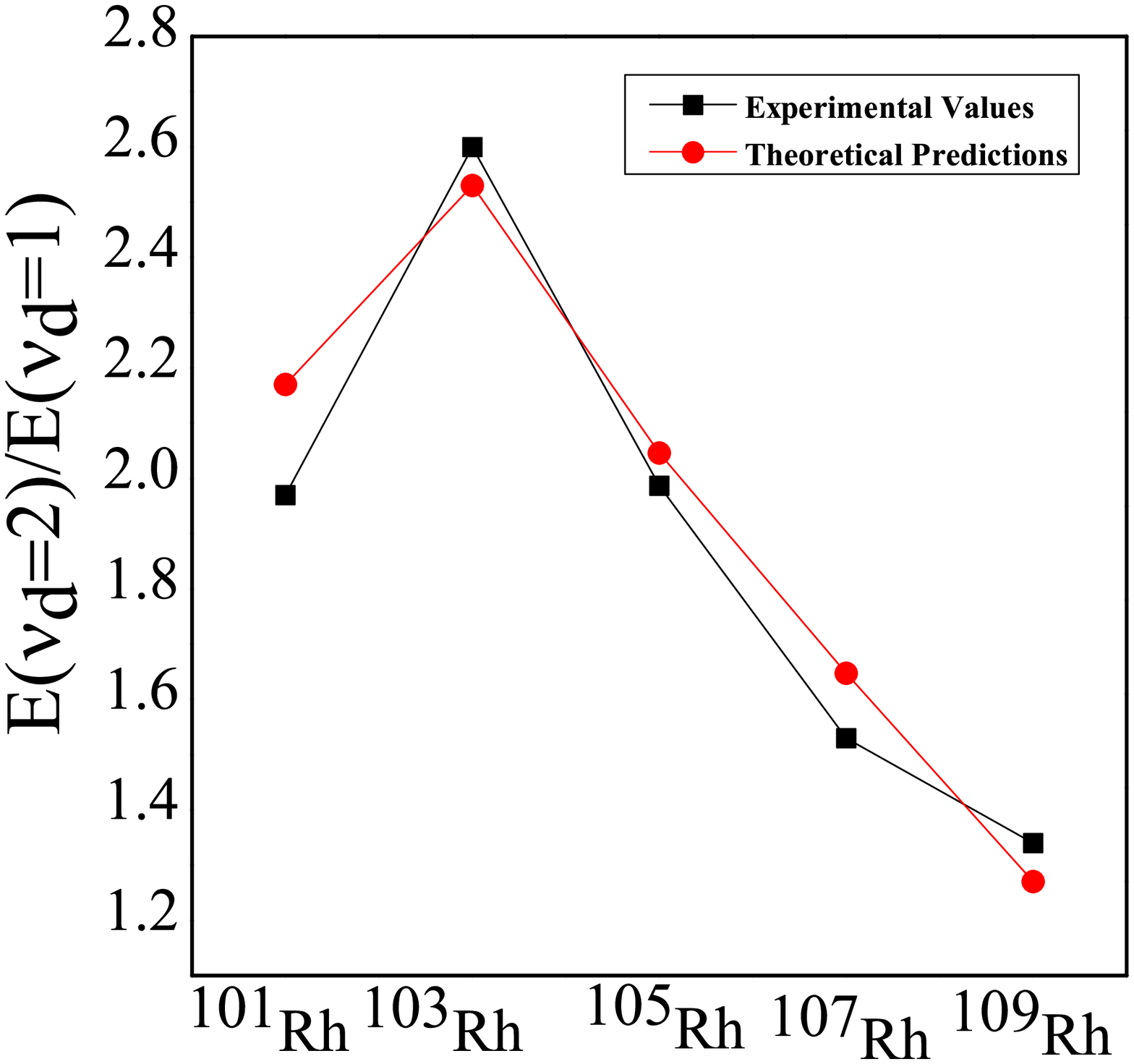}
\caption{$(E( \nu_{d}=2))/(E(\nu_{d}=1))$ prediction values for Ru and Rh
Isotopes. In the experimental spectra, taken from \cite{26,28,30,32,34,36}
\label{fig:13}}
\end{center}
\end{figure}
\begin{figure}
\begin{center}
\includegraphics[height=6cm]{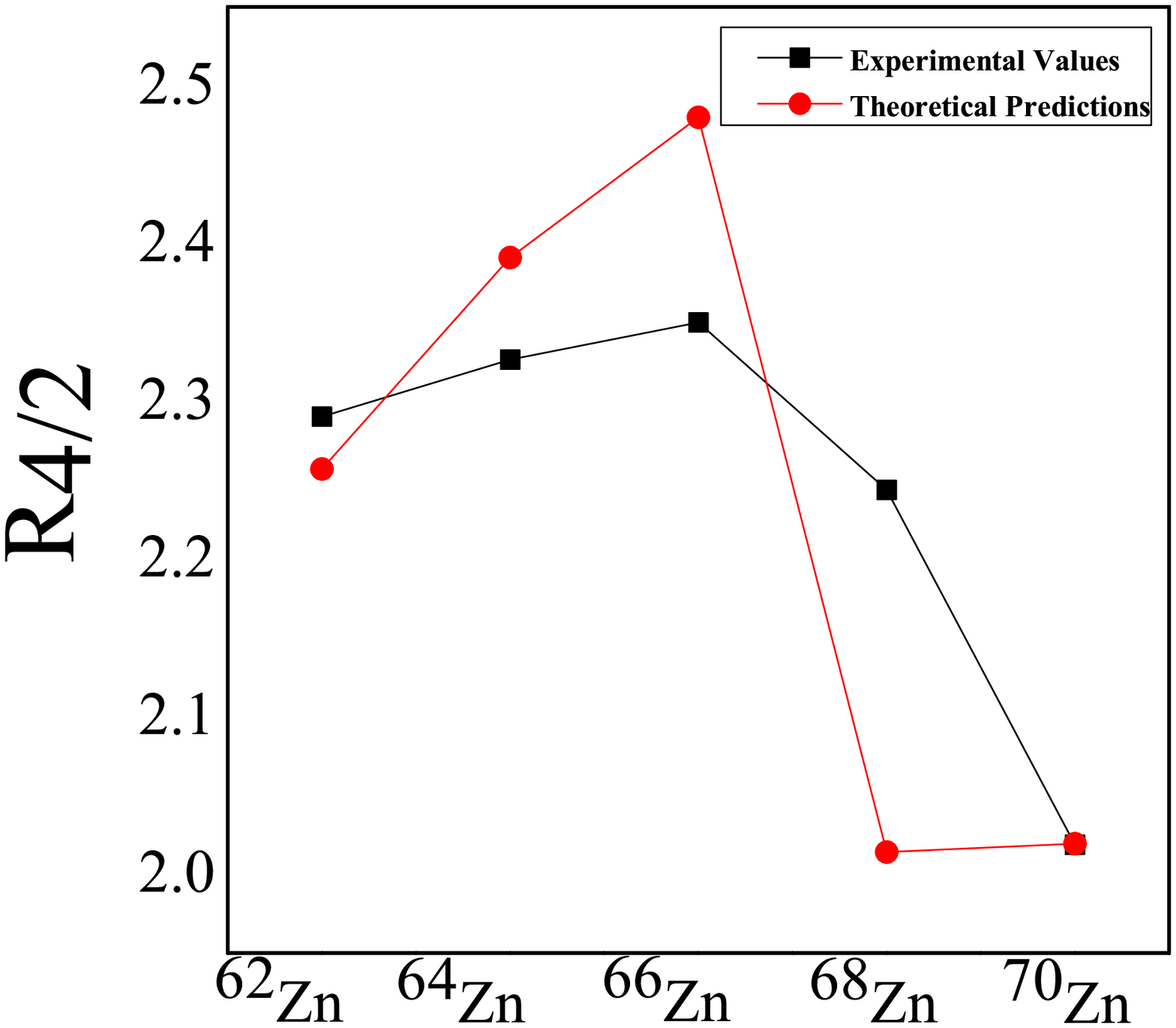}
\includegraphics[height=6cm]{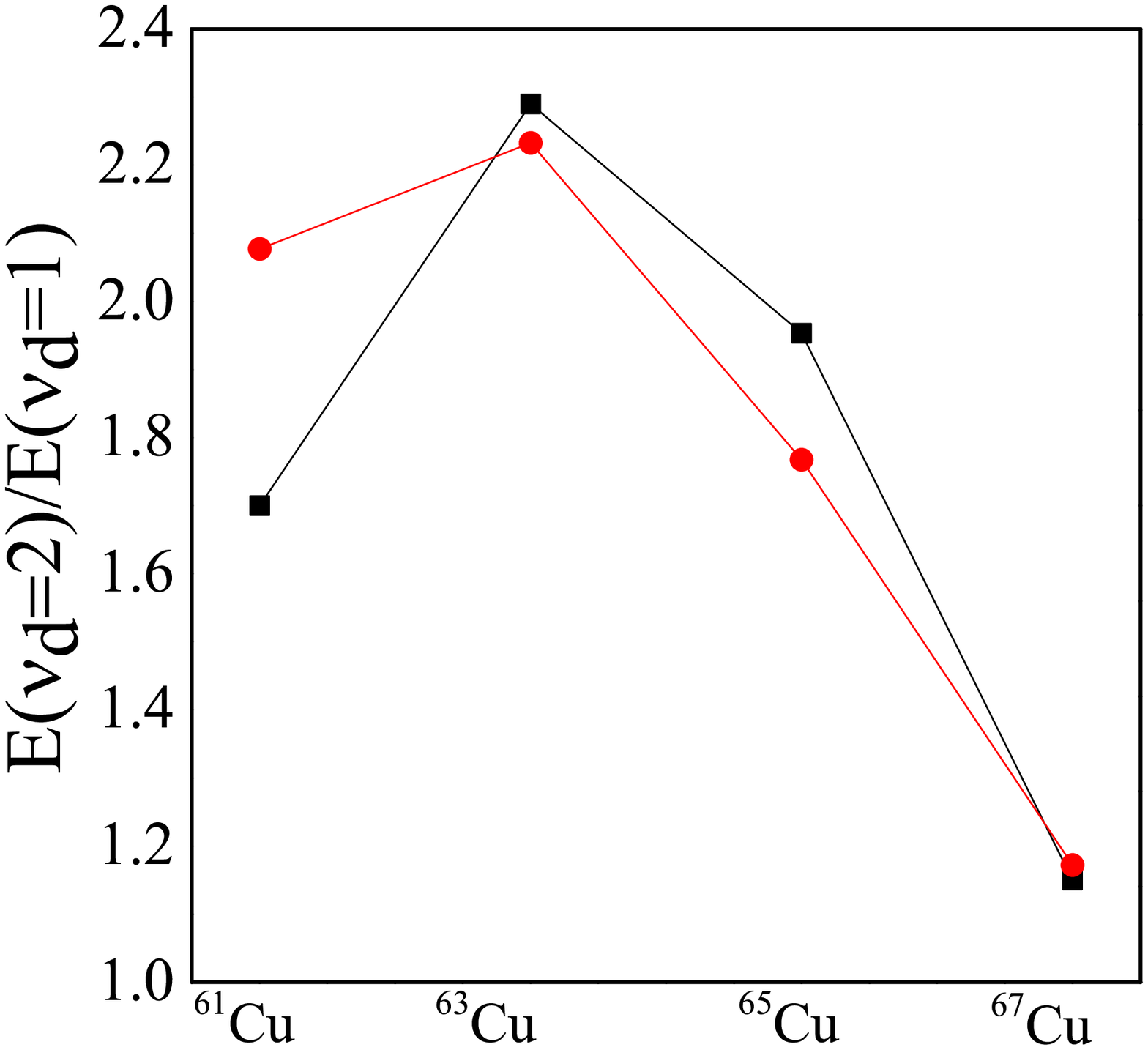}
\caption{$(E( \nu_{d}=2))/(E(\nu_{d}=1))$ prediction values for Zn and Cu
Isotopes. In the experimental spectra, taken from \cite{27,29,31,33,35,36}
\label{fig:14}}
\end{center}
\end{figure}
\begin{figure}
\begin{center}
\includegraphics[height=6cm]{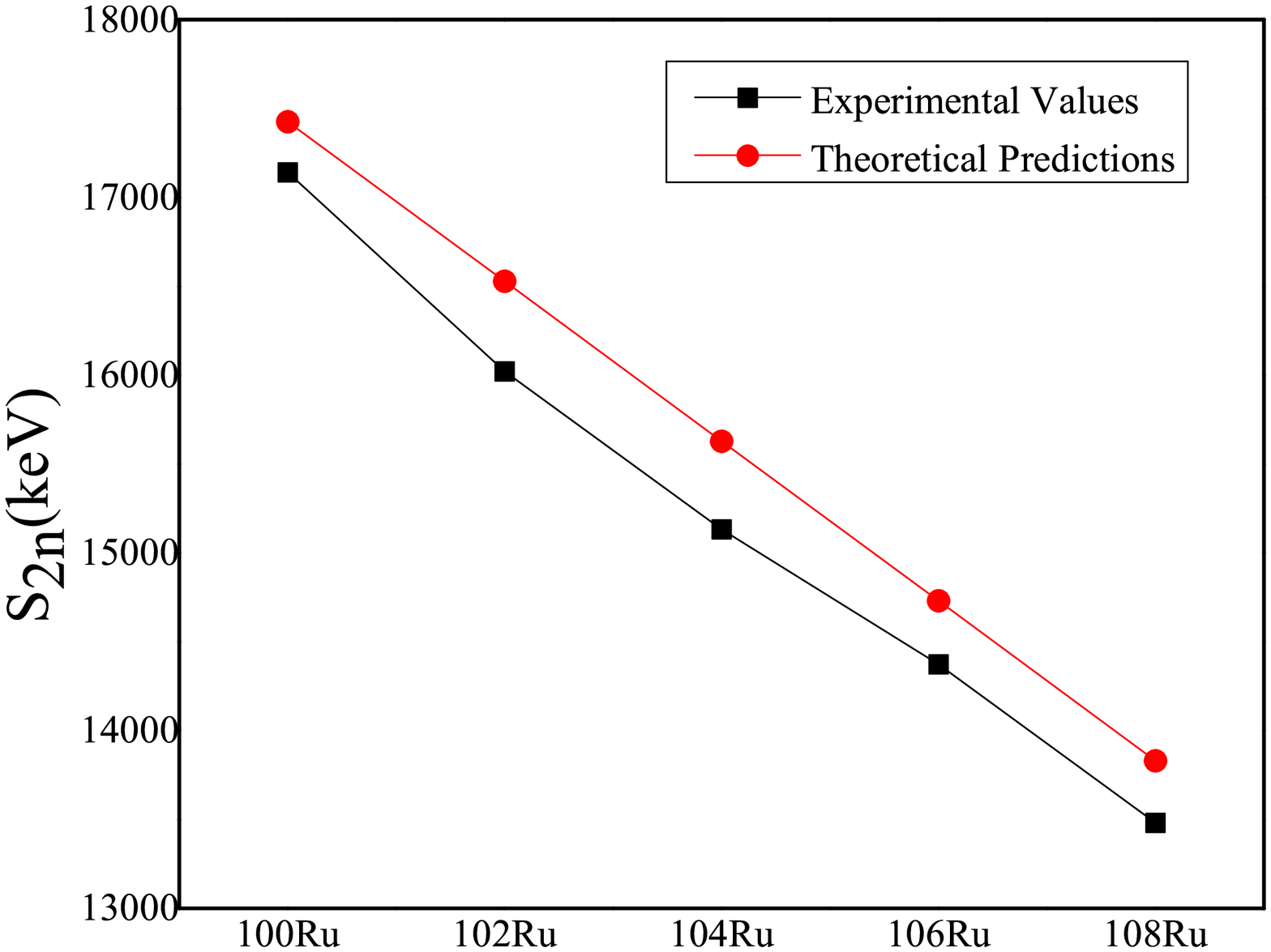}
\includegraphics[height=6cm]{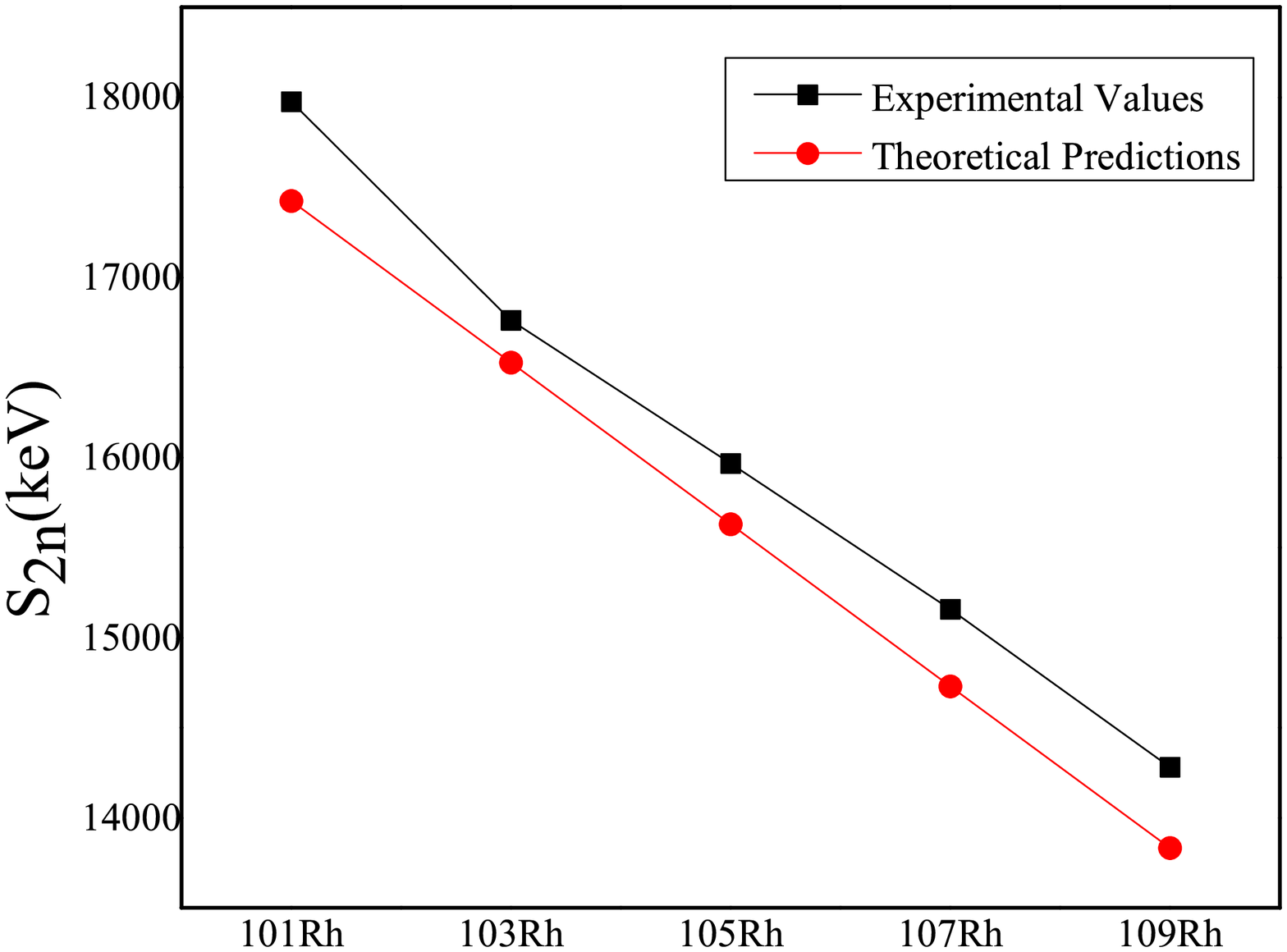}
\caption{A comparison between theoretical and experimental two
neutron separation energies,$S_{2n}$  (in keV) for Ru isotopes
(1eft panel) and Rh isotopes (right panel). Experimental data
from \cite{36}.\label{fig:15}}
\end{center}
\end{figure}
\begin{figure}
\begin{center}
\includegraphics[height=6cm]{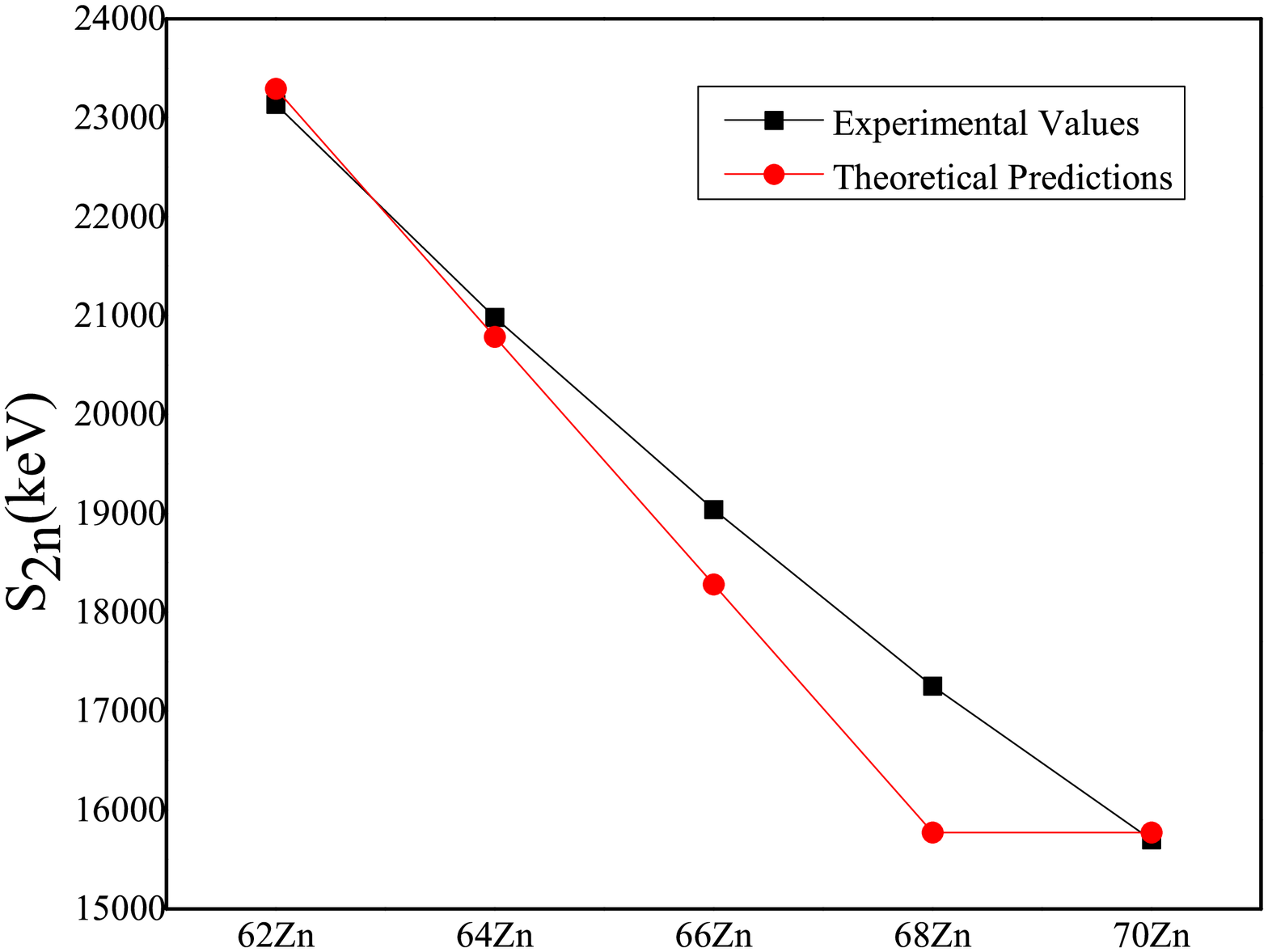}
\includegraphics[height=6cm]{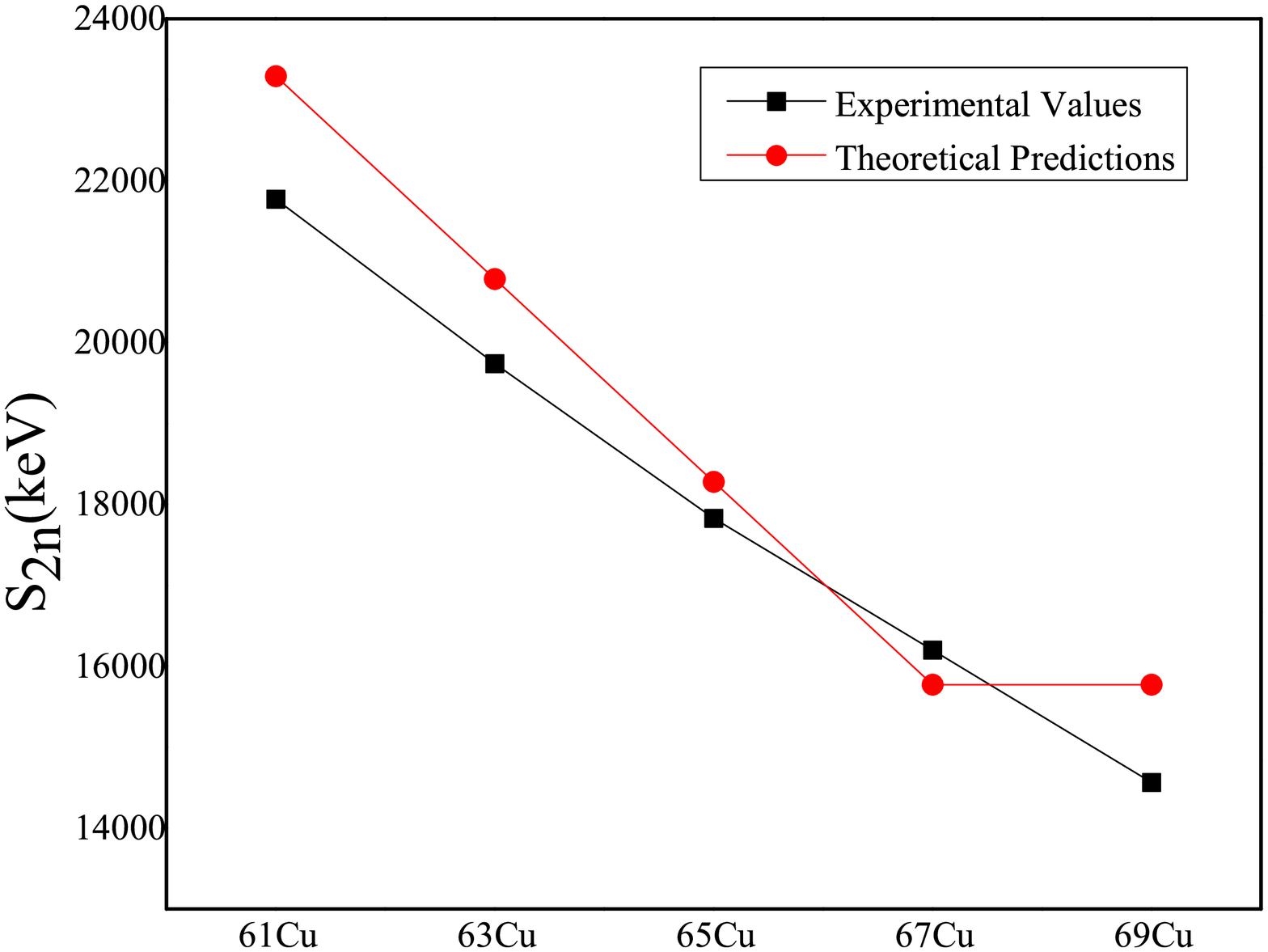}
\caption{A comparison between theoretical and experimental two
neutron separation energies,$S_{2n}$  (in keV) for Zn isotopes
(1eft panel) and Cu isotopes (right panel). Experimental data
from \cite{36}.\label{fig:16}}
\end{center}
\end{figure}
\end{document}